\documentclass[11pt]{article}
\usepackage{latexsym,amssymb,amstext,amsmath}
\usepackage{hyperref}
\allowdisplaybreaks

\def\Slash#1{\rlap{\hbox{$\mskip 3 mu /$}}#1}      
\newcommand{\ft}[2]{{\textstyle\frac{#1}{#2}}}

\begin{document}
%
\begin{titlepage}
\begin{flushright} \small
 ITP-UU-11/23 \\  Nikhef-2011-021
\end{flushright}
\bigskip

\begin{center}
 {\LARGE\bfseries Electric and magnetic charges
   \\[3mm]
   in N=2 conformal supergravity theories}
\\[10mm]
\textbf{Bernard de Wit$^{a,b}$ and Maaike van Zalk$^a$}\\[5mm]
\vskip 4mm $^a${\em Institute for Theoretical Physics, Utrecht University,}
  \\ {\em  Leuvenlaan 4, 3584 CE Utrecht, The Netherlands}\\[1mm]
$^b${\em Nikhef Theory Group, Science Park 105, 1098 XG Amsterdam, The
  Netherlands}\\[3mm]
{\tt  B.deWit@uu.nl}\;,\;{\tt M.vanZalk@uu.nl}
\end{center}

\vspace{3ex}

\begin{center}
{\bfseries Abstract}
\end{center}
\begin{quotation} \noindent General Lagrangians are constructed for
  N=2 conformal supergravity theories in four space-time dimensions
  involving gauge groups with abelian and/or non-abelian electric and
  magnetic charges. The charges are encoded in the gauge group
  embedding tensor. The scalar potential induced by the gauge
  interactions is quadratic in this tensor, and, when the embedding
  tensor is treated as a spurionic quantity, it is formally covariant
  with respect to electric/magnetic duality. This work establishes a
  general framework for studying any deformation induced by gauge
  interactions of matter-coupled N=2 supergravity theories. As an
  application, full and residual supersymmetry realizations in
  maximally symmetric space-times are reviewed. Furthermore, a general
  classification is presented of supersymmetric solutions in
  $\mathrm{AdS}_2\times S^2$ space-times. As it turns out, these
  solutions allow either eight or four supersymmetries. With four
  supersymmetries, the spinorial parameters are Killing spinors of
  $\mathrm{AdS}_2$ that are constant on $S^2$, so that they carry no
  spin, while the bosonic background is rotationally invariant.
\end{quotation}

\vfill

\end{titlepage}

\section{Introduction}
\label{sec:introduction} \setcounter{equation}{0}
In four space-time dimensions, Lagrangians with abelian gauge fields
have generically less symmetry than their corresponding equations of
motion. The full invariance group of the combined field equations and
Bianchi identities in principle involves a subgroup of the
electric/magnetic duality group, $\mathrm{Sp}(2n,\mathbb{R})$ for $n$
vector fields, suitably combined with transformations of the matter
fields. Subgroups of the symmetry group of the Lagrangian can be
gauged in the conventional way by introducing covariant derivatives
and covariant field strengths. Introducing gauge groups which involve
elements of the electric/magnetic duality group that do not belong to
the symmetry group of the Lagrangian, are not possible in this way.

To circumvent this problem, one may therefore first convert the
Lagrangian by an electric/magnetic equivalence transformation to a
different, but equivalent, Lagrangian that has the desired gauge group
as a symmetry.  However, this procedure is cumbersome. One reason for
this is that the gauge fields in the old and in the new
electric/magnetic duality frame are not generically related by local
field redefinitions. The effect of changing the duality frame is
therefore not straightforward, and it is by no means trivial to
explicitly obtain the new Lagrangian (see e.g.
\cite{deWit:2001pz}). A related aspect is that, when the gauge fields
belong to supermultiplets, their relation with other fields of the
multiplet will be affected by changes of the duality frame, unless one
simultaneously performs corresponding redefinitions of these fields as
well.\footnote{
  One way to circumvent this is by describing the scalar fields in
  terms of sections whose parametrization is linked to a specific
  frame (see, for instance,
  \cite{Andrianopoli:1996cm}). } 
The modern embedding tensor approach circumvents all these problems by
introducing, from the start, both electric and magnetic gauge fields
as well as tensor gauge fields. In this approach the gauge group is
not restricted to a subgroup of the invariance group of the
Lagrangian, but it must only be a subgroup of the symmetry group of
field equations and Bianchi identities.  The formalism is
straightforwardly applicable to any given Lagrangian, and the gauge
group is only restricted by two group-theoretical constraints on the
embedding tensor \cite{deWit:2005ub}.

In this paper we study general gaugings of $N=2$ supergravity theories
based on vector supermultiplets and hypermultiplets.  Because these
theories can generally be studied by means of the superconformal
multiplet calculus \cite{deWit:1980tn,deWit:1984pk,deWit:1984px}, it
suffices to understand the embedding tensor framework in the context
of conformal supergravity. This study is facilitated by the fact that
the embedding tensor framework has already been considered for rigid
$N=2$ supersymmetric gauge theories \cite{de Vroome:2007zd}, without
paying particular attention to the class of superconformally invariant
models. The purpose of the present paper is to fill this gap by
presenting a comprehensive treatment of the embedding tensor method in
the context of locally supersymmetric $N=2$ theories.

Theories with $N=2$ supersymmetry are special with respect to
electric/magnetic duality. For $N=1$ supersymmetry the
transformations of the matter fields under electric/magnetic
duality, and thus under the gauge group, are not a priori defined,
and will depend on the details of the model. On the other hand, in
theories with $N>2$ supersymmetries all of the matter fields are
closely linked to the vector fields, because they belong to common
supermultiplets. Theories with $N=2$ supersymmetries are
exceptional in that they exhibit both of these characteristic
features. The complex scalars belonging to the vector multiplets
transform in a well-defined way under electric/magnetic duality so
that the Lagrangian will retain its standard form expressed in
terms of a holomorphic function, while the scalars of the
hypermultiplets have no a priori defined transformations under
electric/magnetic duality. Prior to switching on the gauging, the
hypermultiplets are invariant under some rigid symmetry group that
is independent of the electric/magnetic duality group. Once the
gauge group has been embedded in the latter group, then one has to
separately specify its embedding into the symmetry group
associated with the hypermultiplets.

The embedding tensor approach of \cite{deWit:2005ub} makes use of both
electric and magnetic charges and their corresponding gauge
fields. The charges are encoded in terms of an embedding tensor, which
specifies the embedding of the gauge group into the full rigid
invariance group. This embedding tensor is treated as a spurionic
object (a quantity that is treated as a dynamical field, but that is
frozen to a constant at the end of the calculation), so that the
electric/magnetic duality structure of the ungauged theory is
preserved when the charges are turned on. Besides introducing a set of
dual magnetic gauge fields, also tensor gauge fields are required
transforming in the adjoint representation of the rigid invariance
group. These extra fields carry additional off-shell degrees of
freedom, but the number of physical degrees of freedom remains the
same owing to extra gauge transformations.  Prior to
\cite{deWit:2005ub} it had already been discovered that magnetic
charges tend to be accompanied by tensor fields. An early example of
this was presented in \cite{Louis:2002ny}, and subsequently more
theories with magnetic charges and tensor fields were constructed, for
instance, in \cite{Dall'Agata:2003yr,Sommovigo:2004vj,DAuria:2004yi},
mostly in the context of abelian gauge groups. The embedding tensor
approach has already been explored for many supersymmetric theories in
four space-time dimensions.  For instance, it was successfully applied
to $N=4$ supergravity \cite{Schon:2006kz} and to $N=8$ supergravity
\cite{de Wit:2007mt}. More recently it has also been discussed for
$N=1$ supergravity \cite{Hartong:2009az}.  In \cite{de Vroome:2007zd}
some applications to $N=2$ supergravity were already presented, under
the assumption that the conformal multiplet calculus
\cite{deWit:1980tn,deWit:1984pk,deWit:1984px} is applicable. As it
turned out, the results of the embedding tensor approach confirm
and/or clarify various previous results in the literature, especially
for abelian gaugings \cite{Antoniadis:1996,Behrndt:2001mx}. The
embedding tensor is ideally suited for the study of flux
compactifications in string theory (for a review, see
\cite{Grana:2005jc}). Recently it was successfully employed in a study
of partial breaking of $N=2$ to $N=1$ supersymmetry
\cite{Louis:2009xd,Louis:2010ui}.

The supersymmetric Lagrangians derived in this paper incorporate
gaugings in both the vector and hypermultiplet sectors. The vector
multiplets are initially defined as off-shell multiplets, but the
presence of the magnetic charges causes a breakdown of off-shell
supersymmetry. Of course, conventional hypermultiplets based on a
finite number of fields will not constitute an off-shell
representation of the supersymmetry algebra irrespective of the
presence of charges. We refer to a more in-depth discussion of the
off-shell aspects of the embedding tensor method in \cite{de
  Vroome:2007zd}, where a construction was presented in which the
tensor fields associated with the magnetic charges were contained in a
tensor supermultiplet.

Besides giving a comprehensive treatment of the embedding tensor
formalism in the context of local $N=2$ supersymmetric theories, we
also present two applications to illustrate how the embedding tensor
formalism can be used to obtain rather general results about
realizations of $N=2$ gauged supergravities. One concerns the
supersymmetric realizations in maximally symmetric spaces. In flat
Minkowski space, it was established that residual supersymmetry is
only possible in the presence of magnetic charges
\cite{Cecotti:1984rk,Cecotti:1985sf,Ferrara:1995xi,
  Ferrara:1995gu,Fre:1996js}. Here, we therefore briefly review the
situation in the context of the embedding tensor approach, where it is
natural to have both electric and magnetic charges.

A second application deals with supersymmetric solutions in
$\mathrm{AdS}_2\times S^2$ space-times. Here we establish that there
exist only two classes of supersymmetric solutions. One concerns fully
supersymmetric solutions. It contains the solutions described in
\cite{Hristov:2009uj} as well as the near-horizon solution of ungauged
supergravity that appears for BPS black holes. The other class
exhibits four supersymmetries and these solutions may appear as
near-horizon geometries of BPS black holes in $N=2$ gauged
supergravity. Interestingly enough, solutions in $\mathrm{AdS}_2\times
S^2$ with only two supersymmetries are excluded. The spinor parameters
associated with the four supersymmetries are $\mathrm{AdS}_2$ Killing
spinors that are constant on $S^2$, so that they carry no
spin. Nevertheless the bosonic background is rotationally invariant.
The spin assignments change in this background, because the spin
rotations associated with the $S^2$ isometries become entangled with
R-symmetry transformations, a phenomenon that is somewhat similar to
what happens for magnetic monopole solutions where the rotational
symmetry becomes entangled with gauge transformations
\cite{Hasenfratz:1976gr}.  In the superconformal perspective, these
solutions have R-symmetry connections living on $S^2$, and this
explains the geometric origin of the entanglement. It is to be
expected that the near-horizon geometry of a recently presented
static, spherically symmetric, black hole solution
\cite{Dall'Agata:2010gj,Hristov:2010ri} will coincide with one of the solutions
described in this paper. The results of this paper then imply that
this black hole solution must exhibit supersymmetry enhancement at the
horizon.

This paper is organized as follows. In section
\ref{vector-multiplets-duality} we recall the relevant features of
$N=2$ vector multiplets and electric/magnetic duality in the context
of conformal supergravity, and we introduce the electric and magnetic
gauge fields. Hypermultiplets, hyperk\"ahler cones and their
isometries are introduced in a superconformal setting in section
\ref{sec:intro-hypermultiplets}. In section \ref{sec:lagrangian} we
present the relevant Lagrangians for matter fields coupled to
conformal supergravity. Section \ref{sec:gauge-transf-charges}
contains a discussion of the possible gauge transformations, the
electric and magnetic charges, and the embedding tensor. In section
\ref{sec:gauge-hierarchy} we describe the introduction of tensor
fields, needed in the presence of general charge assignments.  Section
\ref{sec:rest-supersymm-non} deals with the algebra of superconformal
transformations in the presence of a gauging. It presents the extra
masslike terms and the scalar potential in the vector multiplet and
hypermultiplet Lagrangians that are induced by these
gaugings. Finally, in section \ref{sec:summary-applications} we
summarize our results and review two applications. Readers who are not
primarily interested in the more technical details of the embedding
tensor formalism, can proceed directly to this section. We have
refrained from collecting additional information in an appendix and
refer instead to the appendices presented in \cite{deWit:2010za}.

\section{Superconformal vector multiplets and electric/magnetic
  duality}
\label{vector-multiplets-duality} \setcounter{equation}{0}
Vector supermultiplets in four space-time dimensions with $N=2$
supersymmetry can be defined in a superconformal background.
Consider $n+1$ of these multiplets, labeled by indices
$\Lambda=0,1,\ldots,n$. Vector supermultiplets comprise complex
scalar fields $X^\Lambda$, gauge fields $W_\mu{}^\Lambda$, and
Majorana spinors which are conveniently decomposed into chiral and
anti-chiral components: spinors $\Omega_i{}^\Lambda$ have
positive, and spinors $\Omega^{i\Lambda}$ have negative chirality
(so that $\gamma^5 \Omega_i{}^\Lambda= \Omega_i{}^\Lambda$ and
$\gamma^5 \Omega^{i\Lambda}= -\Omega^{i\Lambda}$). The spinors
carry indices $i=1,2$, and transform as doublets under the
R-symmetry group $\mathrm{SU}(2)$. This group is realized locally
with gauge fields belonging to the superconformal background, as
we shall discuss below. Furthermore there are auxiliary fields
$Y_{ij}{}^\Lambda$, which satisfy the pseudo-reality constraint
$(Y_{ij}{}^\Lambda)^\ast = \varepsilon^{ik}\varepsilon^{jl}
Y_{kl}{}^\Lambda$, so that they transform as real vectors under
$\mathrm{SU}(2)$. The tensors $F^{\pm}_{\mu\nu}{}^\Lambda$ are the
(anti-)selfdual (complex) components of the field strengths, which
will be expressed in terms of vector fields $W_\mu{}^\Lambda$. The
supersymmetry transformations of these fields will depend on the
superconformal background.

Before presenting the supersymmetry transformations of the vector
multiplets, we first specify the superconformal background fields,
which comprise the so-called Weyl supermultiplet, and their relation
to the superconformal transformations. The latter contains the
generators of general-coordinate, local Lorentz, dilatation, special
conformal, chiral $\mathrm{SU}(2)$ and $\mathrm{U}(1)$, supersymmetry
(Q) and special supersymmetry (S) transformations.  The gauge fields
associated with general-coordinate transformations ($e_\mu{}^a$),
dilatations ($b_\mu$), chiral symmetry ($\mathcal{V}_\mu{}^i{}_j$ and
$A_\mu$) and Q-supersymmetry ($\psi_\mu{}^i$) are independent
fields. The remaining gauge fields associated with the Lorentz
($\omega_\mu{}^{ab}$), special conformal ($f_\mu{}^a$) and
S-supersymmetry transformations ($\phi_\mu{}^i$) are dependent
fields. They are composite objects, which depend on the independent
fields of the multiplet
\cite{deWit:1980tn,deWit:1984pk,deWit:1984px}. The corresponding
supercovariant curvatures and covariant fields are contained in a
tensor chiral multiplet, which comprises $24+24$ off-shell degrees of
freedom. In addition to the independent superconformal gauge fields,
it contains three other fields: a Majorana spinor doublet $\chi^i$, a
scalar $D$, and a selfdual Lorentz tensor $T_{abij}$, which is
anti-symmetric in $[ab]$ and $[ij]$. We refer to the appendices in
\cite{deWit:2010za} for an extended summary of the superconformal
transformations of the Weyl multiplet fields, the expressions for the
curvatures and other useful details.

The transformations of the vector multiplet fields under dilatations
and chiral transformations are given in table
\ref{tab:weights-vm}. Under local Q- and S-supersymmetry they are as
follows \cite{deWit:1980tn},
\begin{align}
  \label{eq:susyr}
  \delta X^{\Lambda}  = &\, \bar{\epsilon}^i \Omega_i^{\; \Lambda}\,,
  \,\nonumber\\
  \delta W_{\mu}{}^{\Lambda}  = &\, \varepsilon^{ij} \bar{\epsilon}_i
  (\gamma_{\mu} \Omega_j{}^{\Lambda}+2\,\psi_{\mu j} X^\Lambda)
  + \varepsilon_{ij}
  \bar{\epsilon}^i (\gamma_{\mu} \Omega^{j\, \Lambda} +2\,\psi_\mu{}^j
  \bar X^\Lambda)   \,,\nonumber\\
  \delta \Omega_i{}^{\Lambda} = &\, 2 \Slash{D}
  X^{\Lambda} \epsilon_i + \ft12 \gamma^{\mu \nu}
  \hat F^-_{\mu\nu}{}^\Lambda \varepsilon_{ij} \epsilon^j +
  Y_{ij}{}^{\Lambda} \epsilon^j + 2\, X^\Lambda \eta_i  \,,
  \nonumber\\
  \delta Y_{ij}{}^{\Lambda} = &\, 2\, \bar{\epsilon}_{(i}
  \Slash{D}\Omega_{j)}{}^{\Lambda} + 2\, \varepsilon_{ik}
  \varepsilon_{jl}\, \bar{\epsilon}^{(k} \Slash{D}\Omega^{l)
  \Lambda} \,.
\end{align}
Here $\epsilon^i$ and $\epsilon_i$ denote the spinorial parameters
of Q-supersymmetry and $\eta^i$ and $\eta_i$ those of
S-supersymmetry. The field strengths $F_{\mu\nu}{}^\Lambda=
2\,\partial_{[\mu}W_{\nu]}{}^\Lambda$ are contained in the
supercovariant combination,
\begin{align}
  \label{eq:f-hat}
  \hat F_{\mu\nu}{}^\Lambda =&\,  F^+_{\mu\nu}{}^\Lambda +
  F^-_{\mu\nu}{}^\Lambda
  - \varepsilon^{ij} \bar{\psi}_{[\mu\,i}
  (\gamma_{\nu]} \Omega_j{}^{\Lambda}+ \psi_{\nu] j} X^\Lambda)
  -  \varepsilon_{ij}
  \bar{\psi}_{[\mu}{}^i (\gamma_{\nu]} \Omega^{j\, \Lambda}
  +\psi_{\nu]}{}^j \bar X^\Lambda) \nonumber\\
  &\,
  - \tfrac14(X^\Lambda\, T_{\mu\nu ij}\,\varepsilon^{ij} + \bar X^\Lambda\,
  T_{\mu\nu}{}^{ij}\,\varepsilon_{ij}  )\,.
\end{align}
The full superconformally covariant derivatives are denoted by
$D_\mu$, while $\mathcal{D}_\mu$ will denote a covariant
derivative with respect to Lorentz, dilatation, chiral
$\mathrm{U}(1)$, and $\mathrm{SU}(2)$ transformations. As an
example of the latter, we note the definitions,
\begin{align}
  \label{eq:ex-cov-der}
  \mathcal{D}_\mu X^\Lambda=&\, \big(\partial_\mu -b_\mu + \mathrm{i}
    A_\mu \big) X^\Lambda \, ,\nonumber\\
    \mathcal{D}_\mu \Omega_i{}^\Lambda =&\, \big(\partial_\mu -\ft14
      \omega_\mu{}^{ab}\gamma_{ab} -\ft32 b_\mu +\ft12 \mathrm{i}
      A_\mu\big)\Omega_i{}^\Lambda-\ft12 \mathcal{V}_\mu{}^j{}_i\,
    \Omega_j{}^\Lambda \, .
\end{align}

\begin{table}[t]
\begin{center}\begin{tabular}{|c||cccc|}
 \hline field & $X^M$ & $\Omega_i{}^M$ & $W_{\mu}{}^M$ &
 $Y_{ij}{}^\Lambda$ \\\hline\hline
  w & $1$ & $\frac{3}{2}$ & 0 & 2  \\\hline
  c & $-1$ & $-\frac{1}{2}$ & 0 & 0  \\ \hline
\end{tabular}
\end{center}\caption{Weyl and chiral weights of the vector multiplet
  fields.}\label{tab:weights-vm}\end{table}

We now assume an holomorphic function $F(X)$ of the fields
$X^\Lambda$, which is homogeneous of second degree, i.e.
$F(\lambda X)= \lambda^2 F(X)$, for any complex parameter
$\lambda$. As is well known \cite{de Wit:1983rz,deWit:1984pk},
such a function can be used to write down a consistent action for
the vector multiplets in the superconformal background provided by
the Weyl multiplet fields. Rather than to determine this action,
we first consider an extension of the field representation that
will facilitate the treatment of electric/magnetic duality in the
presence of non-zero gauge charges. Since this duality ultimately
involves the equations of motion, it will be essential that the
action exists, but for the purpose of this section it is not
necessary to display its precise form.

In the absence of charged fields, abelian gauge fields
$W_\mu{}^\Lambda$ appear exclusively through the field strengths,
${F}_{\mu\nu}{}^\Lambda = 2\,\partial_{[\mu}W_{\nu]}{}^\Lambda$.
The field equations for these fields and the Bianchi identities
for the field strengths comprise $2(n+1)$ equations,
\begin{equation}
  \label{eq:eom-bianchi}
  \partial_{[\mu} {F}_{\nu\rho]}{}^\Lambda
   = 0 = \partial_{[\mu} {G}_{\nu\rho]\,\Lambda} \,,
\end{equation}
where
\begin{equation}
  \label{eq:def-G}
  {G}_{\mu\nu\,\Lambda} =\mathrm{i} e\,
  \varepsilon_{\mu\nu\rho\sigma}\,
  \frac{\partial \mathcal{L}}{\partial{F}_{\rho\sigma}{}^\Lambda}
  \;.
\end{equation}
At this point we cannot give the form of $G_{\mu\nu\Lambda}$,
because we have not yet specified the action. Instead, we will
extract its definition below by using supersymmetry.

It is convenient to combine the tensors $F_{\mu\nu}{}^\Lambda$ and
$G_{\mu\nu\Lambda}$ into a $(2n+2)$-dimensional vector,
\begin{equation}
  \label{eq:GM}
  G_{\mu\nu}{}^M= \begin{pmatrix}{F}_{\mu\nu}{}^\Lambda\cr
    \noalign{\vskip 1.5mm}
  {G}_{\mu\nu\Lambda}
\end{pmatrix} \,,
\end{equation}
so that (\ref{eq:eom-bianchi}) reads $\partial_{[\mu}
{G}_{\nu\rho]}{}^M = 0$. Obviously these $2(n+1)$ equations are
invariant under real $2(n+1)$-dimensional rotations of the tensors
$G_{\mu\nu}{}^M$,
\begin{equation}
  \label{eq:em-duality}
  \begin{pmatrix} {F}^\Lambda\cr \noalign{\vskip 1.5mm} {G}_\Lambda
  \end{pmatrix}
  \longrightarrow
  \begin{pmatrix} U^\Lambda{}_\Sigma & Z^{\Lambda\Sigma} \cr
    \noalign{\vskip 1.5mm}
    W_{\Lambda\Sigma} & V_\Lambda{}^\Sigma
    \end{pmatrix}
  \begin{pmatrix}{F}^\Sigma\cr \noalign{\vskip 1.5mm} {G}_\Sigma
  \end{pmatrix}
    \,.
\end{equation}
Half of the rotated tensors can be adopted as new field strengths
defined in terms of new gauge fields, and the Bianchi identities on
the remaining tensors can then be interpreted as field equations
belonging to some new Lagrangian expressed in terms of the new field
strengths. In order that such a Lagrangian exists, the real matrix in
(\ref{eq:em-duality}) must belong to the group ${\rm
  Sp}(2n+2;\mathbb{R})$. This group consists of real matrices that
leave the skew-symmetric tensor $\Omega_{MN}$ invariant,
\begin{equation}
  \label{eq:omega}
  \Omega = \left( \begin{array}{cc}
0 & {\bf 1}\\ \!-{\bf 1} & 0
\end{array} \right)  \;.
\end{equation}
The conjugate matrix $\Omega^{MN}$ is defined by
$\Omega^{MN}\Omega_{NP}= - \delta^M{}_P$. Here we employ an
$\mathrm{Sp}(2n+2;\mathbb{R})$ covariant notation for the
$2(n+1)$-dimensional symplectic indices $M,N,\ldots$, such that
$Z^M= (Z^\Lambda, Z_\Sigma)$. Likewise we use vectors with lower
indices according to $Y_M= (Y_\Lambda,Y^\Sigma)$, transforming
according to the conjugate representation so that $Z^M\,Y_M$ is
invariant.

The ${\rm Sp}(2n+2;\mathbb{R})$ transformations are known as
electric/magnetic dualities, which also act on electric and magnetic
charges (for a review of electric/magnetic duality, see
\cite{deWit:2001pz}). The Lagrangian depends on the electric/magnetic
duality frame and is therefore not unique.  Different Lagrangians
related by electric/magnetic duality lead to equivalent field
equations and thus belong to the same equivalence class. These
alternative Lagrangians remain supersymmetric but because the field
strengths (and thus the underlying gauge fields) have been redefined,
the standard relation between the various fields belonging to the
vector supermultiplet, encoded in \eqref{eq:susyr}, is lost. However,
upon a suitable redefinition of the other vector multiplet fields
(possibly up to terms that will vanish subject to equations of motion)
this relation can be preserved. It is to be expected that the new
Lagrangian is again encoded in terms of a holomorphic homogeneous
function, expressed in terms of the redefined scalar fields. Just as
the Lagrangian changes, this function will change as well. Hence,
different functions $F(X)$ can belong to the same equivalence class.
The new function is such that the vector $X^M=(X^\Lambda,F_\Lambda)$
transforms under electric/magnetic duality according to
\begin{equation}
  \label{eq:em-duality-X}
  \begin{pmatrix} {X}^\Lambda\cr   \noalign{\vskip 1.5mm} {F}_\Lambda
  \end{pmatrix}
  \longrightarrow
  \begin{pmatrix} \tilde{X}^\Lambda\cr \noalign{\vskip 1.5mm}
    {\tilde F}_\Lambda \end{pmatrix} =
  \begin{pmatrix} U^\Lambda{}_\Sigma & Z^{\Lambda\Sigma} \cr
    \noalign{\vskip 1.5mm}
    W_{\Lambda\Sigma} & V_\Lambda{}^\Sigma
  \end{pmatrix}
  \begin{pmatrix} {X}^\Sigma\cr \noalign{\vskip 1.5mm} {F}_\Sigma
  \end{pmatrix} \,.
\end{equation}
The new function $\tilde F(\tilde X)$ of the new scalars $\tilde
X^\Lambda$ follows from integration of (\ref{eq:em-duality-X}) and
takes the form
\begin{align}
  \label{eq:new-F}
  \tilde F(\tilde X)=&\, F(X) -\ft12 X^\Lambda F_\Lambda(X)
  + \ft12 (U^\mathrm{T} W)_{\Lambda\Sigma} X^\Lambda X^\Sigma  \nonumber
  \\
  &\,
  + \ft12 (U^\mathrm{T} V+ W^\mathrm{T} Z)_\Lambda{}^\Sigma  X^\Lambda
  F_\Sigma(X)
  + \ft12 (Z^\mathrm{T} V)^{\Lambda\Sigma} F_\Lambda(X) F_\Sigma(X)\,.
\end{align}
There are no integration constants in this case because the
function must remain homogeneous of second degree.

In general it is not easy to determine $\tilde F(\tilde X)$ from
(\ref{eq:new-F}) as it involves the inversion of $\tilde X^\Lambda
= U^\Lambda{}_\Sigma X^\Sigma + Z^{\Lambda\Sigma} F_\Sigma(X)$. As
we emphasized in section \ref{sec:introduction}, this is the
reason why one prefers to avoid changing the electric/magnetic
duality frame. The duality transformations on higher derivatives
of $F(X)$ follow by differentiation and we note the results,
\begin{align}
  \label{eq:dual-higher-F-der}
  \tilde F_{\Lambda\Sigma}(\tilde X) =&\, (V_\Lambda{}^\Gamma
  F_{\Gamma\Xi} + W_{\Lambda\Xi} )\, [\mathcal{S}^{-1}]^\Xi{}_\Sigma\,,
  \nonumber\\
  \tilde F_{\Lambda\Sigma\Gamma}(\tilde X) =&\, F_{\Xi\Delta\Omega}\,
  [\mathcal{S}^{-1}]^\Xi{}_\Lambda\,[\mathcal{S}^{-1}]^\Delta{}_\Sigma\,
  [\mathcal{S}^{-1}]^\Omega{}_\Gamma\,,
\end{align}
where
\begin{equation}
  \label{eq:def-S}
  \mathcal{S}^\Lambda{}_\Sigma = \frac{\partial \tilde X^\Lambda}{\partial
  X^\Sigma} = U^\Lambda{}_\Sigma + Z^{\Lambda\Gamma}F_{\Gamma\Sigma} \,.
\end{equation}
It is also convenient to introduce the symmetric real matrix,
\begin{equation}
  \label{eq:def-N}
  N_{\Lambda\Sigma} = -\mathrm{i} F_{\Lambda\Sigma} + \mathrm{i}\bar
  F_{\Lambda\Sigma} \,,
\end{equation}
whose inverse will be denoted by $N^{\Lambda\Sigma}$, and which
transforms under electric/magnetic duality according to
\begin{equation}
  \label{eq:dual-N}
  \tilde N_{\Lambda\Sigma}(\tilde X,\tilde{\bar X}) =
  N_{\Gamma\Delta}\, [\mathcal{S}^{-1}]^\Gamma{}_\Lambda\,
  [\bar{\mathcal{S}}^{-1}]^\Delta{}_\Sigma \,.
\end{equation}

To determine the action of the dualities on the fermion fields, we
consider supersymmetry transformations of the symplectic vector
$X^M= (X^\Lambda, F_\Lambda)$, which can be written as $\delta X^M
= \bar\epsilon^i \Omega_i{}^M$, thus defining an
$\mathrm{Sp}(2n+2;\mathbb{R})$ covariant fermionic vector,
$\Omega_i{}^M$,
\begin{equation}
  \label{eq:symplectic-fermion}
  \Omega_i{}^M = \begin{pmatrix} \Omega_i{}^\Lambda \cr
    \noalign{\vskip 1.5mm}
    F_{\Lambda\Sigma}\, \Omega_i{}^\Sigma
  \end{pmatrix}  \;.
\end{equation}
Complex conjugation leads to a second vector, $\Omega^i{}^M$, of
opposite chirality. From (\ref{eq:symplectic-fermion}) one derives
that, under electric/magnetic duality,
\begin{equation}
  \label{eq:em-fermion}
  \tilde\Omega_i{}^\Lambda =
  \mathcal{S}^\Lambda{}_\Sigma\,\Omega_i{}^\Sigma\,.
\end{equation}
Note the identity
\begin{equation}
  \label{eq:X-Omega}
  \Omega_{MN} \,X^M \Omega_i{}^N=0\,,
\end{equation}
which also implies that supersymmetry variations of $\Omega_i{}^M$ are
subject to $\Omega_{MN} \,X^M \,\delta\Omega_i{}^N=0$ up to terms
quadratic in the vector multiplet spinors. This observation explains
some of the identities that we will encounter in due course.

The supersymmetry transformation of $\Omega_i{}^M$ follows from
\eqref{eq:susyr}, and we decompose it into the following form,
\begin{equation}
  \label{eq:delta-Omega-M}
  \delta\Omega_i{}^M = 2\Slash{D}X^M\epsilon_i + \ft12
  \gamma^{\mu\nu}\hat G^-_{\mu\nu}{}^M\varepsilon_{ij}\,\epsilon^j +
  Z_{ij}{}^M \epsilon^j + 2\, X^M\eta_i \,.
\end{equation}
From this the existence follows of a symplectic vector of
anti-selfdual supercovariant field strengths,
\begin{equation}
  \label{eq:symplectic-field-strength}
  \hat G^-_{\mu\nu}{}^M = \begin{pmatrix} \hat G^-_{\mu\nu}{}^\Lambda \cr
    \noalign{\vskip 1.5mm}
    \hat G^-_{\mu\nu\Lambda}
  \end{pmatrix}  \;.
\end{equation}
where $\hat G^-_{\mu\nu}{}^\Lambda= \hat F^-_{\mu\nu}{}^\Lambda$,
with $\hat F^-_{\mu\nu}{}^\Lambda$ defined in \eqref{eq:f-hat},
and $\hat G^-_{\mu\nu\Lambda}$  is defined by,
\begin{equation}
\label{eq:def-G-hat-magn}
\hat{G}_{\mu\nu\Lambda}^-=F_{\Lambda\Sigma}\hat{F}_{\mu\nu}^{-}{}^{\Sigma}-\tfrac18
F_{\Lambda\Sigma\Gamma} \, \bar \Omega_i{}^{\Sigma}
  \gamma_{\mu\nu} \Omega_j{}^{\Gamma}\,\varepsilon^{ij}\,.
\end{equation}

We can also define a second symplectic array of anti-selfdual field
strengths,
\begin{equation}
  \label{eq:symplectic-field-strength}
   G^-_{\mu\nu}{}^M = \begin{pmatrix}  G^-_{\mu\nu}{}^\Lambda \cr
    \noalign{\vskip 1.5mm}
     G^-_{\mu\nu\Lambda}
  \end{pmatrix}  \;,
\end{equation}
with $G_{\mu\nu}{}^\Lambda= F_{\mu\nu}{}^\Lambda$. The second
component, $G_{\mu\nu\Lambda}$, then follows from the identification
(compare to the decomposition \eqref{eq:f-hat}),
\begin{align}
  \label{eq:G-hat}
  \hat G_{\mu\nu}{}^M =&\, G^+_{\mu\nu}{}^M + G^-_{\mu\nu}{}^M
  - \varepsilon^{ij} \bar{\psi}_{[\mu\,i}
  (\gamma_{\nu]} \Omega_j{}^M+ \psi_{\nu] j} X^M)
  -  \varepsilon_{ij}
  \bar{\psi}_{[\mu}{}^i (\gamma_{\nu]} \Omega^{j\, M} +\psi_{\nu]}{}^j
  \bar X^M) \nonumber\\[1mm]
  &\,
  - \tfrac14(X^M\, T_{\mu\nu ij}\,\varepsilon^{ij} + \bar X^M\,
  T_{\mu\nu}{}^{ij}\,\varepsilon_{ij}  )\,.
\end{align}
This implies the following decomposition for $G^-_{\mu\nu\Lambda}$
(and likewise for $G^+_{\mu\nu\Lambda}$),
\begin{equation}
  \label{eq:def-G-magn}
  G^-_{\mu\nu\Lambda} = F_{\Lambda\Sigma} F^-_{\mu\nu}{}^\Sigma -
  2\mathrm{i} \mathcal{O}^-_{\mu\nu\Lambda}\,,
\end{equation}
with
\begin{align}
  \label{eq:def-O}
  \mathcal{O}^-_{\mu\nu\Lambda} =&\,
  -\tfrac1{16}\mathrm{i} F_{\Lambda\Sigma\Gamma} \, \bar
  \Omega_i{}^{\Sigma} \gamma_{\mu\nu}
  \Omega_j{}^{\Gamma}\,\varepsilon^{ij}
  -\tfrac18 N_{\Lambda\Sigma} \varepsilon_{ij}
  \bar\psi_\rho{}^i\gamma_{\mu\nu} \gamma^\rho\Omega^{j\Sigma}
  \nonumber \\[1mm]
  &\,
  -\tfrac18 N_{\Lambda\Sigma}\bar X^\Sigma\,
    \varepsilon_{ij} \bar\psi_\rho{}^i
  \gamma^{\rho\sigma}\gamma_{\mu\nu}  \psi_\sigma{}^j
   +\tfrac1{8} N_{\Lambda\Sigma}
  \bar X^\Sigma \,
  T_{\mu\nu}{}^{ij}\varepsilon_{ij} \,.
\end{align}
Note that the homogeneity of $F(X)$ is crucial for deriving these
results. The definition (\ref{eq:G-hat}) shows that also
$(F_{\mu\nu}{}^\Lambda, G_{\mu\nu\Sigma})$ transforms as a
symplectic vector under electric/magnetic duality.

Consistency requires that the field strengths $G_{\mu\nu}{}^M$
satisfy a Bianchi identity. While $G_{\mu\nu}{}^\Lambda$ clearly
does, it is not obvious for the field strengths
$G_{\mu\nu\Lambda}$. The latter Bianchi identity can, however, be
provided by the field equation for the vector fields following
from some supersymmetric action. In that case $G_{\mu\nu\Lambda}$
will coincide with (\ref{eq:def-G}).  We shall verify in section
\ref{sec:lagrangian} that this is indeed the case for the action
encoded in the holomorphic function $F(X)$. It should be obvious
that also the field strengths $\hat G_{\mu\nu}{}^M$ satisfy a
Bianchi-type identity, but of a more complicated form. Identities of
this type have been presented in \cite{deWit:1980tn} for $\hat
G_{\mu\nu}{}^\Lambda$.

To summarize, both the fields strengths $\hat G_{\mu\nu}{}^M$ and
$G_{\mu\nu}{}^M$ transform as a symplectic vector under duality, and
they differ in their fermionic terms and in terms proportional to the
selfdual and anti-selfdual tensor fields $T_{abij}$ and
$T_{ab}{}^{ij}$, respectively. The supercovariant field strengths
$\hat G_{\mu\nu}{}^M$ appear in the supersymmetry transformation rules
of the fermions, while the field strengths $G_{\mu\nu}{}^M$, when
constrained by the standard Bianchi identities, imply that
$F_{\mu\nu}{}^\Lambda$ can be expressed in terms of a vector potential
$W_\mu{}^\Lambda$, and is subject to corresponding field equations.

Regarding the quantities $Z_{ij}{}^M$, that also follow from
(\ref{eq:delta-Omega-M}), we have a similar situation. They are
defined by
\begin{equation}
  \label{eq:symplectic-Z}
  Z_{ij}{}^M = \begin{pmatrix} Y_{ij}{}^\Lambda \cr
    \noalign{\vskip 1.5mm}
  F_{\Lambda\Sigma}\, Y_{ij}{}^\Sigma -\ft12 F_{\Lambda\Sigma\Gamma}
    \,\bar\Omega_i{}^\Sigma\Omega_j{}^\Gamma
  \end{pmatrix}  \;,
\end{equation}
which suggests that $Z_{ij}{}^M$ transforms under
electric/magnetic duality as a symplectic vector. However, this is
only possible provided we impose a pseudo-reality condition on
$Z_{ij\Lambda}$. This constraint can also be understood as the
result of field equations associated with a supersymmetric action,
whose Lagrangian will be presented in the next section
\ref{sec:lagrangian}.

From the fact that the field strengths $G_{\mu\nu\Lambda}$ are
subject to a Bianchi identity, it follows that they can be
expressed in terms of magnetic duals $W_{\mu\Lambda}$. Hence we
introduce these magnetic gauge fields, whose role will eventually
become clear in the context of the embedding tensor formalism
which will be introduced in due course. Together with the electric
gauge fields $W_\mu{}^\Lambda$, the magnetic duals constitute a
symplectic vector, $W_\mu{}^M= (W_\mu{}^\Lambda, W_{\mu\Lambda})$,
where $G_{\mu\nu}{}^M=2\,\partial_{[\mu}W_{\nu]}{}^M$. As we shall
see, this relationship is, however, not exact and the
identification is subject to certain equations of
motion. The supersymmetry transformations of $W_\mu{}^M$ are
conjectured to take a duality covariant form,
\begin{equation}
  \label{eq:susy-W}
  \delta W_\mu{}^M= \varepsilon^{ij} \bar{\epsilon}_i
  (\gamma_{\mu} \Omega_j{}^M +2\,\psi_{\mu j} X^M)
  + \varepsilon_{ij}
  \bar{\epsilon}^i (\gamma_{\mu} \Omega^{j\, M} +2\,\psi_\mu{}^j
  \bar X^M)   \, .
\end{equation}
Observe that, with this transformation rule, the field strengths
$\hat G_{\mu\nu}{}^M$ are supercovariant. As mentioned above,
$G_{\mu\nu\Lambda}$ and $2\,\partial_{[\mu} W_{\nu] \Lambda}$ are
not identical! This can be seen by calculating the supersymmetry
variation of $2\,\partial_{[\mu} W_{\nu]\Lambda}$ and showing that
it only coincides with the supersymmetry variation of
(\ref{eq:def-G-magn}) up to equations of motion. In the presence of gauge charges in the context of embedding tensor formalism, the
Lagrangian can depend simultaneously on electric and magnetic gauge
fields, as is described in later sections.

The consistency, up to equations of motion, of introducing dual gauge
fields $W_{\mu\Lambda}$ is also confirmed when considering the closure
of the supersymmetry algebra, based on (\ref{eq:susy-W}). Although we
started with an off-shell definition of the vector multiplets, so that
all superconformal transformations will close under commutation
without imposing the equations of motion, this is not necessarily the
case for the newly introduced gauge field $W_{\mu\Lambda}$. Before
discussing this in detail we present the decomposition of the
commutator of two infinitesimal Q-supersymmetry transformations, with
parameters $\epsilon_1$ and $\epsilon_2$,
\begin{equation}
  \label{eq:QQ-commutator}
  [\delta(\epsilon_1),\delta(\epsilon_2)] =
  \xi^{\mu} {D}_{\mu} +
  \delta_M(\varepsilon)+\delta_K(\Lambda_K)+
  \delta_S(\eta)+\delta_{\text{gauge}}(\Lambda^M)\,,
\end{equation}
where the parameters of the various infinitesimal transformations
on the right-hand side are given by
\begin{align}
  \label{eq:convparungauged}\
  \xi^{\mu}  = &\, 2 \,\bar{\epsilon}_2{}^i\gamma^{\mu}\epsilon_{1i} +
  \text{h.c.}\,,\nonumber\\
  \varepsilon^{ab} = &\, \bar{\epsilon}_1{}^i\epsilon_2{}^j\,
  T^{ab}{}_{ij} + \text{h.c.}\,,\nonumber\\
  \Lambda^a_K = &\, \bar\epsilon_1{}^i \epsilon_2{}^j\,
  {D}_bT^{ba}{}_{ij}-\tfrac{3}{2}\,
  \bar\epsilon_2{}^i\gamma^a\epsilon_{1i}\,D+\text{h.c.}\,,\nonumber\\
  \eta^i =&\,
  6\,\bar{\epsilon}_{[1}{}^i\epsilon_{2]}{}^j\,\chi_j\,,\nonumber\\
  \Lambda^M =&\, 4\,\bar X^M \,\bar\epsilon_2{}^i\epsilon_1{}^j\,
  \varepsilon_{ij}
  + \text{h.c.}\,,
\end{align}
where the first term proportional to $\xi^\mu$ denotes a
supercovariant translation, i.e. a general coordinate
transformation with parameter $\xi^\mu$, suitably combined with
field-dependent gauge transformations so that the result is
supercovariant. The terms proportional to $\Lambda^M$ denote the
abelian gauge transformation acting on both the electric and the
magnetic gauge fields $W_\mu{}^M$. This result was already known
for all the fields \cite{deWit:1980tn}, except for
$W_{\mu\Lambda}$. The validity of (\ref{eq:QQ-commutator}) on
$W_{\mu\Lambda}$ can be derived in direct analogy with the
calculation of the commutation relation on $W_\mu{}^\Lambda$, upon
replacing $G_{\mu\nu\Lambda}$ by
$2\,\partial_{[\mu}W_{\nu]\Lambda}$.

The electric/magnetic duality transformations define equivalence
classes of Lagrangians. A subgroup thereof may constitute an
invariance of the theory, meaning that the Lagrangian and its
underlying function $F(X)$ do not change
\cite{deWit:1984pk,Cecotti:1988qn}. More specifically, an invariance
implies
\begin{equation}
  \label{eq:invariant-F}
  \tilde F(\tilde X)= F(\tilde X)\,,
\end{equation}
so that the result of the duality leads to a Lagrangian based on
$\tilde F(\tilde X)$ which is identical to the original
Lagrangian. Because $\tilde F(\tilde X)\not = F(X)$, as is obvious
from (\ref{eq:new-F}), $F(X)$ is not an invariant {\it function}.
Instead the above equation implies that the substitution
$X^\Lambda\to \tilde X^\Lambda$ into the function $F(X)$ and its
derivatives, induces precisely the duality transformations. For
example, we obtain,
\begin{align}
  \label{eq:dual-symm-F-der}
  F_\Lambda (\tilde X) =&\,
  V_\Lambda{}^\Sigma F_\Sigma(X) + W_{\Lambda\Sigma} X^\Sigma \,,
  \nonumber\\
  F_{\Lambda\Sigma}(\tilde X)=&\, (V_\Lambda{}^\Gamma
  F_{\Gamma\Xi} + W_{\Lambda\Xi} )\, [\mathcal{S}^{-1}]^\Xi{}_\Sigma\,,
  \nonumber\\
  F_{\Lambda\Sigma\Gamma}(\tilde X)=&\, F_{\Xi\Delta\Omega}\,
  [\mathcal{S}^{-1}]^\Xi{}_\Lambda\,[\mathcal{S}^{-1}]^\Delta{}_\Sigma\,
  [\mathcal{S}^{-1}]^\Omega{}_\Gamma\,.
\end{align}
Another useful transformation rule is,
\begin{equation}
  \label{eq:dual-O}
  \tilde{\mathcal{O}}_{\mu\nu\Lambda}^- =
  \mathcal{O}_{\mu\nu\Sigma}^-\,
  [\mathcal{S}^{-1}]^\Sigma{}_\Lambda \,.
\end{equation}
In section~\ref{sec:gauge-transf-charges} we are precisely
interested in this subclass of electric/magnetic duality
transformations, as these are the ones that can be gauged.

\section{Superconformal hypermultiplets}
\label{sec:intro-hypermultiplets} \setcounter{equation}{0}
In this section we give a brief description of hypermultiplets and
their isometries, following the framework of \cite{deWit:1999fp}. The
$n_{\mathrm{H}}+1$ hypermultiplets are described by
$4(n_{\mathrm{H}}+1)$ real scalars $\phi^A$, $2(n_{\mathrm{H}}+1)$
positive-chirality spinors $\zeta^{\bar\alpha}$ and
$2(n_{\mathrm{H}}+1)$ negative-chirality spinors $\zeta^\alpha$.
Hence target-space indices $A,B,\ldots$ take values $1,2,\ldots,
4(n_{\mathrm{H}}+1)$, and the indices $\alpha,\beta, \ldots$ and
$\bar\alpha,\bar\beta, \ldots$ run from 1 to
$2(n_{\mathrm{H}}+1)$. The chiral and anti-chiral spinors are related
by complex conjugation (as we are dealing with $2(n_{\mathrm{H}}+1)$
Majorana spinors) under which indices are converted according to
$\alpha\leftrightarrow \bar\alpha$.

For superconformally invariant Lagrangians, the scalar fields of the
hypermultiplets parametrize a $4(n_{\mathrm{H}}+1)$-dimensional
hyperk\"ahler cone
\cite{deWit:1998zg,Gibbons:1998xa,deWit:1999fp,deWit:2001bk}. Such a
cone has a homothetic conformal Killing vector $\chi^A$,
\begin{equation}
 \label{eq:hom-kil-vec}
D_A \chi^B= \delta_A{}^B \,,
\end{equation}
which, locally, can be expressed in terms of a hyperk\"ahler potential
$\chi$ (in later sections denoted by $\chi_\mathrm{hyper}$),
\begin{equation}
  \label{eq:1}
  \chi_A = \partial_A \chi\,.
\end{equation}
The cone metric can thus be written as
$g_{AB}=D_A\partial_B\chi$. This relation does not define the metric
directly, because of the presence of the covariant derivative which
contains the Christoffel connection. We also note the relation
\begin{equation}
  \label{eq:hyp-pot}
  \chi=\ft12 g_{AB}\,\chi^A\chi^B\,.
\end{equation}
Hyperk\"ahler spaces have three hermitian, covariantly constant complex
structures $J_{ij}=J_{ji}$, satisfying the algebra of quaternions,
\begin{equation}
  \label{eq:quat-algebra}
  J_{ijAB} \equiv (J^{ij}{}_{AB})^* = \varepsilon_{ik}\varepsilon_{jl}
  J^{kl}{}_{AB}\,, \quad\quad J^{ij}{}_A{}^C\, J^{kl}{}_{CB} = \ft12
  \varepsilon^{i(k}\varepsilon^{l)j}\, g_{AB}+ \varepsilon^{(i(k}\,
  J^{l)j)}{}_{AB}\,.
\end{equation}
As it turns out, the hyperk\"ahler potential serves as a K\"ahler
potential for each of the complex structures.

Hyperk\"ahler cones have $\mathrm{SU}(2)$ isometries; the
corresponding Killing vectors are expressed in terms of the complex
structures and the homothetic Killing vector,
\begin{equation}
  \label{eq:spkilling}
  k_{ij}{}^A = J_{ij}{}^{AB}\, \chi_B\,,
\end{equation}
from which it follows that
\begin{equation}
  \label{eq:su2vector}
  D_A k^{ij}{}_B = - J^{ij}{}_{AB}\,.
\end{equation}
From the above results, it follows that the homothetic Killing vector
$\chi^A$ and the three $\mathrm{SU}(2)$ Killing vectors $k^{ijA}$ are
mutually orthogonal,
\begin{equation}
  \label{eq:orthogonal}
  \chi^A \chi_A = 2\chi\,, \qquad\qquad k_{ij}{}^A\,k^{kl}{}_A =
  \delta_{(i}{}^k\,\delta_{j)}{}^l \,\chi\,,\qquad\qquad \chi^A\, k^{ij}{}_A
  = 0 \,.
\end{equation}

The hypermultiplet fields transform under dilations, associated
with the homothetic Killing vector, and the
$\mathrm{SU}(2)\times\mathrm{U}(1)$ transformations of the
superconformal group, with parameters $\Lambda_\mathrm{D}$,
$\Lambda_{\mathrm{SU}(2)}$ and $\Lambda_{\mathrm{U}(1)}$, respectively,
\begin{align}
  \label{eq:DandRtransfo}
  \delta \phi^A =&\, \Lambda_{\mathrm{D}}\, \chi^A +
  \Lambda_{\mathrm{SU(2)}}{}^i{}_k \,\varepsilon^{jk}\,
  k_{ij}{}^A\,,\nonumber\\
  \delta \zeta^\alpha+\delta\phi^A\,
  \Gamma_{A}{}^{\!\alpha}{}_{\!\beta}\, \zeta^\beta =&\, \big(\ft32
    \Lambda_{\mathrm{D}}-\ft12
    \mathrm{i} \Lambda_{\mathrm{U(1)}}\big) \zeta^\alpha\,.
\end{align}
Here $\Gamma_A{}^\alpha{}_\beta$ denote the connections
associated with field-dependent reparametrizations of the fermions of
the form $\zeta^\alpha \to S^\alpha{}_{\!\beta} (\phi)\, \zeta^\beta$.
Naturally the conjugate connections
$\bar{\Gamma}_A{}^{\bar\alpha}{}_{\bar\beta}$ are associated with the
reparametrizations $\zeta^{\bar\alpha}\to\bar
S^{\bar\alpha}{}_{\!\bar\beta} (\phi)\,\zeta^{\bar\beta}$. These
tangent-space reparametrizations act on all quantities carrying
indices $\alpha$ and $\bar\alpha$. The corresponding curvatures
$R_{AB}{}^\alpha{}_\beta$ and $\bar{R}_{AB}{}^{\bar\alpha}{}_{\bar\beta}$
take their values in $\mathrm{sp}(n_\mathrm{H}+1) \cong
\mathrm{usp}(2n_\mathrm{H}+2;\mathbb{C})$. These curvatures are
linearly related to the Riemann curvature $R_{ABC}{}^D$ of the target
space, as we shall see later.

Before turning to the supersymmetry transformations, it is of interest
to discuss possible additional isometries of hyperk\"ahler cones that
commute with supersymmetry. They are characterized by Killing vectors
$k^A{}_{\sf m}(\phi)$, labeled by indices $\sf m,n,p$, etcetera. They
generate a group of motions, denoted by $\mathrm{G}_\mathrm{hyper}$,
that leaves the complex structures invariant so that they are called
tri-holomorphic. Furthermore, they commute with $\mathrm{SU}(2)$ and
dilatations. These three properties are reflected in the following
equations,
\begin{align}
  \label{eq:triholo-and-su(2)comm}
  k^C{}_{\sf m}\, \partial_C J^{ij}{}_{AB} - 2 \partial_{[A}k^C{}_{\sf m}
  \,J^{ij}{}_{B]C} =&\, 0\,, \nonumber\\
  k_{ij}{}^B\, D_B k^A{}_{\sf m}= D_B k_{ij}{}^A \, k^B{}_{\sf m} =&\,
  J_{ij}{}^A{}_B\, k^B{}_{\sf m}\nonumber\\
  \chi_A\,k^A{}_{\sf m} =&\, 0\,.
\end{align}
Such tri-holomorphic isometries can be gauged by coupling to the
(electric and/or magnetic) gauge fields belonging to the vector
multiplets, as we shall discuss in due course.\footnote{ 
  As always, the dilatations and the
  $\mathrm{SU}(2)\times\mathrm{U}(1)$ symmetries will be gauged when
  coupling to the corresponding gauge fields of conformal
  supergravity. } 
The total isometry group of the hyperk\"ahler space is thus the
product of $\mathrm{SU}(2)$ times the group
$\mathrm{G}_\mathrm{hyper}$ generated by the Killing vectors
$k^A{}_{\sf m}$. The structure constants of the latter are denoted by
$f_{\sf mn}{}^{\sf p}$, and follow from the Lie bracket
relation,\footnote{
  We note that derivatives of Killing vectors are constrained by the
  Killing equation, which induces constraints on multiple derivatives,
  as is shown below,
\begin{equation}
  \label{eq:killing-eq-and-DDk}
  D_Ak_{B} + D_B k_{A}=0\,, \qquad  D_{A}D_{B} k_{C}
  =  R_{BCAE}\,  k^{\,E}   \ .
\end{equation}
} 
\begin{equation}
  \label{eq:killingclosure}
  k^B{}_{\sf m}\,\partial_Bk^A{}_{\sf n}-k^B{}_{\sf
    n}\,\partial_Bk^A{}_{\sf m}
  = -f_{\sf mn}{}^{\sf p}\, k^A{}_{\sf p} \, .
\end{equation}

The infinitesimal transformations act on the hypermultiplet fields
according to
\begin{align}
  \label{eq:delta-phi-Killing}
  \delta\phi^A=&\,g\,\Lambda^{\sf m}\, k^A{}_{\sf m}(\phi)\,,\nonumber \\
  \delta\zeta^\alpha +\delta\phi^A \Gamma_A{}^{\!\alpha}{}_{\!\beta}
  \,\zeta^\beta =&\, g\, \Lambda^{\sf m} \,{t_{\sf m}}^{\alpha}{}_{\!\beta}(\phi)
  \,\zeta^\beta \,,
\end{align}
where we introduced a generic coupling constant $g$ and
$\phi$-dependent matrices ${t_{\sf m}}^{\alpha}{}_{\!\beta}(\phi)$
which take values in $\mathrm{sp}(n_\mathrm{H}+1)$, and are
proportional to $D_A k^B{}_{\sf m}$. Explicit definitions will be
given later, but we already note that they satisfy the following
relations,
\begin{align}
  \label{eq:t-der-comm}
  D_A t_{\sf m}{}^{\alpha}{}_{\!\beta}  =&\,
  R_{AB}{}^{\!\alpha}{}_{\!\beta} \,k^B{}_{\sf m} \,,\nonumber\\
  {} [\,t_{\sf m} ,\,t_{\sf n}\,]^\alpha{}_{\!\beta}   = &\, f_{\sf
    mn}{}^{\sf p}\,
  (t_{\sf p})^\alpha{}_{\!\beta}  + k^A{}_{\sf m}\,k^B{}_{\sf n}\,
  R_{AB}{}^{\!\alpha}{}_{\!\beta} \,.
\end{align}
This result is consistent with the Jacobi identity. The above results
can be summarized by noting that the linear combinations, $X_{\sf
  m}{}^\alpha{}_\beta= \delta^\alpha{}_\beta\, k^A{}_{\sf m} D_A -
t_{\sf m}{}^\alpha{}_\beta$, close under commutation according
to\footnote{
  To be precise, the $X_{\sf m}$ are the generators acting on
  $\phi$-dependent tangent-space tensors (provided the matrix $t_{\sf
    m}$ is replaced by the appropriate generator for the corresponding
  tensor representation). } 
\begin{equation}
  \label{eq:X-closure}
  {}[ X_{\sf m}, X_{\sf n} ]^\alpha{}_\beta = - f_{\sf mn}{}^{\sf p}
  \,X_{\sf p}{}^\alpha{}_\beta\,.
\end{equation}

One can show that the curl of $J^{ij}{}_{AB}\, k^B{}_{\sf m}$ vanishes,
so that these vectors can be solved in terms of the derivative of the
so-called Killing potentials, or moment maps, denoted by
$\mu^{ij}{}_{\sf m}$. On the hyperk\"ahler cone there are no
integration constants, and one can explicitly determine these
potentials,
\begin{equation}
\label{eq:expr-moment-maps}
\mu^{ij}{}_{\sf m} =- \ft12 k^{ij}{}_A \, k^A{}_{\sf m}\,.
\end{equation}
This can easily be verified by showing that
$\partial_A\mu^{ij}{}_{\sf m}=J^{ij}{}_{AB}\,k^B{}_{\sf m}$, making use of
\eqref{eq:triholo-and-su(2)comm} and the Killing equation given in
\eqref{eq:killing-eq-and-DDk}. Using also \eqref{eq:killingclosure}
one derives the so-called equivariance condition,
\begin{equation}
  \label{eq:equivariance}
  J^{ij}{}_{AB}\,k^A{}_{\sf m}\,k^B{}_{\sf n}= -f_{\sf mn}{}^{\sf
    p}\,\mu^{ij}{}_{\sf p} \ .
\end{equation}
The Killing potentials scale with weight $w=2$ under dilatations and
transform covariantly under the isometries and $\mathrm{SU}(2)$
transformations,
\begin{align}
  \label{k-potential-isometry}
  \delta\mu^{ij}{}_{\sf m}  =&\, \big(g\,\Lambda^{\sf n} \,k^A{}_{\sf n}  +
  \Lambda_{\mathrm{SU(2)}}{}^k{}_m \,\varepsilon^{lm}\,
  k_{kl}{}^A \big) \,\partial_A \mu^{ij}{}_{\sf m} \nonumber\\
   =&\,\big( -g\,\Lambda^{\sf n} \, f_{\sf nm}{}^{\sf p}
   \,\mu^{ij}{}_{\sf p}
   +2\,\Lambda_{\mathrm{SU(2)}}{}^{(i}{}_k \,\mu^{j)k}{}_{\sf m}  \big) \, .
\end{align}

So far, supersymmetry played a central role, as most of the above
results are implied by the superconformal algebra imposed on the
hypermultiplet fields. We refer the reader to \cite{deWit:1999fp} for
a full derivation along these lines. To define the supersymmetry
transformations one needs the notion of quaternionic vielbeine, which
can convert the $4(n_\mathrm{H}+1)$ target-space indices $A,B,\ldots$
to the tangent-space indices $\alpha,
\beta,\ldots,\bar\alpha,\bar\beta\ldots$ carried by the fermions. All
quantities of interest can be expressed in terms of these
vielbeine. For instance, the scalar fields transform as follows under
supersymmetry,
\begin{equation}
  \label{eq:2Q-tr-phi}
    \delta\phi^A= 2(\gamma^A_{i\bar\alpha} \,\bar\epsilon^i
    \zeta^{\bar \alpha} + \bar\gamma^{Ai}_{\alpha} \,\bar\epsilon_i
    \zeta^\alpha )\,,
\end{equation}
where the pseudoreal quantity $\gamma^A_{i\bar\alpha}(\phi)$
corresponds to the $(4n_\mathrm{H}+4)\times(4n_\mathrm{H}+4)$ inverse
quaternionic vielbein. Its inverse is the vielbein denoted by $\bar
V_A^{i\bar\alpha}$, which is needed for writing down the supersymmetry
transformation of the fermions. So we have,
\begin{align}
  \label{eq:3}
    \bar V^{i\bar \alpha}_A \, \gamma^A_{j\bar \beta} = &\,\delta^i{}_j\,
  \delta^{\bar \alpha}{}_{\bar \beta}\,,\nonumber\\
  \gamma^A_{i\bar{\alpha}} \bar{V}_{B}^{j\bar{\alpha}}
  +\bar{\gamma}^{Aj}_\alpha V^\alpha_{Bi} =&\,\delta_i{}^j\, 
  \delta^A{}_{\!B}\,.
\end{align}
Here we emphasize that we use a notation (as elsewhere in this paper)
where $\mathrm{SU}(2)$ indices are raised and lowered by complex
conjugation. The quaternionic vielbeine are covariantly constant,
e.g.,
\begin{equation}
  \label{eq:cov-const-gamma}
  D_A \gamma^B_{i\bar\alpha} = \partial_A\gamma^B_{i\bar\alpha}
  +\Gamma_{AC}{}^B \gamma^C_{i\bar\alpha} -
  \bar\Gamma_A{}^{\bar\beta}{}_{\bar\alpha} \,\gamma^B_{i\bar\beta}
  =0 \,.
\end{equation}
Observe that it is not necessary to introduce a $\mathrm{SU}(2)$
connection here. When coupling to the superconformal fields, the
$\mathrm{SU}(2)$ symmetry will be realized locally and a connection
will be provided by the gauge field $\mathcal{V}_\mu{}^i{}_j$ of the
Weyl multiplet. The fact that the vielbeine are covariantly constant
provides a relation between the Riemann curvature $R_{ABC}{}^D$ and
the tangent-space curvature $\bar R_{AB}{}^{\bar\alpha}{}_{\bar\beta}$,
\begin{equation}
  \label{eq:sp-integrability}
  R_{ABC}{}^D \, \gamma_{i\bar\alpha}^C - \bar R_{AB}{}^{\bar
    \beta}{}_{\bar\alpha} \,\gamma^D_{i\bar\beta} =0\,.
\end{equation}
Both curvatures can actually be written in terms of
\begin{equation}
  \label{eq:defW}
   W_{\bar\alpha\beta\bar\gamma\delta} = \ft12 R_{ABCD}
   \,\gamma^A_{i\bar\alpha}\,\bar\gamma^{iB}_\beta\,\gamma^C_{j\bar\gamma}
   \,\bar\gamma^{jD}_\delta \, ,
\end{equation}
which appears as the coefficient of the four-spinor term in the
supersymmetric Lagrangian (cf. \eqref{eq:hyper}).

A typical feature of the superconformal hypermultiplets is that they
can be formulated in terms of local sections $A_i{}^\alpha(\phi)$ of
an $\mathrm{Sp}(n_\mathrm{H}+1)\times\mathrm{Sp}(1)$
bundle.\footnote{
  The existence of such an associated quaternionic bundle was
  established based on a general analysis of quaternion-K\"ahler
  manifolds \cite{Swann:1991}. Here $\mathrm{Sp}(1) \cong
  \mathrm{SU}(2)$   denotes the corresponding R-symmetry subgroup of
  the $N=2$ superconformal group.  } 
This section is provided by
\begin{equation}
  \label{eq:bundle 2}
  A_i{}^\alpha(\phi) \equiv \chi^B(\phi) \, V_{Bi}^\alpha (\phi) \,.
\end{equation}
Obviously the vielbeine can be re-obtained from these sections, as we
easily derive,
\begin{equation}
  \label{eq:DA-V}
  D_B A_i{}^\alpha = V_{Bi}^\alpha\,.
\end{equation}
We note a few relevant equations,
\begin{align}
  \label{eq:DA-eqs}
    g^{AB}\,D_AA_i{}^\alpha\, D_BA_j{}^\beta =&\,
    \varepsilon_{ij}\,\Omega^{\alpha\beta} \,,\nonumber\\
     g^{AB}\,D_AA_i{}^\alpha\, D_BA^{j\bar\beta} =&\,
    \delta_i{}^j \,G^{\alpha\bar\beta} \,,
\end{align}
which defines two tensors, $\Omega^{\alpha\beta}$ and
$G^{\alpha\bar\beta}$, which are skew symmetric and hermitian,
respectively. Obviously both tensors are covariantly constant. We also
note the following relations,
\begin{align}
  \label{eq:fer-metric}
   G_{\bar\alpha\beta}\,V^\beta_{A\,i}=&\,
   \varepsilon_{ij}\,\Omega_{\bar\alpha\bar\beta}\,\bar
  V^{j\bar\beta}_A = g_{AB}
  \,\gamma^B_{i\bar\alpha} \,,
  \nonumber \\
  G_{\bar{\gamma}\alpha}
  \bar{\Omega}^{\bar{\gamma}\bar{\delta}}G_{\bar{\delta}\beta} =&\,
  \bar{\Omega}_{\alpha\beta}\,,\nonumber\\
  \Omega_{\bar\alpha\bar\beta} \bar\Omega^{\bar\beta\bar\gamma}= &\,-
  \delta_{\bar\alpha}{}^{\bar\gamma}\,,\nonumber\\
  \bar\Omega_{\alpha\beta} \,A_i{}^\alpha A_j{}^\beta =&\,
  \varepsilon_{ij} \chi \,.
\end{align}
The first one establishes the fact that the quaternionic vielbein
$V_{Ai}^\alpha$ is pseudoreal. Furthermore we note
\begin{align}
  \label{eq:DA-eqs2}
  \bar\Omega_{\alpha\beta} A_i{}^\alpha \,D_B A_j{}^\beta =&\, \tfrac12
  \varepsilon_{ij} \chi_B + k_{ij B} \,,\nonumber \\
  \bar\Omega_{\alpha\beta} \,D_A A_i{}^\alpha \,D_B A_j{}^\beta =&\, \tfrac12
  \varepsilon_{ij} \,g_{AB} - J_{ij\,AB} \,,\nonumber\\
  A^{i\bar \alpha} \equiv  (A_i{}^\alpha)^\ast =&\, \varepsilon^{ij}\,
  \bar\Omega^{\bar\alpha\bar\beta} \,G_{\bar\beta\gamma}\,A_j{}^\gamma\, .
\end{align}

Let us now introduce the local Q- and S-supersymmetry transformations
of the hypermultiplet fields, employing the sections $A_i{}^\alpha$
\begin{align}
  \label{eq:4dsusy}
  \delta A_i{}^\alpha+ \delta\phi^B
  \Gamma_B{}^\alpha{}_\beta A_i{}^\beta = &\,
  2\,\bar\epsilon_i\zeta^\alpha +2\,\varepsilon_{ij}
  G^{\alpha\bar\beta}\Omega_{\bar\beta\bar\gamma}\,\bar\epsilon^j
  \zeta^{\bar\gamma}
  \,,\nonumber\\
  \delta\zeta^\alpha +\delta\phi^A\,
  \Gamma_{A}{}^{\!\alpha}{}_{\!\beta}\, \zeta^\beta =&\,
  \Slash{D} A_i{}^\alpha\,\epsilon^i   +A_i{}^\alpha \,\eta^i\,,
  \nonumber\\
  \delta\zeta^{\bar \alpha}+\delta\phi^A\,
  \bar\Gamma_{A}{}^{\!\bar\alpha}{}_{\!\bar \beta} \,\zeta^{\bar\beta}
  =&\, \Slash{D}A^{i\bar \alpha}\, \epsilon_i  +
    A^{i\bar\alpha} \,\eta_i \,.
\end{align}
The Weyl and chiral weights of these sections and the fermion fields
are listed in table \ref{tab:weights-hm}. The reader can easily verify
that these weight assignments are consistent with the above
supersymmetry transformations. The bosonic parts of the covariant
derivatives on the scalar and fermion fields is given by,
\begin{align}
  \label{eq:cov-phi-zeta}
  \mathcal{D}_{\mu}\phi^A =&\, \partial_\mu
  \phi^A - b_\mu \,\chi^A +\ft12 \mathcal{V}_{\mu}{}^i{}_k\,
  \varepsilon^{jk}\, k^A_{ij}\,,\nonumber\\
  \mathcal{D}_\mu A_i{}^\alpha =&\, \partial_\mu A_i{}^\alpha  -b_\mu
  A_i{}^\alpha +\tfrac12 \mathcal{V}_{\mu i}{}^j A_j{}^\alpha
  +\partial_\mu\phi^A \Gamma_A{}^\alpha{}_\beta A_i{}^\beta \,, \nonumber\\
  \mathcal{D}_\mu \zeta^\alpha =\,& \partial_\mu \zeta^\alpha -\ft14
  \omega_{\mu}{}^{ab}\gamma_{ab} \, \zeta^\alpha
  - \ft32 b_\mu\zeta^\alpha+ \ft12 \mathrm{i} A_\mu \zeta^\alpha
  + \partial_\mu\phi^A\, \Gamma_{A}{}^{\!\alpha}{}_{\!\beta} \,\zeta^\beta\,,
\end{align}
where we have now introduced the superconformal gauge fields, in
addition to the target-space connections. The covariantization of the
above derivatives with respect to Q- and S-supersymmetry follows
immediately from \eqref{eq:4dsusy}.

\begin{table}[tb]
\begin{center}\begin{tabular}{|c||cc|}
 \hline field & $A_i{}^\alpha$ & $\zeta{}^\alpha$  \\\hline\hline
  w & $1$ & $\frac{3}{2}$    \\\hline
  c & $0$ & $-\frac{1}{2}$  \\ \hline
\end{tabular}\end{center}\caption{Weyl and chiral weights of the
hypermultiplet fields. }   \label{tab:weights-hm}\end{table}

An expression for the generators $t_{\sf m}$ associated with the
tri-holomorphic Killing vectors follows from requiring the invariance
of the quaternionic vielbeine $V_{Ai}^\alpha$ up to a target-space
rotation,
\begin{equation}
  \label{eq:fermion-t}
  (t_{\sf m})^{\alpha}{}_{\!\beta} = \ft12 V_{Ai}^{\alpha} \,
  \bar\gamma^{Bi}_{\beta}\; D_B k^A{}_{\sf m}\,.
\end{equation}
The invariance implies that target-space scalars satisfy algebraic
identities such as
\begin{equation}
  \label{eq:t-G-Omega}
  \bar t_{\sf m}{}^{\bar\gamma}{}_{\!\bar\alpha} \, G_{\bar\gamma\beta}
  +t_{\sf m}{}^{\gamma}{}_{\!\beta} \,
  G_{\bar\alpha\gamma}= 0= \bar t_{\sf m}{}^{\bar\gamma}{}_{\![\bar\alpha} \,
  \Omega_{\bar\beta]\bar\gamma} \,,
\end{equation}
which confirm that the matrices $t_{\sf m}{}^\alpha{}_\beta$ take
values in $\mathrm{sp}(n_{\mathrm{H}}+1)$. Furthermore we note the
relations,
\begin{align}
  \label{eq:repr-t-mu}
  k^A{}_{\sf m} \,V_{Ai}^\alpha =&\, k^A{}_{\sf m}\, D_A A_i{}^\alpha = t_{\sf
    m}{}^\alpha{}_\beta\, A_i{}^\beta \,, \nonumber\\
  \mu_{ij\sf m} =&\, -\tfrac12 k_{Aij}\,k^A{}_{\sf m} = -\tfrac12
  \bar\Omega_{\alpha\beta}\,A_i{}^\alpha \,t_{\sf m}{}^\beta{}_\gamma
  A_j{}^\gamma \,.
\end{align}
For a more complete list of identities we refer to \cite{deWit:1999fp}.

\section{Lagrangians}
\label{sec:lagrangian} \setcounter{equation}{0}
In this section we consider the various matter Lagrangians that are
superconformally invariant. All these Lagrangians can be found in the
literature (see, e.g.,
\cite{deWit:1980tn,deWit:1984pk,deWit:1984px,deWit:1999fp}),
including some of the terms quartic in the fermions. We have not
eliminated any auxiliary fields, so that the results pertain to fully
off-shell couplings, with the exception of the hypermultiplets. In the
formula below, we have substituted the explicit expressions for
the dependent gauge fields associated with Lorentz transformations,
conformal boosts and S-supersymmetry. For these expressions we refer
to the appendices in \cite{deWit:2010za}.

All Lagrangians given below can be viewed as matter Lagrangians in a
given superconformal supergravity background.  However, the conformal
supergravity background represents dynamical degrees of freedom which
will mix with the matter degrees of freedom. For the Lagrangian of the
vector multiplets, physical fields can be identified that are
invariant under scale transformations and S-supersymmetry, so that we
will be dealing with supergravity coupled to only $n$ vector
supermultiplets. The remaining vector multiplet acts as a compensating
field: its scalar and spinor degrees of freedom are not physical and
only the vector field and the corresponding triplet of auxiliary
fields remain. For the hypermultiplet Lagrangians, a similar
rearrangement of degrees of freedom will take place. One of the
hypermultiplets will play the role of a compensator with respect to
the local $\mathrm{SU}(2)$. The precise choice of the compensator
multiplets is irrelevant, and the resulting theories remain gauge
equivalent.\footnote{ 
  The hypermultiplet compensator can be replaced by a tensor
  multiplet, but this option will not be considered here.
} 
Therefore it is best to not make any particular choice
for the compensating multiplets at this stage and keep the formulae in
their most symmetric form. At the end one may then select fields that
are invariant under certain local superconformal transformations, so
that the compensating fields decouple from the Lagrangian, or one may
simply adopt a convenient gauge choice.

The Lagrangian for the vector multiplets is decomposed into four
separate parts,
\begin{equation}
  \label{eq:vector-lagrangian-decom}
  \mathcal{L}_\mathrm{vector}
  =\mathcal{L}_\mathrm{kin}^{(1)}+\mathcal{L}_\mathrm{kin}^{(2)}
  +\mathcal{L}_\mathrm{aux}+\mathcal{L}_\mathrm{conf}  \,,
\end{equation}
which are each separately consistent with electric/magnetic
duality. We stress that this is not a invariance property. Under
generic electric/magnetic duality, one obtains in general a different
Lagrangian based on a function $\tilde{F}(\tilde{X})$ that is not
identical to the original function. Only the subgroup that satisfies
(\ref{eq:invariant-F}) constitutes an invariance. The only terms that
have been suppressed in \eqref{eq:vector-lagrangian-decom} are quartic
in the fermion fields and separately consistent with respect to
electric/magnetic duality.

The first term in \eqref{eq:vector-lagrangian-decom} contains  the
kinetic terms of the scalar and spinor fields,
\begin{align}
\label{eq:lagrangian-matter-cov}
  e^{-1} \mathcal{L}_\mathrm{kin}^{(1)} =&\, -\mathrm{i} \Omega_{MN}\,
  \mathcal{D}_\mu X^M \,\mathcal{D}^\mu \bar X^N +
  \ft14\mathrm{i}\Omega_{MN}\left[\bar
   \Omega^{iM} \Slash{\mathcal{D}} \Omega_i{}^N
  -\bar \Omega_i{}^M \Slash{\mathcal{D}} \Omega^{iN} \right]
  \nonumber\\[.6ex]
  &\,-\tfrac12\mathrm{i} \Omega_{MN} \left[\bar\psi_\mu{}^i
  \Slash{\mathcal{D}} \bar X^M \gamma^\mu\, \Omega_i{}^N -
  \bar\psi_{\mu i}
  \Slash{\mathcal{D}}X^M \gamma^\mu\, \Omega^{i N} \right]\,.
\end{align}

The kinetic terms for the vector fields and their moment couplings
to the tensor and fermion fields are contained in
$\mathcal{L}_\mathrm{kin}^{(2)}$,
\begin{align}
  \label{eq:kinetic-vector}
  e^{-1} \mathcal{L}_\mathrm{kin}^{(2)}=&\,\ft14
  \mathrm{i}\left[F_{\Lambda\Sigma} \,F^{-\,\Lambda}_{\mu\nu}
  F^{-\,\mu\nu\Sigma} -\bar F_{\Lambda\Sigma} \,F^{+\,\Lambda}_{\mu\nu}
  F^{+\mu\nu\Sigma}\right] \nonumber\\[.6ex]
  &\,+\big[\mathcal{O}^-_{\mu\nu\Lambda}
  F^{-\mu\nu\Lambda}-N^{\Lambda\Sigma}\,\mathcal{O}^-_{\mu\nu\Lambda}
  \mathcal{O}^{-\mu\nu}{}_{\Sigma}+\mbox{h.c.}   \big]\,,
\end{align}
with $\mathcal{O}^-_{\mu\nu\Lambda}$ as defined in
(\ref{eq:def-O}). Here we included a term quadratic in the tensors
$\mathcal{O}$, such that the resulting expression is consistent with
respect to electric/magnetic duality.\footnote{ 
  To appreciate the presence of this term, we note that
  \eqref{eq:kinetic-vector} can be written as
  \begin{equation}
    \label{eq:rewrite-kin}
    e^{-1} \mathcal{L}_\mathrm{kin}^{(2)}=\ft14
  \mathrm{i}\big[F^{-\Lambda}_{\mu\nu}\, G^{-\mu\nu}{}_\Lambda +
      \mathrm{h.c.} \big]
  -\mathrm{i}\big[\mathcal{O}^{-\mu\nu}{}_\Sigma\, N^{\Sigma\Lambda}
  \big(G^-_{\mu\nu\Lambda}- \bar F_{\Lambda\Gamma}\,
  F^-_{\mu\nu}{}^\Lambda\big) +\mbox{h.c.}\big] \,.
  \end{equation}
  Modulo the field equation of the vector fields, the first term can
  be written as a total derivative, whereas the second term is
  manifestly consistent with electric/magnetic duality as follows from
  \eqref{eq:dual-N}, \eqref{eq:dual-symm-F-der} and \eqref{eq:dual-O}.
} 
Note that one can explicitly construct the field strength tensors
$G_{\mu\nu\Lambda}$ from \eqref{eq:vector-lagrangian-decom}, according
to definition (\ref{eq:def-G}). The result coincides precisely with
the expression given by (\ref{eq:def-G-magn}), as was claimed
previously.

The terms associated with the auxiliary fields $Y_{ij}{}^\Lambda$
are given in $\mathcal{L}_\mathrm{aux}$ \cite{de Vroome:2007zd},
\begin{align}
\label{eq:vector-aux} e^{-1}\mathcal{L}_\mathrm{aux}=&\,\ft1{8}
     N^{\Lambda\Sigma}\,\left (N_{\Lambda\Gamma}Y_{ij}{}^\Gamma +
     \ft12\mathrm{i} (F_{\Lambda\Gamma\Omega}\,
     \bar\Omega_i{}^\Gamma\Omega_j{}^\Omega - \bar F_{\Lambda\Gamma\Omega}
     \,\bar\Omega^{k\Gamma}\Omega^{l\Omega}\varepsilon_{ik}\varepsilon_{jl}
     )\right)  \nonumber\\
     & {}
    \times\left(N_{\Sigma\Xi}Y^{ij\Xi}  +
     \ft12\mathrm{i} (F_{\Sigma\Xi\Delta}
     \,\bar\Omega_m{}^\Xi\Omega_n{}^\Delta
     \varepsilon^{im}\varepsilon^{jn} -  \bar F_{\Sigma\Xi\Delta}
     \,\bar\Omega^{i\Xi}\Omega^{j\Delta} )  \right)\,.
\end{align}
Note that the field equations for the auxiliary fields
$Y_{ij}{}^\Lambda$ indeed imply the pseudo-reality of $Z_{ij\Lambda}$,
as was claimed below \eqref{eq:symplectic-Z}.  The last part of the
Lagrangian describes the remaining couplings of the vector multiplet
fields to conformal supergravity,
\begin{align}
  \label{eq:vector-conf}
  e^{-1}\mathcal{L}_{\mathrm{conf}} =&\,
   \tfrac{1}{6} \chi_{\mathrm{vector}} \,
  \Big[ R +
     ( e^{-1} \varepsilon^{\mu\nu\rho\sigma}
     \bar{\psi}_\mu{}^i \gamma_\nu \mathcal{D}_\rho \psi_{\sigma i}
     - \bar{\psi}_\mu{}^i \psi_\nu{}^j\, T^{\mu \nu}{}_{ij}
          +\mbox{h.c.}) \Big] \nonumber \\[.6ex]
  &\,
     - \chi_{\mathrm{vector}} \,\Big[
     D + \tfrac{1}{2} \bar{\psi}_\mu{}^i \gamma^\mu \chi_i
    +  \tfrac{1}{2} \bar{\psi}_{\mu i} \gamma^\mu \chi^i \Big]  
    \nonumber \\[.6ex]
   &\,
   - \Big(\frac{\partial\chi_{\mathrm{vector}}}{\partial X^\Lambda}
    \Big[\tfrac1{3}\bar\Omega_i{}^{\Lambda}
    \gamma^{\mu\nu} \mathcal{D}_\mu\psi_\nu{}^i -
    \bar{\Omega}_i{}^\Lambda \chi^i    \Big]
    +\mbox{h.c.} \Big)\nonumber\\[.6ex]
&\,
   - \Big(\frac{\partial\chi_{\mathrm{vector}}}{\partial X^\Lambda}
    \Big[\tfrac14 e^{-1} \varepsilon^{\mu\nu\rho\sigma}
    \bar{\psi}_{\mu i} \gamma_\nu \psi_\rho{}^i \,\mathcal{D}_\sigma
    X^\Lambda
    +\tfrac1{48} \,\bar{\psi}_{i\mu} \gamma^\mu
    \gamma_{\rho\sigma} \Omega_j{}^\Lambda \, T^{ij\rho\sigma}\Big]
    +\mbox{h.c.} \Big)\;,
\end{align}
where $\chi_\mathrm{vector}= \mathrm{i} (X^\Lambda\bar F_\Lambda -
\bar X^\Lambda F_\Lambda) = N_{\Lambda\Sigma} X^\Lambda \bar
X^\Sigma = \mathrm{i} \Omega_{MN} X^M \bar X^N$. Note that
$\partial\chi_\mathrm{vector}/\partial X^\Lambda=
N_{\Lambda\Sigma} \bar X^\Sigma$.

We now exhibit the superconformal Lagrangian for hypermultiplets
\cite{deWit:1999fp,deWit:2001bk},
\begin{align}
  \label{eq:hyper}
  e^{-1}\mathcal{L}_{\mathrm{hyper}} =&\,
  \tfrac{1}{6}\,\chi_{\mathrm{hyper}} \,  \Big[ R +
     ( e^{-1} \varepsilon^{\mu\nu\rho\sigma}
     \bar{\psi}_\mu{}^i \gamma_\nu \mathcal{D}_\rho \psi_{\sigma i}
     - \tfrac{1}{4} \bar{\psi}_\mu{}^i \psi_\nu{}^j\, T^{\mu \nu}{}_{ij}
          +\mbox{h.c.}) \Big] \nonumber \\[.ex]
  &\,
     + \tfrac{1}{2}\,\chi_{\mathrm{hyper}}\,\Big[
     D + \tfrac{1}{2} \bar{\psi}_\mu{}^i \gamma^\mu \chi_i
     +  \tfrac{1}{2} \bar{\psi}_{\mu i} \gamma^\mu \chi^i \Big]  \,,
     \nonumber   \\[.6ex]
    &\,
    - \tfrac{1}{2}  G_{\bar{\alpha} \beta}\,\mathcal{D}_\mu A_i{}^\beta
    \,\mathcal{D}^\mu A^{i\bar{\alpha}} -G_{\bar \alpha \beta}( \bar\zeta^{\bar \alpha}\Slash{\mathcal{D}} \zeta^\beta
  +\bar\zeta^\beta\Slash{\mathcal{D}}\zeta^{\bar\alpha}) -\ft14
  W_{\bar\alpha\beta\bar\gamma\delta}\, \bar \zeta^{\bar \alpha}
  \gamma_\mu\zeta^{\beta}\,\bar \zeta^{\bar \gamma}
  \gamma^\mu\zeta^\delta \nonumber \\[.6ex]
   &\,
   - \frac{\partial\chi_{\mathrm{hyper}}}{\partial \phi^A}  \Big(
   \gamma^A{}_{i\bar\alpha}
    \Big[\tfrac2{3}\bar{\zeta}^{\bar\alpha} \gamma^{\mu\nu}
   \mathcal{D}_\mu\psi_\nu{}^i + \bar{\zeta}^{\bar\alpha}   \chi^i
   - \tfrac1{6}  \,\bar \zeta^{\bar\alpha} \gamma_\mu \psi_{\nu j}
   \, T^{\mu\nu ij}\Big]
    +\mbox{h.c.}  \Big) \nonumber\\
    &\,
    + \Big[\ft1{16}\,\bar{\Omega}_{\alpha
    \beta}\,\bar{\zeta}^\alpha
    \gamma^{\mu\nu} T_{\mu\nu ij}\varepsilon^{ij}\zeta^\beta
    -\ft12\, \bar{\zeta}^\alpha \gamma^\mu\gamma^\nu \psi_{\mu
    i}\left(\bar{\psi}_\nu{}^i\,G_{\alpha \bar{\beta}}\,\zeta^{\bar{\beta}}+\varepsilon^{ij}\,\bar{\Omega}_{\alpha \beta}
    \,\bar{\psi}_{\nu j}\zeta^\beta\right)\nonumber\\
    &\,
    +G_{\bar{\alpha} \beta}\,\bar{\zeta}^{\beta}\gamma^\mu
    \Slash{\mathcal{D}}A^{i\bar{\alpha}} \psi_{\mu i}
    -\ft14 e^{-1}\epsilon^{\mu\nu\rho\sigma}G_{
    \bar{\alpha}\beta}\,\bar{\psi}_{\mu}{}^i
    \gamma_\nu \psi_{\rho j}\, A_i{}^\beta
    \mathcal{D}_\sigma A^{j\bar{\alpha}}+ \mbox{h.c.}\Big] \;,
\end{align}
where $W_{\bar\alpha\beta\bar\gamma\delta}$ was defined in
\eqref{eq:defW}, and the hyperk\"ahler potential was introduced in
section \ref{sec:intro-hypermultiplets}.  Since this Lagrangian is
superconformally invariant, the target-space geometry is that of a
hyperk\"ahler cone, which is a cone over a so-called tri-Sasakian
manifold. The latter is a fibration of $\mathrm{Sp}(1)$ over a
$4n_\mathrm{H}$-dimensional quaternion-K\"ahler manifold
$\mathbb{Q}^{4n_\mathrm{H}}$. Hence the hyperk\"ahler cone can be
written as $R^+\times (\mathrm{Sp}(1)\times \mathbb{Q}^{4n_\mathrm{H}})$.

Also tensor multiplets can be coupled to conformal supergravity (see,
e.g. \cite{deWit:2006gn}), but since those multiplets are not involved
in the gaugings they will not be considered here.

\section{Gauge invariance, electric and magnetic charges,
and the embedding tensor} \label{sec:gauge-transf-charges}
\setcounter{equation}{0}
Possible gauge groups must be embedded into the rigid invariance group
$\mathrm{G}_{\mathrm{rigid}}$ of the theory. In the context of this
paper, we are in principle dealing with a product group,
$\mathrm{G}_{\mathrm{rigid}}=\mathrm{G}_\mathrm{symp} \times
\mathrm{G}_\mathrm{hyper}$, where $\mathrm{G}_\mathrm{symp}$ refers to
the invariance group of the electric/magnetic dualities, which acts
exclusively on the vector multiplets, and $\mathrm{G}_\mathrm{hyper}$
refers to the possible invariance group of the hypermultiplet sector
generated by the tri-holomorphic Killing
vectors.\footnote{ 
  Observe that the R-symmetry group,
  $\mathrm{SU}(2)\times\mathrm{U}(1)$, does not play a role here, as
  this group is already realized locally in the coupling to the
  superconformal background.
} 
Here we first concentrate on the gauge group embedded into
$\mathrm{G}_\mathrm{symp}$, which constitutes a subgroup of the
electric/magnetic duality group $\mathrm{Sp}(2n+2;\mathbb{R})$ related
to the matrices considered in (\ref{eq:em-duality}).  The
corresponding gauge group generators thus take the form of
$(2n+2)$-by-$(2n+2)$ matrices $T_M$. Since we are assuming the
presence of both electric and magnetic gauge fields, these generators
decompose according to $T_M=(T_{\Lambda},T^\Lambda)$.  Obviously the
gauge-group generators $T_{M N}{}^P$ must generate a subalgebra of the
Lie algebra associated with $\mathrm{Sp}(2n+2;\mathbb{R})$, which
implies,
\begin{equation}
  \label{eq:sp-constraint}
  T_{M[N}{}^Q\,\Omega_{P]Q} =0\,,
\end{equation}
or, in components,
\begin{align}
  \label{eq:sympl-generators}
  T_{M\Lambda}{}^\Sigma =- T_M{}^\Sigma{}_\Lambda\,,\qquad
  T_{M[\Lambda\Sigma]} = 0= T_M{}^{[\Lambda\Sigma]} \,.
\end{align}
Denoting the gauge group parameters by $\Lambda^M$, infinitesimal
variations of generic $2(n+1)$-dimensional ${\rm
Sp}(2n+2;\mathbb{R})$ vectors $Y^M$ and $Z_M$ thus take the form
\begin{equation}
  \label{eq:gauge-tr-Y-Z}
  \delta Y^M = -g \Lambda^N \,T_{NP}{}^M \,Y^P\,,\qquad
    \delta Z_M = g \Lambda^N \,T_{NM}{}^P \,Z_P\,,
\end{equation}
where $g$ denotes a universal gauge coupling
constant.\footnote{
  The generators follow by expanding the symplectic matrix appearing
  in \eqref{eq:em-duality} and \eqref{eq:em-duality-X} about the
  identity. Comparing with \eqref{eq:gauge-tr-Y-Z}, one establishes
  the correspondence, $U^\Lambda{}_\Sigma \approx
  \delta^\Lambda{}_\Sigma - g\Lambda^M T_{M\Sigma}{}^\Lambda$,
  $V_\Lambda{}^\Sigma \approx
  \delta_\Lambda{}^\Sigma + g\Lambda^M T_{M\Lambda}{}^\Sigma$,
  $Z^{\Lambda\Sigma} \approx -g\Lambda^M T_M{}^{\Lambda\Sigma}$,
  $W_{\Lambda\Sigma} \approx -g\Lambda^M T_{M\Lambda\Sigma}$.
} 
Covariant derivatives can easily be constructed, and
read\footnote{
  In this section and in section \ref{sec:gauge-hierarchy}, we
  suppress the covariantization with respect to superconformal
  symmetries. Starting with section \ref{sec:rest-supersymm-non} the
  derivative $\mathcal{D}_\mu$ will indicate covariantization with
  respect to Lorentz, dilatation, and chiral symmetries, and with the
  newly introduced gauge symmetries associated with the fields
  $W_\mu{}^M$.  }, 
\begin{align}
  \label{eq:cov-derivative}
  \mathcal{D}_\mu Y^M =&\, \partial_\mu Y^M + g W_\mu{}^N\, T_{NP}{}^M\,Y^P
  \nonumber \\
  =&\, \partial_\mu Y^M + g W_\mu{}^\Lambda\, T_{\Lambda P}{}^M\,Y^P +
  g W_{\mu\Lambda}\, T^\Lambda{}_{P}{}^M\,Y^P \,,
\end{align}
and similarly for $\mathcal{D}_\mu Z_M$. The gauge fields then
transform according to
\begin{equation}
  \label{eq:gauge-tr-A}
  \delta W_\mu{}^M = \mathcal{D}_\mu\Lambda^M= \partial_\mu \Lambda^M + g\,
  T_{PQ}{}^M  W_\mu{}^P\, \Lambda^Q \,.
\end{equation}
Note that, for constant parameters $\Lambda^M$,
\eqref{eq:gauge-tr-A} will only be consistent with
\eqref{eq:gauge-tr-Y-Z} provided that $T_{MN}{}^P$ is
antisymmetric in $[MN]$.  Nevertheless, as we shall see,
antisymmetry of $T_{MN}{}^P$ is not necessary in the general case.
Rather, it is sufficient that the $T_{MN}{}^P$ are subject to the
so-called representation constraint \cite{deWit:2005ub},
\begin{equation}
  \label{eq:lin}
  T_{(MN}{}^{Q}\,\Omega_{P)Q} =0
  \Longrightarrow  \left\{
    \begin{array}{l}
      T^{(\Lambda\Sigma\Gamma)}=0\,,\\[.2ex]
      2\,T^{(\Gamma\Lambda)}{}_{\Sigma}=
      T_{\Sigma}{}^{\Lambda\Gamma}\,, \\[.2ex]
      T_{(\Lambda\Sigma\Gamma)}=0\,,\\[.2ex]
      2\,T_{(\Gamma\Lambda)}{}^{\Sigma}=
      T^{\Sigma}{}_{\Lambda\Gamma}\,.
    \end{array}
  \right.
\end{equation}
which does not imply antisymmetry of $T_{MN}{}^P$ in $[M,N]$.
However, for the conventional electric gaugings, where the
magnetic gauge fields $A_{\mu\Lambda}$ decouple and where
$T^\Lambda{}_N{}^P=0$ and $T_\Lambda{}^{\Sigma\Gamma}= 0$,
(\ref{eq:lin}) does imply that $T_{\Gamma\Sigma}{}^\Lambda$ is
antisymmetric in $[\Gamma\Sigma]$.

Note that full covariance of the derivative defined in
(\ref{eq:cov-derivative}) has not yet been established to order
$g^2$, since we have not discussed the closure of the gauge group
generators. This point will be addressed later in this section.

Let us first consider some generic features of the infinitesimal
transformations (\ref{eq:gauge-tr-Y-Z}). Combining the two
equations (\ref{eq:new-F}) and (\ref{eq:invariant-F}) leads to an
expression for $F(\tilde X)- F(X)$, which, for an infinitesimal
symmetry transformation $\delta X^\Lambda=- g\,\Lambda^M
T_{MN}{}^\Lambda\,X^N$, yields
\begin{equation}
  \label{eq:dlta-F}
  F_\Lambda\, \delta X^\Lambda = -\ft12 g\Lambda^M \Big( T_{M\Lambda\Sigma}
  X^\Lambda X^\Sigma
  +T_M{}^{\Lambda\Sigma} F_\Lambda F_\Sigma\Big) \,.
\end{equation}
Substituting the expression for $\delta X^\Lambda$ then leads to
the condition \cite{deWit:1984pk},
\begin{equation}
  \label{eq:symplectic-invariance}
  T_{MN}{}^Q \Omega_{PQ} \,X^NX^P =
  T_{M\Lambda\Sigma} X^\Lambda X^\Sigma -2 T_{M\Lambda}{}^\Sigma X^\Lambda
  F_\Sigma - T_M{}^{\Lambda\Sigma}F_\Lambda F_\Sigma =0\,.
\end{equation}
which must hold for general $X^\Lambda$. The solution of this
condition will specify all continuous symmetries of the
Lagrangian. There are two more useful identities that follow from
it. First one takes the derivative of
(\ref{eq:symplectic-invariance}) with respect to $X^\Lambda$,
\begin{equation}
  \label{eq:derivative-invariance}
  T_{MN\Lambda} X^N = F_{\Lambda\Sigma} \,T_{MN}{}^\Sigma X^N \,,
\end{equation}
and subsequently applies a supersymmetry transformation leading
to,
\begin{equation}
  \label{eq:fermion-der-invariance}
  T_{MN\Lambda} \Omega_i{}^N = F_{\Lambda\Sigma} \,T_{MN}{}^\Sigma
    \Omega_i{}^N +F_{\Lambda\Sigma\Gamma}\,\Omega_i{}^\Sigma
    \,T_{MN}{}^\Gamma X^N   \,.
\end{equation}
The latter two identities show that the gauge covariantization of the
kinetic term for the scalars and spinors in
(\ref{eq:lagrangian-matter-cov}) will not involve
$T_{M\Lambda\Sigma}$. We refer to \cite{de Vroome:2007zd} for further
details about these covariant derivatives.

By introducing a vector $U^M = (U^\Lambda, F_{\Lambda\Sigma}
U^\Sigma)$, it is possible to cast (\ref{eq:derivative-invariance}) in
the symplectically covariant form, $T_{MN}{}^Q\,\Omega_{PQ} \,
X^NU^P=0$. This equation can be rewritten by making use of the
representation constraint \eqref{eq:lin}. Note, for instance, the
following identities,
\begin{align}
  \label{eq:cov-der-invariance}
   T_{(MN)}{}^P \,X^M \,U^N=&\, 0\,,\nonumber\\
   T_{MN}{}^Q \,\Omega_{PQ}\, \bar X^MX^N X^P=&\,T_{MN}{}^Q \,\Omega_{PQ}
   \, \bar X^MX^N \bar X^P =0 \,, \nonumber\\
   T_{MN}{}^\Lambda \,X^M\bar X^N \,N_{\Lambda\Sigma} \,X^\Sigma
   =\,0\,.
\end{align}

As a side remark we note that the Killing potential (or moment
map) associated with the isometries considered above, is related
to
\begin{equation}
  \label{eq:U(1)-moment-map}
  \nu_M = g\, T_{MN}{}^Q \Omega_{PQ} \bar X^N X^P \,.
\end{equation}
Its derivative takes the form $\partial_\Lambda \nu_M = \mathrm{i}
N_{\Lambda\Sigma} \,\delta\bar X^\Sigma$, as follows from making
use of (\ref{eq:derivative-invariance}).

Finally we return to the gauge transformations of the auxiliary
fields $Y_{ij}{}^\Lambda$, which can be derived by requiring that
$\mathcal{L}_\mathrm{aux}$ written in (\ref{eq:vector-aux}) is
gauge invariant. A straightforward calculation leads to the
following result,
\begin{equation}
  \label{eq:gauge-Y}
  \delta Y_{ij}{}^\Lambda= - \ft12 g \Lambda^M T_{MN}{}^\Lambda (
  Z_{ij}{}^N+ \varepsilon_{ik}\varepsilon_{jl} \,Z^{klN}) \,,
\end{equation}
where $Z_{ij}{}^M$ was defined in (\ref{eq:symplectic-Z}). Note
that this result is in accord with the electric/magnetic dualities
suggested for $Z_{ij}{}^M$.

In the remainder of this section we consider the gauge group
embedding in more detail. The embedding into the rigid invariance
group $\mathrm{G}_{\mathrm{rigid}}=\mathrm{G}_\mathrm{symp} \times
\mathrm{G}_\mathrm{hyper}$ is encoded in a so-called embedding
tensor. This tensor must be specified separately for the vector
multiplet and for the hypermultiplet sector, so that we have the
following definitions,
\begin{align}
  \label{eq:T-into-t-decom}
  T_{MN}{}^P =&\, \Theta_M{}^{\sf a} \,t_{{\sf a}N}{}^P\,, \nonumber\\
  k^A{}_M=&\, \Theta_M{}^{\sf m} \,k^A{}_{\sf m} \,,\qquad
  T_M{}^\alpha{}_\beta =\Theta_M{}^{\sf m} \,t_{\sf m}{}^\alpha{}_\beta
  \,,
\end{align}
where the $t_{\sf a}$ denote the generators of
$\mathrm{G}_\mathrm{symp}$, and $k^A{}_{\sf m}$ and $t_{\sf m}$ the
tri-holomorphic Killing vectors and the corresponding matrices of the
group $\mathrm{G}_\mathrm{hyper}$. Because these generators belong to
different groups and act on different multiplets, they carry different
indices (namely, indices $M,N,\ldots$ for the vector multiplets and
indices $\alpha,\beta,\ldots$ for the hypermultiplets). The embedding
tensor can be further decomposed into electric and magnetic
components, according to $\Theta_M{}^{\sf a} = (\Theta_\Lambda{}^{\sf
  a}, \Theta^{\Lambda\,{\sf a}})$, and $\Theta_M{}^{\sf m} =
(\Theta_\Lambda{}^{\sf m}, \Theta^{\Lambda\,{\sf m}})$. With these
definitions, we can now also present the gauge-covariant derivatives
on the hypermultiplet fields (we remind the reader that in this
section and in the next one, we suppress the covariantization with
respect to the superconformal symmetries),
\begin{align}
  \label{eq:cov-hypermultiplet-der}
  {\cal D}_\mu \phi^A =&\, \partial_\mu \phi^A - g W_\mu{}^M \,k^A{}_M
  \,, \nonumber\\
  \mathcal{D}_\mu A_i{}^\alpha =&\,\partial_\mu A_i{}^\alpha 
  +   \partial_\mu\phi^A\,\Gamma_A{}^{\!\alpha}{}_{\!\beta}\,
  A_i{}^\beta 
  -g W_\mu{}^M \,T_M{}^\alpha{}_\beta A_i{}^\beta    \,, \nonumber\\
  {\cal D}_\mu\zeta^\alpha =&\,\partial_\mu \zeta^\alpha+
  \partial_\mu\phi^A\,\Gamma_A{}^{\!\alpha}{}_{\!\beta}\,
  \zeta^\beta -gW_\mu{}^M {T_M}^\alpha{}_{\!\beta}\,\zeta^\beta\, . \;
\end{align}
In particular the covariant derivative of the spinor field is not
entirely straightforward, in view of the fact that matrices ${t_{\sf
    m}}^\alpha{}_\beta$ depend on the fields $\phi^A$. However,
because the Jacobi identity is satisfied on these matrices, there are
no further complications associated with this feature (see \eqref{eq:t-der-comm}).

The gauge group generators $T_M$ should close under commutation
for both representations. This leads to two equations that depend
quadratically on the embedding tensor  \cite{deWit:2003hr},
\begin{align}
  \label{eq:clos}
  f_{\sf ab}{}^{\sf c}\, \Theta_{M}{}^{\sf a}\,\Theta_{N}{}^{\sf b}
  +(t_{{\sf a}})_{N}{}^{P}\,\Theta_{M}{}^{\sf a}\Theta_{P}{}^{\sf c}
  =&\, 0\,,  \nonumber \\[1ex]
  f_{\sf mn}{}^{\sf p}\, \Theta_{M}{}^{\sf m}\,\Theta_{N}{}^{\sf n}
  +(t_{{\sf a}})_{N}{}^{P}\,\Theta_{M}{}^{\sf a}\Theta_{P}{}^{\sf p}
  =&\, 0\,,
\end{align}
where $f_{\sf ab}{}^{\sf c}$ and $f_{\sf mn}{}^{\sf p}$ are the
structure constants of $\mathrm{G}_\mathrm{symp}$ and
$\mathrm{G}_\mathrm{hyper}$, respectively.\footnote{ 
  For convenience we have ignored that the matrices $t_{\sf m}$ depend
  on the scalar fields (see, \eqref{eq:X-closure}, and the preceding
  text). } 
The above equations imply that the gauge algebra generators close
according to
\begin{equation}
  \label{eq:closure}
  {}[T_{M},T_{N}] = -T_{MN}{}^{P}\,T_{P} \;, \quad\quad\quad  k^B{}_M\partial_Bk^A{}_N-k^B{}_N\partial_Bk^A{}_M = T_{MN}{}^P\,
  k^A{}_P \ ,
\end{equation}
so that the structure constants of the gauge group are contained in
$-T_{MN}{}^{P}\equiv -\Theta_{M}{}^{{\sf a}} \,(t_{{\sf
    a}})_{N}{}^{P}$, as is required by the gauge group embedding in
$G_\mathrm{symp}$. This observation was in fact used as input when
deriving (\ref{eq:clos}). Note, however, that the gauge group
structure constants are not necessarily identical to $-T_{MN}{}^{P}$,
as they may differ by terms that vanish upon contraction with the
embedding tensor $\Theta_P{}^{\sf a}$ or $\Theta_P{}^{\sf m}$. This
explains why the $T_{MN}{}^P$ are not necessarily antisymmetric in
$M,N$.

Here and henceforth, the embedding tensor will be regarded as a
spurionic object which we allow to transform under the rigid
invariance group $\mathrm{G}_\mathrm{rigid}$, so that the Lagrangian
and transformation rules will remain formally invariant. Therefore the
embedding tensor can be assigned to a (not necessarily irreducible)
representation of $\mathrm{G}_{\mathrm{rigid}}$.  Eventually the
embedding tensor will be frozen to a constant, so that the invariance
under $\mathrm{G}_\mathrm{rigid}$ will be broken. In this context, it
is relevant to note that (\ref{eq:clos}) implies that the embedding
tensor is invariant under the gauge group. The gauge group is thus
contained in the corresponding stability subgroup of
$\mathrm{G}_{\mathrm{rigid}}$. From symmetrizing the first constraint
(\ref{eq:clos}) in $(MN)$ and making use of the linear conditions
(\ref{eq:lin}) and (\ref{eq:sp-constraint}), one further derives that
$\Omega^{MN}\,\Theta_{M}{}^{{\sf a}}\Theta_{N}{}^{{\sf b}}\, (t_{{\sf
    b}})_{P}{}^{Q} $ must vanish. Hence,
\begin{equation}
  \label{eq:locally-em}
  \Omega^{MN}\,\Theta_{M}{}^{\sf a}\Theta_{N}{}^{\sf b}~=~0
  \;\;\Longleftrightarrow\; \;
  \Theta^{\Lambda\,[\sf a}\Theta_{\Lambda}{}^{{\sf b}]} =0 \;,
\end{equation}
which implies that the charges in the vector multiplet sector are
mutually local, so that an electric/magnetic duality must exist that
converts all the charges to electric ones. Likewise, one derives from
the second constraint (\ref{eq:clos}),
\begin{equation}
  \label{eq:locally-em-2}
  \Omega^{MN}\,\Theta_{M}{}^{\sf a}\Theta_{N}{}^{\sf m}~=~0
  \;\;\Longleftrightarrow\; \;
  \Theta^{\Lambda\,[\sf a}\Theta_{\Lambda}{}^{{\sf m}]} =0 \;,
\end{equation}
which implies that the charges in the hypermultiplet sector are
mutually local with the vector multiplet charges. It is clear that
gauge fields that couple exclusively to charges associated to
hypermultiplets are not restricted by (\ref{eq:locally-em}) and
(\ref{eq:locally-em-2}). Their corresponding gauge groups are
necessarily abelian. To ensure that those charges are also mutually
local, we must impose an additional constraint,
\begin{equation}
  \label{eq:locally-em-3}
   \Omega^{MN}\,\Theta_{M}{}^{\sf m}\Theta_{N}{}^{\sf n}~=~0
  \;\;\Longleftrightarrow\; \;
  \Theta^{\Lambda\,[\sf m}\Theta_{\Lambda}{}^{{\sf n}]} =0 \;,
\end{equation}
which is obviously not related to the closure of the gauge algebra. As
it turns out, the relations (\ref{eq:locally-em}),
(\ref{eq:locally-em-2}) and (\ref{eq:locally-em-3}) play an crucial role
when discussing the Lagrangian.

Generically only a subset of the gauge fields will be involved in
the gauging, so that the embedding tensor will project out a
restricted set of (linear combinations of) gauge fields; the rank
of the tensor determines the dimension of the gauge group, up to
possible central extensions associated with abelian factors.

As stressed before, the generators $T_{MN}{}^P$ are not required
to be antisymmetric in $M,N$. The symmetric part can be written as
follows,
\begin{equation}
  \label{eq:Z-d}
  T_{(MN)}{}^{P}= Z^{P,{\sf a}}\, d_{{\sf a} \,MN} \;,
\end{equation}
with
\begin{eqnarray}
  \label{eq:def-Z-d}
d_{{\sf a}\, MN} &\equiv& (t_{\sf a})_M{}^P\, \Omega_{NP}\,,\nonumber\\
Z^{M,{\sf a}}&\equiv&\ft12\Omega^{MN}\Theta_{N}{}^{{\sf a}} \quad
\Longrightarrow \quad \left\{
\begin{array}{rcr}
Z^{\Lambda{\sf a}} &\!\!=\!\!& \ft12\Theta^{\Lambda{\sf a}} \,,\\[1ex]
Z_{\Lambda}{}^{{\sf a}} &\!\!=\!\!& -\ft12\Theta_{\Lambda}{}^{{\sf a}} \,,
\end{array}
\right.
\end{eqnarray}
so that $d_{{\sf a}\,MN}$ defines an
$\mathrm{Sp}(2n+2,\mathbb{R})$-invariant tensor symmetric in
$(MN)$. Likewise one can introduce a similar tensor $Z^{M,\sf m}$,
relevant for the hypermultiplets, by
\begin{equation}
  \label{eq:def-Z-m}
  Z^{M,{\sf m}}\equiv \ft12\Omega^{MN}\Theta_{N}{}^{{\sf m}} \quad
  \Longrightarrow \quad \left\{
    \begin{array}{rcr}
      Z^{\Lambda{\sf m}} &\!\!=\!\!& \ft12\Theta^{\Lambda{\sf m}} \,,\\[1ex]
      Z_{\Lambda}{}^{{\sf m}} &\!\!=\!\!& -\ft12\Theta_{\Lambda}{}^{{\sf m}} \,,
    \end{array}
  \right.
\end{equation}
Subsequently we note that the constraints (\ref{eq:locally-em}),
(\ref{eq:locally-em-2}) and \eqref{eq:locally-em-3} can now be written
as,
\begin{equation}
  \label{eq:Z-Theta}
  Z^{M,{\sf a}} \,\Theta_M{}^{\sf b} =0 = Z^{M,{\sf a}} \,
  \Theta_M{}^{\sf m}\,, \qquad Z^{M,\sf m}\,\Theta_M{}^{\sf a} = 0=
  Z^{M,{\sf m}} \,\Theta_M{}^{\sf n} \,.
\end{equation}
The latter implies that $Z^{M,{\sf a}}$ and $Z^{M,{\sf m}}$ vanish
when contracted with the gauge-group generators $T_{M}$. Because of
these constraints, only the antisymmetric part of $T_{MN}{}^P$ will
appear in the commutation relation (\ref{eq:closure}). What remains is
to consider the Jacobi identity on the generators $T_M$. Explicit
calculation based on (\ref{eq:closure}) leads to
\begin{equation}
  \label{eq:Jacobi-X}
   T_{[NP}{}^R\,T_{Q]R}{}^M =
   \ft23 Z^{M,{\sf a}}\,d_{{\sf a} R[ N}\,T_{PQ]}{}^R \,,
\end{equation}
which shows that the Jacobi identity holds up to terms that vanish
upon contraction with the embedding tensor. In the following section
we will describe how to introduce a consistent gauging in this
non-standard situation.
\section{The gauge hierarchy}
\label{sec:gauge-hierarchy} \setcounter{equation}{0}
To compensate for the lack of closure noted in the previous
section, and, at the same time, to avoid unwanted degrees of
freedom, the strategy is to introduce an extra gauge invariance for
the gauge fields, in addition to the usual nonabelian gauge
transformations,
\begin{equation}
  \label{eq:A-var}
  \delta W_\mu{}^M =  \mathcal{D}_\mu\Lambda^M-
  g\big[Z^{M,{\sf a}}\,\Xi_{\mu\,{\sf a}} +Z^{M,{\sf
      m}}\,\Xi_{\mu\,{\sf m }}\big] \,,
\end{equation}
where the $\Lambda^M$ are the gauge transformation parameters and the
covariant derivative reads, $\mathcal{D}_\mu\Lambda^M
=\partial_\mu\Lambda^M + g\, T_{PQ}{}^M\,W_\mu{}^P\Lambda^Q$. The
transformations proportional to $\Xi_{\mu\,{\sf a}}$ and
$\Xi_{\mu\,{\sf m}}$ enable one to gauge away those vector fields that
are in the sector where the Jacobi identity is not satisfied (this
sector is perpendicular to the embedding tensor by virtue of
(\ref{eq:Z-Theta})). Note that the covariant derivative is invariant
under the transformations parametrized by $\Xi_{\mu\,{\sf a}}$ and
$\Xi_{\mu\,{\sf m}}$, because of the contraction of the gauge fields
$W_\mu{}^M$ with the generators $T_M$. However, gauge transformations
do no longer form a group by themselves, as is reflected in the
commutation relation,
\begin{equation}
  \label{eq:gauge-commutators}
  {}[\delta(\Lambda_1),\delta(\Lambda_2)] = \delta(\Lambda_3) +
  \delta(\Xi_{{\sf a}\,3}) \,,
\end{equation}
where
\begin{align}
  \label{eq:gauge-parameters}
  \Lambda_3{}^M =&\,  g\,T_{[NP]}{}^M \Lambda_1^N\Lambda_2^P\,,
  \nonumber\\
  \Xi_{3 \mu\,{\sf a}} =&\,   d_{{\sf a} NP}( \Lambda_1^N
  \mathcal{D}_\mu\Lambda_2^P - \Lambda_2^N \mathcal{D}_\mu
  \Lambda_1^P) \,,
\end{align}
with $T_{M{\sf a}}{}^{\sf b}= -\Theta_M{}^{\sf c} f_{{\sf c}{\sf
    a}}{}^{\sf b}$ the gauge group generators in the adjoint
representation of $\mathrm{G}_{\mathrm{symp}}$. As it turns out, this
commutation relation forms the beginning of a full hierarchy of vector
and tensor gauge fields that form a closed algebra
\cite{deWit:2005hv,deWit:2008ta}. Other commutators involving
$\delta(\Lambda)$, $\delta(\Xi_{\sf a})$ and $\delta(\Xi_{\sf m})$
vanish on the gauge fields $W_\mu{}^\Lambda$, so that those can only be
uncovered for the higher-rank tensor gauge fields that we will
introduce shortly.

Non-abelian field strengths associated with the gauge fields
$W_\mu{}^M$ follow from the Ricci identity, $[D_\mu,D_\nu]= - g
\mathcal{F}_{\mu\nu}{}^M\,T_M$, and depend only on the
antisymmetric part of $T_{MN}{}^P$,
\begin{equation}
  \label{eq:field-strength}
  {\cal  F}_{\mu\nu}{}^M =\partial_\mu W_\nu{}^M -\partial_\nu W_\mu{}^M + g\,
  T_{[NP]}{}^M \,W_\mu{}^N W_\nu{}^P \,.
\end{equation}
Because of the lack of closure expressed by (\ref{eq:Jacobi-X}),
these field strengths do not satisfy the Palatini identity,
\begin{equation}
  \label{eq:Palatini}
  \delta\mathcal{F}_{\mu\nu}{}^M = 2\, \mathcal{D}_{[\mu}\delta W_{\nu]}{}^M -
  2 g\, T_{(PQ)}{}^M \,W_{[\mu}{}^P \,\delta W_{\nu]}{}^Q\,,
\end{equation}
under arbitrary variations $\delta W_\mu{}^M$, because of the last
term, which cancels upon multiplication with the generators $T_M$.
The result (\ref{eq:Palatini}) shows in particular that
$\mathcal{F}_{\mu\nu}{}^M$ transforms under the combined gauge
transformations \eqref{eq:A-var} as
\begin{equation}
  \label{eq:delta-cal-F}
  \delta\mathcal{F}_{\mu\nu}{}^M= g\, \Lambda^P T_{NP}{}^M
  \,\mathcal{F}_{\mu\nu}{}^N - 2 g\, Z^{M,{\sf a}} \big(\mathcal{D}_{[\mu}
  \Xi_{\nu]{\sf a}} +d_{{\sf a} PQ} \,W_{[\mu}{}^P\,\delta
  W_{\nu]}{}^Q\big) - 2 g\, Z^{M,{\sf m}} \,\mathcal{D}_{[\mu}
  \Xi_{\nu]{\sf m}}  \,,
\end{equation}
and is therefore not covariant. In deriving this one makes use of the
fact that the tensors $Z^{M,{\sf a}}$ and $Z^{M,{\sf m}}$ are
invariant under the gauge group. The covariant derivative on
$\Xi_{\nu{\sf a}}$ is defined by $\mathcal{D}_\mu \Xi_{\nu{\sf
    a}}= \partial_\mu \Xi_{\nu{\sf a}} - g W_\mu{}^M T_{M{\sf
    a}}{}^{\sf b} \Xi_{\nu{\sf b}}$, and similarly for $\Xi_{\nu{\sf m}}$. These tensor fields belong to the
adjoint representation of the group $G_\mathrm{symp}$.

The standard strategy is therefore to define modified field
strengths,
\begin{equation}
  \label{eq:modified-fs}
  {\cal H}_{\mu\nu}{}^M= {\cal F}_{\mu\nu}{}^M
  + g\big[ Z^{M,{\sf a}} \,B_{\mu\nu\,{\sf a}} + Z^{M,{\sf m}}
  \,B_{\mu\nu\,{\sf m}}\big] \;,
\end{equation}
by introducing new tensor fields $B_{\mu\nu\,{\sf a}}$ and
$B_{\mu\nu\,{\sf m}}$ with suitably chosen gauge transformation
rules, so that covariant results are obtained. This implies that
the variation of the tensor fields should in any case absorb the
unwanted non-covariant terms in \eqref{eq:delta-cal-F}.  At this
point we recall that the invariance transformations in the
ungauged case transform on the field strengths $G_{\mu\nu}{}^M$,
defined in \eqref{eq:GM}, according to a subgroup of
$\mathrm{Sp}(2n+2,\mathbb{R})$ (cf. (\ref{eq:em-duality})). The
field strengths $G_{\mu\nu}{}^M$ consist of the abelian field
strengths $F_{\mu\nu}{}^\Lambda$ and the dual field strengths
$G_{\mu\nu\Lambda}$. The latter were decomposed in
\eqref{eq:def-G-magn} in the form
$G^-_{\mu\nu\Lambda}=F_{\Lambda\Sigma}\,F^-_{\mu\nu}{}^\Sigma
-2\mathrm{i} \mathcal{O}^-_{\mu\nu\Lambda}$. Obviously, in the
presence of the non-abelian gauge interactions, the abelian field
strengths $F_{\mu\nu}{}^\Lambda$ should now be replaced by
\eqref{eq:modified-fs}. Hence it is natural to define new
covariant field strengths according to
\begin{equation}
  \label{eq:nonabelian-fs-array}
  \mathcal{G}_{\mu\nu}{}^M = \begin{pmatrix}
  \mathcal{H}_{\mu\nu}{}^\Lambda \\
  \mathcal{G}_{\mu\nu\Lambda}
  \end{pmatrix}
\end{equation}
with
\begin{eqnarray}
  \label{eq:cal-G}
  \mathcal{G}^-_{\mu\nu}{}^\Lambda &=& \mathcal{H}^-_{\mu\nu}{}^\Lambda\,,
  \nonumber\\
  \mathcal{G}^-_{\mu\nu\Lambda} &=& F_{\Lambda\Sigma}
  \,\mathcal{H}^-_{\mu\nu}{}^\Sigma - 2\mathrm{i}
  \mathcal{O}^-_{\mu\nu\Lambda} \,.
\end{eqnarray}
Just as in section \ref{vector-multiplets-duality}, there exist
corresponding supercovariant field strengths
$\hat{\mathcal{G}}_{\mu\nu}{}^M$ that will appear in the supersymmetry
transformations of the vector multiplet fermion fields. Those will be
discussed in the next section. Just as before, the field strengths
$\hat{\mathcal{G}}_{\mu\nu}{}^M$ and $\mathcal{G}_{\mu\nu}{}^M$ will
only differ by fermionic bilinears and by terms proportional to the
tensor field of the Weyl multiplet.

Following \cite{deWit:2005ub} we subsequently introduce the following
transformation rule for $B_{\mu\nu{\sf a}}$ and $B_{\mu\nu{\sf m}}$
(contracted with $Z^{M,{\sf a}}$ and $Z^{M,{\sf m}}$, respectively,
because only these combinations will appear in the Lagrangian),
\begin{align}
  \label{eq:B-transf-0}
   Z^{M,{\sf a}}\, \delta B_{\mu\nu\,{\sf a}} =&\, 2\,Z^{M,{\sf a}}
   \big(\mathcal{D}_{[\mu} \Xi_{\nu]{\sf a}} + d_{{\sf a}\,NP} W_{[\mu}{}^N \delta
   W_{\nu]}{}^P\big)  - 2\,T_{(NP)}{}^M  \Lambda^P
   \mathcal{G}_{\mu\nu}{}^N    \,,\nonumber\\
   Z^{M,{\sf m}}\, \delta B_{\mu\nu\,{\sf m}} =&\, 2\,Z^{M,{\sf m}}
   \, \mathcal{D}_{[\mu} \Xi_{\nu]{\sf m}}\,.
\end{align}
Note that $B_{\mu\nu\, {\sf a}}$ has variations proportional to
$\Xi_{\mu {\sf m}}$ through the term $\delta W_{\mu}{}^M$
(cf. \eqref{eq:A-var}). As a result of \eqref{eq:B-transf-0} the
modified field strengths (\ref{eq:modified-fs}) are invariant under
tensor gauge transformations. Under the vector gauge transformations
we derive the following result,
\begin{align}
  \label{eq:delta-G/H}
  \delta \mathcal{G}^-_{\mu\nu}{}^\Lambda =&\, - g\,\Lambda^P
  T_{PN}{}^\Lambda \,\mathcal{G}^-_{\mu\nu}{}^N  - g\,\Lambda^P
  T^\Gamma{}_P{}^\Lambda \,
  (\mathcal{G}^-_{\mu\nu} - \mathcal{H}^-_{\mu\nu})_\Gamma \,,
  \nonumber\\
  \delta \mathcal{G}^-_{\mu\nu\Lambda} =&\, - g\,\Lambda^P
  T_{PN\Lambda} \, \mathcal{G}^-_{\mu\nu}{}^N  -g \,
  F_{\Lambda\Sigma}\,\Lambda^P T^\Gamma{}_P{}^\Sigma\,
  (\mathcal{G}^-_{\mu\nu} - \mathcal{H}^-_{\mu\nu})_\Gamma\,,
  \nonumber\\
  \delta(\mathcal{G}^-_{\mu\nu} - \mathcal{H}^-_{\mu\nu})_\Lambda
  =&\,  g \, \Lambda^P(  T^\Gamma{}_{P\Lambda} -T^\Gamma{}_P{}^\Sigma
  \, F_{\Sigma\Lambda})\,
  (\mathcal{G}^-_{\mu\nu} - \mathcal{H}^-_{\mu\nu})_\Gamma\,.
\end{align}
Hence $\delta\mathcal{G}_{\mu\nu}{}^M=-g\,\Lambda^PT_{PN}{}^M\,
\mathcal{G}_{\mu\nu}{}^N$, just as the variation of the abelian field
strengths $G_{\mu\nu}{}^M$ in the absence of charges, up to terms
proportional to $\Theta^{\Lambda,{\sf a}}(\mathcal{G}_{\mu\nu}
-\mathcal{H}_{\mu\nu})_\Lambda$.  According to \cite{deWit:2005ub},
the latter terms represent a set of field equations, and so the last
equation of (\ref{eq:delta-G/H}) expresses the well-known fact that,
under a symmetry, field equations transform into field equations.  As
a result the gauge algebra on the tensors $\mathcal{G}{\mu\nu}{}^M$
closes according to (\ref{eq:gauge-commutators}), up to the same field
equations.

In order that the Lagrangian corresponding to
\eqref{eq:vector-lagrangian-decom} becomes invariant under vector and
tensor gauge transformations, we have to make a number of
changes. First of all, we replace the covariant derivatives on the
scalars and spinors by gauge-covariant derivatives. This ensures the
invariance of $\mathcal{L}_{\mathrm{kin}}^{(1)}$,
$\mathcal{L}_{\mathrm{conf}}$ and $\mathcal{L}_\mathrm{hyper}$, given
in (\ref{eq:lagrangian-matter-cov}), (\ref{eq:vector-conf}) and
\eqref{eq:hyper}, respectively.  The Lagrangian for the auxiliary
fields (\ref{eq:vector-aux}) is already gauge-invariant. In the
following we therefore concentrate on $\mathcal{L}_\mathrm{kin}^{(2)}$
(\ref{eq:kinetic-vector}) which depends on the abelian field strengths
$F_{\mu\nu}{}^\Lambda$. These abelian field-strengths are now replaced
by $\mathcal{H}_{\mu\nu}{}^\Lambda$, so that
\begin{equation}
  \label{eq:def-cal-G}
  \mathcal{G}_{\mu\nu\,\Lambda} = \mathrm{i}e \,
  \varepsilon_{\mu\nu\rho\sigma}\,
  \frac{\partial\mathcal{L}_{\mathrm{vector}}}
  {\partial{\mathcal{H}}_{\rho\sigma}{}^\Lambda}   \;.
\end{equation}
The Lagrangian $\mathcal{L}_\mathrm{kin}^{(2)}$ therefore reads,
\begin{align}
  \label{eq:cov-kinetic-vector}
  e^{-1} \mathcal{L}_\mathrm{kin}^{(2)}=&\,\ft14
  \mathrm{i}\left[F_{\Lambda\Sigma} \,\mathcal{H}^{-\Lambda}_{\mu\nu}
  \mathcal{H}^{-\Sigma\,\mu\nu} -\bar F_{\Lambda\Sigma}
  \,\mathcal{H}^{+\Lambda}_{\mu\nu}
  \mathcal{H}^{+\mu\nu\Sigma}\right] \nonumber\\[.6ex]
  &\,+\big[\mathcal{O}^-_{\mu\nu\Lambda}
  \mathcal{H}^{-\mu\nu\Lambda}-N^{\Lambda\Sigma}
  \mathcal{O}^-_{\mu\nu\Lambda}\mathcal{O}^{-\mu\nu}{}_{\Sigma}+\mbox{h.c.}
  \big]\,.
\end{align}
It is separately invariant under the tensor gauge transformations,
because the tensors $\mathcal{H}$ are invariant.

However, the Lagrangian \eqref{eq:vector-lagrangian-decom} is not
invariant under the vector gauge transformations. To establish this
one has to take into account that also the other fields of the vector
multiplets transform under the gauge group. For instance, there are
contributions from infinitesimal gauge transformations of $F_{\Lambda
  \Sigma}$ and $\mathcal{O}_{\mu\nu\Lambda}$, which follow from
\eqref{eq:dual-symm-F-der} and \eqref{eq:dual-O},
\begin{align}
  \label{eq:var-F-and-O}
  \delta F_{\Lambda \Sigma} =&\, g \Lambda^M
  \big( - T_{M\Lambda\Sigma}+2 \,T_{M(\Lambda}{}^\Gamma F_{\Sigma)
    \Gamma} + F_{\Lambda \Gamma} T_M{}^{\Gamma \Xi}
  F_{\Xi\Sigma}\big)\,, \nonumber\\
  \delta \mathcal{O}^{-}_{\mu\nu\Lambda} =&\, g \Lambda^M
  \mathcal{O}^{-}_{\mu\nu\Sigma} \,\big(T_{M\Lambda}{}^{\Sigma}
    +T_M{}^{\Sigma \Gamma} F_{\Gamma \Lambda}\big)\,.
\end{align}
Nevertheless, it was shown in \cite{deWit:2005ub} that this is
still not sufficient for gauge invariance, and it is necessary to
introduce an additional, universal, term to the Lagrangian, equal
to,
\begin{align}
  \label{eq:Ltop}
  {\cal L}_{\rm top} =&\,
  \ft18\mathrm{i} g\, \varepsilon^{\mu\nu\rho\sigma}\,
  \big(\Theta^{\Lambda{\sf a}}\,B_{\mu\nu\,{\sf
      a}}+\Theta^{\Lambda{\sf m}}\,B_{\mu\nu\,{\sf m}}\big)
  \nonumber\\
  &\qquad \times
  \big(2\,\partial_{\rho} W_{\sigma\,\Lambda} + g
  T_{MN\,\Lambda} \,W_\rho{}^M W_\sigma{}^N
  -\ft14g\Theta_{\Lambda}{}^{{\sf b}}B_{\rho\sigma\,{\sf b}}
  -\ft14g\Theta_{\Lambda}{}^{{\sf n}}B_{\rho\sigma\,{\sf n}} \big)
  \nonumber\\[.9ex]
  &\,
  + \ft13\mathrm{i} g\, \varepsilon^{\mu\nu\rho\sigma} T_{MN\,\Lambda}\,
  W_{\mu}{}^{M} W_{\nu}{}^{N}
  \big(\partial_{\rho} W_{\sigma}{}^{\Lambda}
  +\ft14 gT_{PQ}{}^{\Lambda} W_{\rho}{}^{P}W_{\sigma}{}^{Q}\big)
  \nonumber\\[.9ex]
  &\,
  + \ft16 \mathrm{i} g\,\varepsilon^{\mu\nu\rho\sigma}T_{MN}{}^{\Lambda}\,
  W_{\mu}{}^{M} W_{\nu}{}^{N}
  \big(\partial_{\rho} W_{\sigma}{}_{\Lambda}
  +\ft14 gT_{PQ\Lambda} W_{\rho}{}^{P}W_{\sigma}{}^{Q}\big)
  \;.
\end{align}
The first term represents a topological coupling of the
anti-symmetric tensor fields with the magnetic gauge fields; the
last two terms are a generalization of the Chern-Simons-like terms
that were first found in \cite{deWit:1984px}.

Under arbitrary variations of the vector and tensor fields,
\eqref{eq:cov-kinetic-vector} and \eqref{eq:Ltop} yield (up to
total derivative terms),
\begin{align}
  \label{eq:var-L-totaal}
  e^{-1}\left(\delta\mathcal{L}_\mathrm{kin}^{(2)}+
    \delta\mathcal{L}_{\mathrm{top}}\right)
  =&\, - \ft14 \mathrm{i} g\, \left(\mathcal{G}^{+\mu\nu
      M}-\mathcal{H}^{+\mu\nu
  M}\right)
  \,\Theta_M{}^{{\sf a}} (\delta B_{\mu\nu{\sf a}} - 2d_{{\sf a}
  PQ} W_\mu{}^P \delta W_\nu{}^Q) \nonumber\\
  &\, - \ft14 \mathrm{i} g\, \left(\mathcal{G}^{+\mu\nu
      M}-\mathcal{H}^{+\mu\nu
  M}\right) \,\Theta_M{}^{{\sf m}} \,\delta B_{\mu\nu{\sf
  m}}\nonumber\\
 &\, +\mathrm{i}
   \mathcal{G}^{+\mu\nu M}\Omega_{MN} \, \mathcal{D}_\mu\delta
  W_{\nu}{}^N+ \mathrm{h.c.} \,.
\end{align}
Under the tensor gauge transformations this variation becomes
equal to,
\begin{equation}
  \label{eq:var-L-tensor} e^{-1}\big(
  \delta\mathcal{L}_\mathrm{kin}^{(2)}+\delta\mathcal{L}_{\mathrm{top}}\big)
  = \mathrm{i} g \mathcal{H}^{+\mu\nu M}\big[ \Theta_{M}{}^{\sf a}
  \mathcal{D}_{\mu}\Xi_{\nu \sf a} +\Theta_{M}{}^{\sf m}
  \mathcal{D}_{\mu}\Xi_{\nu \sf m} \big] + \mathrm{h.c.}\, .
\end{equation}
We already demonstrated that $\mathcal{L}_\mathrm{kin}^{(2)}$ is
separately invariant under tensor gauge transformations, so that the
above terms originate exclusively from the variation of
$\mathcal{L}_{\mathrm{top}}$.  The expression \eqref{eq:var-L-tensor}
turns out to be equal to a total derivative because there exists a
Bianchi identity,
\begin{equation}
  \label{eq:Bianchi-DCov}
  \mathcal{D}_{[\mu}{\cal H}_{\nu\rho]}{}^{M} = \ft13g\big[Z^{M,{\sf
  a}}\,\mathcal{H}_{\mu\nu \rho \,\sf a} +Z^{M,{\sf
  m}}\,\mathcal{H}_{\mu\nu \rho \,\sf m} \big]\,,
\end{equation}
and because the embedding tensor is gauge invariant. Here the
gauge-covariant field strengths of the tensor fields are defined as,
\begin{align}
  \label{eq:H-3}
  \mathcal{H}_{\mu \nu \rho\, \sf a} =&\, 3\,
  \mathcal{D}_{[\mu}
  B_{\nu\rho]\,\sf a} +6 \,d_{{\sf a}\, NP}\,W_{[\mu}{}^{N}
  \left(\partial_{\nu} W_{\rho]}{}^P+ \ft13 g T_{[RS]}{}^{P}
    W_{\nu}{}^{R}W_{\rho]}{}^{S}+(\mathcal{G}-\mathcal{H})_{\nu
      \rho]}{}^P\right)\,,\nonumber\\
  \mathcal{H}_{\mu \nu \rho\, \sf m} =&\, 3\,
  \mathcal{D}_{[\mu}   B_{\nu\rho]\,\sf m} \,,
\end{align}
where $\mathcal{D}_\mu B_{\nu \rho {\sf a}}= \partial_\mu
B_{\nu\rho{\sf a}} - g W_\mu{}^M T_{M{\sf a}}{}^{\sf b} B_{\nu\rho{\sf
    b}}$, and likewise for $\mathcal{D}_\mu B_{\nu \rho {\sf m}}$. The
fully gauge-covariant derivative of $\mathcal{H}_{\mu\nu}{}^M$ takes
the form,
\begin{align}
  \label{eq:Cov-Der-H}
  \mathcal{D}_{\rho} \mathcal{H}_{\mu \nu}{}^M=&\,
  \partial_{\rho}\mathcal{H}_{\mu \nu}{}^M+gW_{\rho}{}^P\,
  T_{PN}{}^M\,\mathcal{G}_{\mu \nu}{}^N+gW_{\rho}{}^P\,
  T_{NP}{}^M\,(\mathcal{G}-\mathcal{H})_{\mu\nu}{}^N\nonumber\\
  = &\,
  \partial_{\rho}\mathcal{H}_{\mu \nu}{}^M+gW_{\rho}{}^P\,
  T_{PN}{}^M\,\mathcal{H}_{\mu \nu}{}^N
  +2\, gW_{\rho}{}^P\,Z^{M,{\sf a}}d_{{\sf
     a}PN}\,(\mathcal{G}-\mathcal{H})_{\mu\nu}{}^N\,,
\end{align}
Observe that the covariantization proportional to
$(\mathcal{G}-\mathcal{H})_{\mu\nu}{}^N$ is not generated by
partially integrating the right-hand side of
\eqref{eq:var-L-tensor}, but it vanishes upon contraction with the
embedding tensor. So does the right-hand side of
\eqref{eq:Bianchi-DCov}, so that \eqref{eq:var-L-tensor} is indeed
a total derivative.

As was mentioned before, the combined gauge invariance of the vector
and tensor gauge fields are important to ensure that the number of
physical degrees of freedom will not change by the introduction of the
magnetic vector gauge fields and the tensor gauge fields
\cite{deWit:2005ub}. The combined gauge algebra is consistent for the
tensor fields upon projection with the embedding tensor, which is
sufficient because the action depends only on these projected
fields. If this were not the case, new tensor fields of higher rank
would have been required \cite{deWit:2005hv}. The projection with the
embedding tensor will determine in which fields the physical degrees
of freedom can reside. The precise way in which the number of physical
degrees of freedom are accounted for is therefore rather subtle. From
\eqref{eq:var-L-totaal} it is indeed clear that the components of
the tensor fields that are projected to zero by multiplication with
$\Theta^{\Lambda{\sf{a}}}$ or $\Theta^{\Lambda{\sf m}}$, are simply
not present in the action. Their absence can be regarded as the result
of an additional gauge invariance.  In addition, there are
transformations of the tensor fields linear in $(\mathcal{G}
-\mathcal{H})_{\mu\nu\Lambda}$ that leave the Lagrangian invariant
\cite{de Vroome:2007zd,de Wit:2007mt},
\begin{align}
  \label{eq:Delta-invariance}
  \Theta^{\Lambda{\sf a}}\delta B_{\mu\nu{\sf a}} =&\,
  \Delta_1^{[\Lambda\Sigma]}\,
  (\mathcal{G} -\mathcal{H})^+_{\mu\nu\Sigma} + \mathrm{h.c.}
  \,,\nonumber\\
  \Theta^{\Lambda{\sf a}}\delta B_{\mu\nu{\sf a}} =&\,
  \Delta_2^{(\Lambda\Sigma)\rho}{}_{[\mu}\,
  (\mathcal{G} -\mathcal{H})_{\nu]\rho\Sigma}  \,,
\end{align}
where $\Delta_1^{\Lambda\Sigma}$ is an arbitrary complex parameter,
and $\Delta_2^{\Lambda\Sigma\rho}{}_\mu$ is real and
traceless. Similar transformations exist for variations contracted
with $\Theta^{\Lambda\sf m}$.  Often these transformations emerge when
verifying the validity of the supersymmetry algebra, something that we
will discuss in section \ref{sec:rest-supersymm-non}.

A similar situation arises with the magnetic gauge fields
$W_{\mu\Lambda}$. Under variations of the gauge fields $W_\mu{}^M$
one derives,
\begin{equation}
  \label{eq:A-field-eq}
  \delta\mathcal{L}_{\mathrm{kin}}^{(2)}  +
  \delta\mathcal{L}_{\mathrm{top}} =  \ft12 \mathrm{i}\,
  \varepsilon^{\mu\nu\rho\sigma} \,\mathcal{D}_\nu \mathcal{G}_{\rho\sigma}{}^M
  \Omega_{MN} \delta W_\mu{}^N \,,
\end{equation}
where $\mathcal{L}_{\mathrm{kin}}^{(2)}$ was defined in
(\ref{eq:cov-kinetic-vector}), up to a total derivative and up to
terms that vanish as a result of the field equation for
$B_{\mu\nu\,\sf a}$.  Substituting~(\ref{eq:Bianchi-DCov}) we can
rewrite (\ref{eq:A-field-eq}) as follows,
\begin{equation}
  \label{eq:A-field-eq-2}
  \delta\mathcal{L}_{\mathrm{kin}}^{(2)} +
  \delta\mathcal{L}_{\mathrm{top}} = \ft12 \mathrm{i}\,
  \varepsilon^{\mu\nu\rho\sigma} \left[- \mathcal{D}_\nu
  \mathcal{G}_{\rho\sigma\Lambda} \,\delta W_\mu{}^\Lambda + \ft16 g\big(
  \mathcal{H}_{\nu\rho\sigma{\sf a}} \,\Theta^{\Lambda{\sf a}}
  +\mathcal{H}_{\nu\rho\sigma{\sf m}} \,\Theta^{\Lambda{\sf m}}\big) \delta
  W_{\mu\Lambda} \right] \,.
\end{equation}
Because the minimal coupling of the gauge fields to matter fields is
always proportional to the embedding tensor, the full Lagrangian does
not change under variations of the magnetic gauge fields that are
projected to zero by the embedding tensor components
$\Theta^{\Lambda{\sf a}}$ or $\Theta^{\Lambda{\sf m}}$, up to terms
that are generated by the variations of the tensor fields through the
`universal' variation, $\delta B_{\mu\nu{\sf a}}=2\,d_{{\sf a}PQ}
W_{[\mu}{}^P \delta W_{\nu]}{}^Q$.

All these gauge symmetries have a role to play in balancing the
degrees of freedom. In \cite{deWit:2005ub} a precise accounting of
all gauge symmetries was bypassed in the analysis. Observe that
not all these symmetries have a bearing on the dynamical modes of
the theory as they also act on fields that only play an auxiliary
role.

\section{General gaugings: the superconformal algebra and the
  Lagrangian}
\label{sec:rest-supersymm-non}
\setcounter{equation}{0}
When switching on a gauging there are several qualitative changes that
are of interest. First of all, the superconformal algebra will no
longer be realized off shell (i.e. without using the equations of
motion) in the vector multiplet sector, at least for gaugings with
magnetic charges. Only for the Weyl multiplet the closure remains
realized off shell. Naturally a generic gauging induces the presence
of vector multiplet fields into the hypermultiplet supersymmetry
transformations. It is therefore not surprising that also the vector
multiplet transformations will generically acquire terms proportional
to the hypermultiplet fields. In this section we will present the full
transformation rules that include new terms of order $g$, and
subsequently we will re-establish the closure for general gaugings. As
it turns out, additional symmetries such as
\eqref{eq:Delta-invariance}, are relevant for the closure. This
feature is well known from previous applications of the embedding
tensor formalism.

A second, not unrelated, feature is that the Lagrangian must be
modified by including masslike terms for the fermions proportional to
$g$, and a scalar potential proportional to $g^2$. The explicit
expressions for these terms, which are relevant for many applications,
will be presented at the end of this section. These modifications are
familiar from $N=2$ supergravity theories with purely electric charges
\cite{deWit:1980tn,deWit:1984px,deWit:1999fp}.

Rigid $N=2$ supersymmetric theories with both electric and magnetic
charges, have been presented in \cite{de Vroome:2007zd}, and it
remains to complete these results in a fully superconformal
setting. It is clear that the modification of the results derived in
\cite{de Vroome:2007zd} must be relatively minor. The supersymmetry
transformations of the matter fields will now become covariant with
respect to the superconformal symmetries, while at the same time they
should remain in accord with the known results for rigid
theories. Modifications that supersede previous work will therefore
mainly involve terms proportional to the gravitino fields. The most
conspicuous ones are those appearing in the supersymmetry
transformations of the tensor fields $B_{\mu\nu \sf a}$ and $B_{\mu\nu
  \sf m}$.

To exhibit this in more detail, let us first present the full Q- and
S-supersymmetry transformations for the hypermultiplet fields. They
follow straightforwardly upon supercovariantizing the rules presented
in section~\ref{sec:intro-hypermultiplets}, including the terms of
order $g$ that were already found in \cite{de Vroome:2007zd},
\begin{align}
\label{eq:conv-rules-hyp} \delta\phi^A=
&\,2\,(\gamma^A_{i\bar\alpha} \,\bar\epsilon^i
    \zeta^{\bar \alpha} + \bar\gamma^{Ai}_{\alpha} \,\bar\epsilon_i
    \zeta^\alpha )\,,\nonumber\\
\delta A_i{}^\alpha+ \delta\phi
  \Gamma_A{}^\alpha{}_\beta A_i{}^\beta = &\,
  2\,\bar\epsilon_i\zeta^\alpha +2\,\varepsilon_{ij}
  G^{\alpha\bar\beta}\Omega_{\bar\beta\bar\gamma}\,\bar\epsilon^j
  \zeta^{\bar\gamma}
  \,,\nonumber\\
  \delta\zeta^\alpha +\delta\phi^A\,
  \Gamma_{A}{}^{\!\alpha}{}_{\!\beta}\, \zeta^\beta =&\, \Slash{D}
  A_i{}^\alpha\,\epsilon^i +2gX^M \,T_{M}{}^\alpha{}_\beta
  A_i{}^\beta\,\varepsilon^{ij}\epsilon_j +A_i{}^\alpha \,\eta^i\,.
\end{align}
where $D_{\mu}$ denotes the derivative fully covariantized with
respect to all the superconformal transformations and the gauge
symmetries. Likewise we present the full Q- and S-supersymmetry
transformations for the vector multiplet fields,
\begin{align}
  \label{eq:conv-rules}
  \delta X^M =&\, \bar{\epsilon}^i\Omega_i{}^M\,,\nonumber\\[.2ex]
  \delta \Omega_i{}^M =&\,  2 \Slash{D}X^M\epsilon_i
  +\hat{Z}_{ij}{}^M\epsilon^j+ \ft{1}{2}\gamma^{\mu
    \nu}\hat{\mathcal{G}}^-_{\mu\nu}{}^M\varepsilon_{ij}\epsilon^j\nonumber\\
  &\,-2g\,{T_{PN}}^M\bar{X}^{P}X^{N}\varepsilon_{ij}\epsilon^j
  +2\,\mathrm{i} g \Omega^{MN}\mu_{ijN}\epsilon^j
  +2X^M\eta_i\,,\nonumber\\[.2ex]
  \delta W_{\mu}{}^M =& \, \varepsilon^{ij} \bar{\epsilon}_i
  (\gamma_{\mu} \Omega_j{}^M +2\,\psi_{\mu j} X^M)
  + \varepsilon_{ij}
  \bar{\epsilon}^i (\gamma_{\mu} \Omega^{j\, M} +2\,\psi_\mu{}^j
  \bar X^M)  \,,\nonumber\\[.2ex]
   \delta Y_{ij}{}^{\Lambda}  =&\,
   2\,\bar{\epsilon}_{(i}\Slash{D}\Omega_{j)}{}^{\Lambda}
   +2\,\varepsilon_{ik}\varepsilon_{jl}\bar{\epsilon}^{(k}
   \Slash{D}\Omega^{l)\Lambda}\nonumber\\
   &\,   -4g\,{T_{MN}}^{\Lambda}\big[\bar{\Omega}_{(i}{}^{M}
     \epsilon^k\varepsilon_{j)k}\,\bar{X}^{N}
     -\bar{\Omega}^{kM}\epsilon_{(i}\varepsilon_{j)k}\,X^{N}\big]\nonumber\\
   &\, +4\,\mathrm{i} g\,k^{A\Lambda}
  \big[\varepsilon_{k(i}\,
  \gamma_{j)\bar\alpha A} \bar\epsilon^k\zeta^{\bar\alpha} +
  \varepsilon_{k(i}\, \bar\epsilon_{j)} \zeta^{\alpha} \,
  \bar\gamma^{k}_{\alpha A}   \big]    \,.
\end{align}
Here the moment maps are defined by,
\begin{equation}
  \label{eq:moment-maps-theta}
  \mu_{ij M}= \Theta_M{}^{\sf m} \mu_{ij{\sf m}}\,,
\end{equation}
and the symplectic vector $\hat{Z}_{ij}{}^M$ appearing in $\delta
\Omega_i{}^M$ is given by,
\begin{equation}
  \label{eq:symplectic-Z-hat}
  \hat{Z}_{ij}{}^M =\begin{pmatrix} Y_{ij}{}^\Lambda \cr
    \noalign{\vskip 1.5mm}
    F_{\Lambda\Sigma}\, Y_{ij}{}^\Sigma -\ft12 F_{\Lambda\Sigma\Gamma}
    \,\bar\Omega_i{}^\Sigma\Omega_j{}^\Gamma+2\,\mathrm{i}g
    [\mu_{ij\Lambda}+F_{\Lambda\Sigma}\,\mu_{ij}{}^\Sigma]
  \end{pmatrix}\,.
\end{equation}
This expression differs from the previous one for the ungauged theory,
given in \eqref{eq:symplectic-Z}, by the presence of the moment maps
originating from the hypermultiplet sector. This implies that the
original pseudo-reality condition on $Z_{ij\Lambda}$ must be replaced
by a pseudo-reality condition on $\hat{Z}_{ij\Lambda}$. As this
condition was previously imposed by invoking the field equations for
the auxiliary fields, it follows that those field equations must now
receive modifications proportional to the moment maps, as we shall
confirm later in this section. Note that, in \eqref{eq:conv-rules}, we
refrained from giving the supersymmetry transformation of $\hat
Z_{ij\Lambda}$, which is not an independent field.

Another tensor appearing in $\delta\Omega_i{}^M$, a modification of
the tensor \eqref{eq:symplectic-field-strength}, is the supercovariant
field strength $\hat{\mathcal{G}}_{\mu\nu}{}^M$, which coincides with
the field strengths \eqref{eq:nonabelian-fs-array} up to fermion
bilinears and terms proportional to the tensor field of the Weyl
multiplet. These supercovariant field strengths are defined by,
\begin{align}
  \label{eq:hat-G-gauged}
  \hat{\mathcal{G}}_{\mu\nu}^-{}^\Lambda =&\,
  \hat{\mathcal{H}}_{\mu\nu}^-{}^\Lambda \,, \nonumber \\
  \hat{\mathcal{G}}_{\mu\nu\Lambda}^- =&\,
  F_{\Lambda\Sigma}\,\hat{\mathcal{H}}_{\mu\nu}^-{}^\Sigma -\tfrac18
  F_{\Lambda\Sigma\Gamma} \,\bar\Omega_i{}^\Sigma\gamma_{\mu\nu}
  \Omega_j{}^\Gamma \,\varepsilon^{ij}\,.
\end{align}
where $\hat{\mathcal{H}}_{\mu\nu}{}^\Lambda$ is the supercovariant
extension of \eqref{eq:modified-fs}. In view of (\ref{eq:f-hat}), we
expect the following decomposition for
$\hat{\mathcal{H}}_{\mu\nu}{}^\Lambda$,
\begin{align}
  \label{eq:hat-H}
    \hat{\mathcal{H}}_{\mu\nu}{}^\Lambda =&\, \mathcal{H}_{\mu\nu}{}^\Lambda
  - \varepsilon^{ij} \bar{\psi}_{[\mu\,i}
  (\gamma_{\nu]} \Omega_j{}^\Lambda+ \psi_{\nu] j} X^\Lambda)
  -  \varepsilon_{ij}
  \bar{\psi}_{[\mu}{}^i (\gamma_{\nu]} \Omega^{j\, \Lambda} +\psi_{\nu]}{}^j
  \bar X^\Lambda) \nonumber\\[1mm]
  &\,
  - \tfrac14(X^\Lambda\, T_{\mu\nu ij}\,\varepsilon^{ij} + \bar X^\Lambda\,
  T_{\mu\nu}{}^{ij}\,\varepsilon_{ij}  )\,.
\end{align}
However, in the presence of a gauging, this expression leads to
supersymmetry variations proportional to the gravitini fields induced
by the terms in $\delta\Omega_i{}^\Lambda$ of order $g$. As it turns
out, by suitably adjusting the supersymmetry transformations of the
tensor fields, $\delta B_{\mu\nu \sf a}$ and $\delta B_{\mu\nu \sf
  m}$, one can ensure that the $\hat{\mathcal{H}}_{ab}{}^\Lambda$ will
still transform covariantly under Q- and S-supersymmetry,
\begin{align}
  \label{eq:trasf-QS-hat-cal}
  \delta \hat{\mathcal{H}}_{ab}{}^\Lambda =&\, -2 \,
  \varepsilon_{ij}\,\bar{\epsilon}^i\gamma_{[a}D_{b]}\Omega^{j\Lambda}-2 g\,
  T_{(NP)}{}^\Lambda\bar{X}^N\,\bar{\Omega}_i{}^P\gamma_{ab}\epsilon^i
  \nonumber\\
  &\,
  - 2 \mathrm{i} g \,k^{A \Lambda} \;\gamma_{Ai\bar\alpha}
  \,\bar\zeta^{\bar\alpha} \gamma_{ab}\epsilon^i-\varepsilon^{ij}
  \bar\eta_i\gamma_{ab}\Omega_j{}^\Lambda +\text{h.c.}\,.
\end{align}
As a result the combined transformations of the
tensor fields, $B_{\mu\nu \sf a}$ and $B_{\mu\nu {\sf m}}$, under
tensor and vector gauge transformations and Q- and S-supersymmetry,
now read as follows,
\begin{align}
  \label{eq:Q-S-tensors}
   Z^{M,{\sf a}}\, \delta B_{\mu\nu\,{\sf a}} =&\, 2\,Z^{M,{\sf a}}
   \mathcal{D}_{[\mu} \Xi_{\nu]{\sf a}} + 2\, T_{(NP)}{}^M \big[
   W_{[\mu}{}^N \delta W_{\nu]}{}^P  -  \Lambda^N
   \mathcal{G}_{\mu\nu}{}^P  \big]   \nonumber\\
   &\, - 2\,T_{(NP)}{}^M \big[
  \bar{X}^N\bar{\Omega}_i{}^P\gamma_{\mu\nu}\epsilon^i +
  X^N\bar{\Omega}^{iP}\gamma_{\mu\nu}\epsilon_i 
  + 2\,\bar{X}^N X^P \big(\bar{\epsilon}^i\gamma_{[\mu}\psi_{\nu]i}
  +\bar{\epsilon}_i\gamma_{[\mu}\psi_{\nu]}{}^i\big)
  \big] \,,\nonumber\\[2mm]
   Z^{M,{\sf m}}\, \delta B_{\mu\nu\,{\sf m}} =&\, 2\,Z^{M,{\sf m}}
   \, \mathcal{D}_{[\mu} \Xi_{\nu]{\sf m}}    - 2\mathrm{i}\Omega^{MN}
   k^A{}_N \, \big[ \gamma_{Ai\bar\alpha}
  \,\bar\zeta^{\bar\alpha} \gamma_{\mu\nu}\epsilon^i -
  \bar\gamma^i_{A\alpha}
  \,\bar\zeta^{\alpha} \gamma_{\mu\nu}\epsilon_i  \big] \nonumber\\
  &\,+4\,\mathrm{i}\, \Omega^{MN} \mu_{jkN}\,\varepsilon^{ij}
  \big[\bar{\psi}_{i[\mu}\gamma_{\nu]}\epsilon^k
  +\bar{\psi}^k{}_{[\mu}\gamma_{\nu]}\epsilon_i\big]\,.
\end{align}
Note that the tensors transform covariantly under diffeomorphisms, and
are scale invariant.  As was already alluded to, the moment maps
$\mu_{ijM}$ enter the transformation rules of the vector multiplet
fields. In fact, only the magnetic moment maps $\mu_{ij}{}^\Lambda$
appear in these transformation rules.\footnote{
  The reader may verify that the contribution to $\Omega_i{}^M$
  proportional to $\mu_{ij\Lambda}$ vanishes against a similar
  contribution contained in $\hat Z_{ij}{}^M$. } 
For purely electric charges and corresponding moment maps
$\mu_{ij\Lambda}$, the supersymmetry transformations
\eqref{eq:conv-rules-hyp} and \eqref{eq:conv-rules} reduce to the
transformations presented in \cite{deWit:1984px} and
\cite{deWit:1999fp}.  The latter transformations still realize the
supersymmetry algebra for the vector multiplet fields (but not for the
hypermultiplet fields) without the need for imposing equations of
motion.

Now that the full supersymmetry transformations have been established,
we consider the superconformal algebra. Its most non-trivial
commutation relation is the one of two Q-supersymme\-tries. This
commutation relation, which was already specified in
\eqref{eq:QQ-commutator}, must now be extended with tensor gauge
transformations. Hence
\begin{align}
  \label{eq:QQ-commutator-gauged}
  [\delta(\epsilon_1),\delta(\epsilon_2)] =&\, \xi^{\mu}D_{\mu} +
  \delta_M(\varepsilon)+\delta_K(\Lambda_K)+ \delta_S(\eta) +
  \delta_{\text{gauge}}(\Lambda^M) \nonumber\\
  &\, +\delta_{\text{tensor}}(\Xi_{\mu\,\sf a}) +
  \delta_{\text{tensor}}(\Xi_{\mu\,\sf m})\,,
\end{align}
and it should hold modulo field equations and some of the spurious
symmetries that we discussed in the previous section.  The various
parameters in (\ref{eq:QQ-commutator-gauged}) have already been
specified in \eqref{eq:convparungauged}, except for the parameters of
the tensor gauge transformations, which read,
\begin{align}
  \label{eq:convparxi}
  \Xi_{\mu \,{\sf a}} =&\, -2\,d_{\textsf{a}\,NP}\bar{X}^NX^P\xi_{\mu}
  \,,\nonumber\\
  \Xi_{\mu{\sf m}}=&\, -8\,\mathrm{i}\,\varepsilon^{ij}\mu_{jk{\sf
      m}}\big(\bar{\epsilon}_{2i}\gamma_{\mu}\epsilon_1{}^k
    +\bar{\epsilon}_2{}^k\gamma_\mu \epsilon_{1i}\big)\,,
\end{align}
up to terms that vanish upon contraction with the embedding
tensor.\footnote{ 
  The result for $\Xi_{\mu{\sf m}}$ given in \eqref{eq:convparxi} is
  new compared to previous work. It is determined by verifying the
  commutator \eqref{eq:QQ-commutator-gauged} on the vector and tensor
  gauge fields, as will be discussed in some detail below. 
} 
The combination $\xi^{\mu} D_{\mu}$ denotes an infinitesimal covariant
general coordinate transformation, which includes contributions from
{\it all} the field-dependent gauge transformations such as a Q- and
S-supersymmetry transformation with parameters $-\tfrac12\xi^\rho
\psi_\rho{}^i$ and $-\tfrac12\xi^\rho \phi_\rho{}^i$, or vector
gauge transformations with parameters
$\Lambda^M=-\xi^{\rho}W_{\rho}{}^M$, such that the combined result
takes a supercovariant form. For the corresponding field-dependent
tensor gauge transformations, the parameters take a slightly more
complicated form \cite{de Vroome:2007zd,de Wit:2007mt},
\begin{align}
  \label{eq:cov-transl-tensor}
  \Xi_{\mu\,\textsf{a}}=&\,
  -\xi^{\rho}\left(B_{\rho\mu\,\textsf{a}}+
    d_{\textsf{a}\,NP}W_{\rho}{}^NW_{\mu}{}^P\right)  \,,\nonumber\\
  \Xi_{\mu\,\textsf{m}} =&\, -\xi^{\rho}B_{\rho\mu\,\textsf{m}}  \,.
\end{align}

In what follows we will verify the validity of
(\ref{eq:QQ-commutator-gauged}) on the auxiliary fields
$Y_{ij}{}^\Lambda$, $W_\mu{}^M$ and the tensor fields $B_{\mu\nu{\sf
    a}}$ and $B_{\mu\nu{\sf m}}$, as these are most susceptible to the
presence of the new gauge transformations, thereby exhibiting a
variety of subtleties that play a role. Many aspects of this
evaluation have their counterpart in a similar evaluation of $N=8$
supergravity, which appeared in \cite{de Wit:2007mt}. At this point we
mention two general identities that are relevant in the present
calculations.  They follow from (\ref{eq:derivative-invariance}),
(\ref{eq:fermion-der-invariance}) and (\ref{eq:cov-der-invariance}),
\begin{align}
  \label{eq:id-Z-G}
  T_{(MN)}{}^P X^M \hat{Z}_{ij}{}^N =&\,\ft{1}{2}\,T_{(MN)}{}^P
  \,\bar{\Omega}_i{}^M\Omega_j{}^N-2\mathrm{i} g T_{(MN)}{}^P
  X^M \Omega^{NQ}\mu_{ijQ}\,,\nonumber \\
  T_{(MN)}{}^P X^M \hat{\mathcal{G}}_{\mu\nu}^-{}^N=&\,
  \ft{1}{8}\,T_{(MN)}{}^P \,\varepsilon^{ij}
  \,\bar{\Omega}_i{}^M\gamma_{\mu\nu}\Omega_j{}^N\,.
\end{align}
Of course, in the calculations we must also take into account that the
superconformal gauge fields, $\omega_\mu{}^{ab}$, $f_\mu{}^a$ and
$\phi_\mu{}^i$, depend on the other superconformal fields.

Let us first consider the supersymmetry commutator
(\ref{eq:QQ-commutator-gauged}) on the auxiliary fields
$Y_{ij}{}^\Lambda$. As it turns out, its validity requires to impose
the field equations associated with the tensor fields, which take the
following form,
\begin{eqnarray}
\label{eq:eom-B}
\Theta^{\Lambda\sf a}\, \mathcal{G}_{\mu\nu\Lambda} =
\Theta^{\Lambda\sf a} \,\mathcal{H}_{\mu\nu\Lambda}\,,\quad\quad \quad\Theta^{\Lambda\sf m}\, \mathcal{G}_{\mu\nu\Lambda} =
\Theta^{\Lambda\sf m} \,\mathcal{H}_{\mu\nu\Lambda}\,,
\end{eqnarray}
and the field equations associated with the magnetic gauge fields,
 \begin{align}
   \label{eq:eqmagcov}
   0=&\, \ft16  e^{-1}
   \varepsilon^{\mu\nu\rho\sigma}\left(Z^{\Lambda,\sf a}
     \mathcal{H}_{\nu\rho\sigma \,\sf a}+Z^{\Lambda,\sf
       m}\mathcal{H}_{\nu\rho \sigma\,\sf m}\right)+ T_{(MN)}{}^\Lambda
   \big(  -2\,\bar{X}^M
   \stackrel{\leftrightarrow}{\mathcal{D}}\!{}^\mu X^N
   \nonumber\\[.2ex]
   &\, + \bar{\Omega}^{iM}\gamma^\mu \Omega_i{}^N+
   \bar{X}^M\bar{\psi}_\nu{}^i\gamma^\mu\gamma^\nu \Omega_i{}^N-
   X^M\bar{\psi}_{\nu i}\gamma^\mu\gamma^\nu \Omega^{iN}-\ft12 e^{-1}
   \varepsilon^{\mu\nu\rho\sigma} \bar{\psi}_{\nu
     i}\gamma_\rho \psi_{\sigma}{}^i\,\bar{X}^M X^N \big)\nonumber\\[.2ex]
   &\, + \mathrm{i} G_{\bar{\alpha}\beta} T^{\Lambda \beta}{}_\gamma
   \big(\ft12 A^{i\bar{\alpha}}
   \stackrel{\leftrightarrow}{\mathcal{D}}\!{}^\mu A_i{}^\gamma
   -2 \bar{\zeta}^{\bar{\alpha}}\gamma^\mu \zeta^\gamma+
   \bar{\psi}_\nu{}^i \gamma^\mu\gamma^\nu \zeta^{\bar{\alpha}}
   A_i{}^\gamma- \bar{\psi}_{\nu i}\gamma^\mu\gamma^\nu \zeta^\gamma
   A^{i\bar{\alpha}}\big)\nonumber\\[.2ex]
   &\,-\mathrm{i} e^{-1}\varepsilon^{\mu\nu\rho\sigma} \bar{\psi}_\nu{}^i
   \gamma_\rho \psi_{\sigma j} \varepsilon^{jk}\mu_{ik}{}^\Lambda \,,
\end{align}
where we made use of the Bianchi identity (\ref{eq:Bianchi-DCov}).

Secondly we evaluate the supersymmetry commutator on the vector fields
$W_\mu{}^M$,
\begin{eqnarray}
\label{eq:susy-comm-vector}
[\delta(\epsilon_1),\delta(\epsilon_2)]W_{\mu}{}^M &=&
\xi^{\rho}\mathcal{G}_{\rho \mu}{}^M+\mathcal{D}_{\mu}\Lambda^M-g\,Z^{M,\sf{a}}\,\Xi_{\mu\,\sf{a}}- g\,Z^{M,\sf{m}}\,\Xi_{\mu\,\sf{m}}\nonumber\\
&&-\xi^{\rho}\left(
\ft{1}{2}\,\varepsilon_{ij}\,\bar{\psi}_{\rho}{}^i\gamma_{\mu}\Omega^{jM}
+\varepsilon_{ij}\bar{X}^M \bar{\psi}_\rho{}^i
\psi_\mu{}^j+\text{h.c.}\right)\,,
\end{eqnarray}
where the parameters $\xi^\mu$, $\Lambda^M$, $\Xi_{\mu\,\sf{a}}$ and
$\Xi_{\mu\,\sf{m}}$ are as in \eqref{eq:QQ-commutator-gauged}. In this
result one can replace $\mathcal{G}_{\mu \nu}{}^M$ by
$\mathcal{H}_{\mu \nu}{}^M$. For the electric gauge fields this is
trivial as $\mathcal{G}_{\mu\nu}{}^\Lambda$ and
$\mathcal{H}_{\mu\nu}{}^\Lambda$ are identical. For the magnetic gauge
fields the replacement is effectively allowed because $W_{\mu
  \Lambda}$ appear in the Lagrangian contracted with the embedding
tensor, as can be seen from \eqref{eq:A-field-eq-2}. Therefore,
without loss of generality, one can safely contract
(\ref{eq:susy-comm-vector}) for the magnetic gauge fields with the
embedding tensors, $\Theta^{\Lambda \sf a}$ or $\Theta^{\Lambda \sf
  m}$, upon which one can replace $\mathcal{G}_{\mu \nu\Lambda}$ with
$\mathcal{H}_{\mu \nu\Lambda}$ by virtue of
\eqref{eq:eom-B}. Finally one uses the following equality,
\begin{align}
  \label{eq:closure-vector}
    \xi^\rho\mathcal{H}_{\rho \mu}{}^M =&\,
    \xi^{\rho}\partial_{\rho}W_{\mu}{}^M+
    \partial_{\mu} \xi^\rho W_\rho{}^M- \mathcal{D}_\mu\left(\xi^\rho
      W_\rho{}^M\right)\nonumber\\
    &\,+\,g Z^{M,\sf a} \xi^\rho \left(B_{\rho \mu\, \sf a}+d_{{\sf a}\,
      NP} W_\rho{}^N W_\mu{}^P\right) +\,g Z^{M,\sf m} \xi^\rho
    B_{\rho \mu\, \sf m}\,.
\end{align}
Substituting this identity into \eqref{eq:susy-comm-vector} shows that
the $\xi^\mu$-dependent terms decompose into a general coordinate
transformation with parameter $\xi^\mu$, a non-abelian gauge
transformation with parameter $-\xi^\mu W_\mu{}^M$, tensor gauge
transformations with parameters
$-\xi^{\rho}\left(B_{\rho\mu\,\textsf{a}}+d_{\textsf{a}\,NP}
  W_{\rho}^NW_{\mu}^P\right)$ and $-\xi^{\rho}B_{\rho\mu\,\textsf{m}}$
and a supersymmetry transformation with parameter $-\ft12
\xi^\mu\psi_{\mu i}$. Together they constitute a covariant general
coordinate transformation with parameter $\xi^\mu$. Consequently the
supersymmetry commutator closes according to
(\ref{eq:QQ-commutator-gauged}).

Subsequently we turn to the supersymmetry commutator on the tensor
fields $B_{\mu\nu\,{\sf a}}$. Here it suffices to consider those fields
contracted with $Z^{\Lambda,\sf a}$ because no other components of the
tensor field appear in the Lagrangian according to
(\ref{eq:var-L-totaal}). Hence, we first evaluate
\begin{align}
  \label{eq:susy-comm-tensor}
  Z^{\Lambda,\sf a}\,[\delta(\epsilon_1),\delta(\epsilon_2)]
  B_{\mu\nu\,\sf a}=  &\, 2\,Z^{\Lambda, \sf
    a}\mathcal{D}_{[\mu} \Xi_{\nu]\sf a}  -2\,T_{(MN)}{}^\Lambda
  \Lambda^M\mathcal{G}_{\mu \nu}{}^N
   \nonumber\\[.2ex]
  &\,+2\,T_{(MN)}{}^\Lambda\,W_{[\mu}{}^M\,
  [\delta(\epsilon_1),\delta(\epsilon_2)]\,W_{\nu]}{}^N
  \nonumber\\[.2ex]
  &\,+T_{(MN)}{}^\Lambda
  \xi^{\rho}\left(\bar{X}^M\bar{\Omega}_i{}^N\gamma_{\mu\nu}\psi_{\rho}{}^i
    -2\bar{\psi}_\rho{}^i\gamma_{[\mu}\psi_{\nu]i}\,
    \bar{X}^MX^N+\text{h.c.}\right)\nonumber\\[.2ex]
  &\, + e\, \varepsilon_{\mu\nu\rho\sigma}\, T_{(MN)}{}^\Lambda \xi^\rho
  \big(-2\,\bar{X}^M \stackrel{\leftrightarrow}{\mathcal{D}}{\!}^\sigma X^N+
  \bar{\Omega}^{iM}\gamma^\sigma \Omega_i{}^N\nonumber\\[.2ex]
  &\, +\bar{X}^M\bar{\psi}_\lambda{}^i\gamma^\sigma\gamma^\lambda \Omega_i{}^N-
  X^M\bar{\psi}_{\lambda i}\gamma^\sigma\gamma^\lambda
  \Omega^{iN} -\tfrac12
  e^{-1}\varepsilon^{\sigma\lambda\tau\omega} \bar{\psi}_{\lambda i}\gamma_\tau
  \psi_{\omega}{}^i \,\bar{X}^M X^N \big)\nonumber\\[.2ex]
  &\, +16\, \mathrm{i}\, g \,T_{(MN)}{}^\Lambda \Omega^{MP} \big( X^N  \,\mu^{ij}{}_P\,\bar{\epsilon}_{2i}\gamma_{\mu\nu}\epsilon_{1j}
  -\bar{X}^N\, \mu_{ijP}\, \bar{\epsilon}^i_{2}\gamma_{\mu\nu}\epsilon^j_{1}\big) \,,
\end{align}
with the parameters $\xi^\mu$, $\Lambda^M$ and $\Xi_{\mu\,\sf{a}}$ as
in \eqref{eq:QQ-commutator-gauged}. The first four terms can
straightforwardly be compared to the variation of $B_{\mu\nu{\sf a}}$
given in the first formula of (\ref{eq:Q-S-tensors}). However, there
is a subtlety regarding the commutator on $W_\nu{}^N$ in the third
term, because this supersymmetry commutator only closes on the gauge
fields, up to a term $\xi^\rho(\mathcal{G} -
\mathcal{H})_{\rho\nu}{}^N$. Therefore the commutator yields the
transformations indicated on the right-hand side of
(\ref{eq:QQ-commutator-gauged}) plus this extra
term.\footnote{
  Upon contraction with $Z^{M\,\sf a}$ this term vanishes and we have
  argued that it could therefore be suppressed in the commutator on
  the gauge fields on $W_\nu{}^N$. See the text preceding
  \eqref{eq:closure-vector}. However, in the case at hand the extra term
  has to be retained. } 
Obviously the commutator on $W_\nu{}^N$ generates also a
diffeomorphism, which will play a role later on in the
calculation. Finally the fourth term represents precisely a
supersymmetry transformation with parameter $\epsilon^i=-\tfrac12
\xi^\rho \psi_\rho{}^i$.

The remaining terms in (\ref{eq:susy-comm-tensor}), however, do not
seem to have a role to play. At this point we note that the Lagrangian
does not depend separately on $Z^{\Lambda,{\sf a}} B_{\mu\nu\, {\sf
    a}}$ and $Z^{\Lambda,{\sf m}} B_{\mu\nu\, {\sf m}}$, but depends
only on the linear combination $Z^{\Lambda,\sf a}\,B_{\mu \nu \,{\sf
    a}}+ Z^{\Lambda,{\sf m}}\,B_{\mu \nu \,{\sf m}}$. Consequently,
the algebra is required to close only on this linear
combination. Therefore we also evaluate the commutator on
$Z^{\Lambda,{\sf m}}\,B_{\mu \nu \,{\sf m}}$,
\begin{align}
  \label{eq:susy-comm-tensor-hyper}
  Z^{\Lambda,\sf
    m}\,[\delta(\epsilon_1),\delta(\epsilon_2)]B_{\mu \nu \,\sf
    m}=&\, 2\, Z^{\Lambda, \sf m} \mathcal{D}_{[\mu} \Xi_{\nu]\sf
    m}\nonumber\\[.2ex]
  &\,+\mathrm{i}\,\xi^\rho\big( k^{A\Lambda} \; \gamma_{Ai\bar\alpha}
  \,\bar\zeta^{\bar\alpha} \gamma_{\mu\nu}\psi_{\rho}{}^i -2\,
  \varepsilon^{ij}\mu_{jk}{}^\Lambda\bar{\psi}_{i[\mu}\gamma_{\nu]}
  \psi_{\rho}{}^k-\mathrm{h.c.}\big)\nonumber\\[.2ex]
  &\,-16\, \mathrm{i} g T_{(MN)}{}^\Lambda \Omega^{MP} \big( X^N
  \,\mu^{ij}{}_P\,\bar{\epsilon}_{2i}\gamma_{\mu\nu}\epsilon_{1j}
  -\bar{X}^N\, \mu_{ijP}\, \bar{\epsilon}^i_{2}\gamma_{\mu\nu}
  \epsilon^j_{1}\big) \nonumber\\[.2ex]
  &\,+ \mathrm{i}e \,\varepsilon_{\mu\nu\rho\sigma}\xi^\rho\big[\,
  G_{\bar{\alpha}\beta} T^{\Lambda \beta}{}_\gamma \,
  \big(\tfrac12 A^{i\bar{\alpha}}
  \stackrel{\leftrightarrow}{\mathcal{D}}{}^{\!\sigma} A_i{}^\gamma
  -2\, \bar{\zeta}^{\bar{\alpha}}\gamma^\sigma \zeta^\gamma \nonumber\\[.2ex]
  &\,+\bar{\psi}_\lambda{}^i \gamma^\sigma\gamma^\lambda \zeta^{\bar{\alpha}}
  A_i{}^\gamma- \bar{\psi}_{\lambda i}\gamma^\sigma\gamma^\lambda \zeta^\gamma
  A^{i\bar{\alpha}}\big)\nonumber\\[.2ex]
  &\,-e^{-1}\varepsilon^{\sigma\lambda\tau\omega} \, \bar{\psi}_{\lambda}{}^i
  \gamma_\tau \psi_{\omega j} \varepsilon^{jk}\mu_{ik}{}^\Lambda
  \big]\,,
\end{align}
with the parameters $\xi^\mu$ and $\Xi_{\mu\,\sf{m}}$ as in
\eqref{eq:QQ-commutator-gauged}. The first line establishes closure
with respect to $\Xi_{\mu\,\sf{m}}$. Furthermore, the next line
correctly reproduces a supersymmetry transformation with parameter
$\epsilon^i =-\tfrac12 \xi^\rho\psi_\rho{}^i$.

When considering the sum of the two variations
\eqref{eq:susy-comm-tensor} and \eqref{eq:susy-comm-tensor-hyper}
there are some cancelations, and on the remaining terms we can impose
the field equation \eqref{eq:eqmagcov}. This leaves the following
terms,
\begin{align}
  \label{eq:tensor-lin-comb-comm}
  [\delta(\epsilon_1),\delta(\epsilon_2)] \big(Z^{\Lambda,\sf a}\,
  B_{\mu\nu\,\sf a}+ Z^{\Lambda,\sf
    m}\,B_{\mu \nu \,\sf m}\big) =&\,
  Z^{\Lambda,\sf a} \, \xi^\rho\mathcal{H}_{\mu\nu\rho \,\sf
    a}+Z^{\Lambda,\sf m}\, \xi^\rho\mathcal{H}_{\mu\nu\rho\,\sf
    m}\nonumber\\
  &\,
  - 2\, T_{(MN)}{}^\Lambda W_{[\mu}{}^M
  \,\xi^\rho(\mathcal{G}-\mathcal{H})_{\nu]\rho}{}^N  + \cdots \,,
\end{align}
where the dots refer to terms that have already been accounted for in
the context of \eqref{eq:QQ-commutator-gauged}. The explicit terms in
\eqref{eq:tensor-lin-comb-comm} contribute to the (covariant)
general coordinate transformation, as follows from the following
identities, which can be derived straightforwardly from
\eqref{eq:H-3},
\begin{align}
  \label{eq:closure-tensor}
  Z^{\Lambda,
    \sf a}\,\xi^{\rho}\,\mathcal{H}_{\rho\mu\nu\,\textsf{a}}=&\,
  Z^{\Lambda, \sf a}\left(\xi^{\rho}\partial_{\rho}B_{\mu \nu\,\sf{a}}
    -2 \,\partial_{[\mu}\xi^\rho
    B_{\nu]\rho\,\textsf{a}}\right)\nonumber\\[.2ex]
  &\,
  +2\, Z^{\Lambda,\sf a} \mathcal{D}_{[\mu} \left(\xi^\rho
    B_{\nu]\rho\, \sf a}-\xi^\rho d_{\textsf{a}\,M N} W_{\nu]}{}^M
    W_{\rho}{}^N\right)\nonumber\\[.2ex]
  &\,+2 \, T_{(MN)}{}^\Lambda \xi^\rho W_\rho{}^M
  \mathcal{G}_{\mu\nu}{}^N\nonumber\\[.2ex]
  &\,
  -2\,T_{(MN)}{}^\Lambda
  W_{[\mu}{}^M\left(\xi^{\rho}\partial_{|\rho|}W_{\nu]}{}^N+
    \partial_{\nu]}\xi^\rho W_\rho{}^N  -2\,\xi^{\rho}
    (\mathcal{G}-\mathcal{H})_{\nu]\rho}{}^N \right)\nonumber\\[.2ex]
  &\,
  -2\, g\, T_{(MN)}{}^\Lambda Z^{M,{\sf m}}\, \xi^\rho \,W_{\rho}{}^N
  B_{\mu\nu\,\sf m}\,, \nonumber\\[.4ex]
 Z^{\Lambda, \sf m}\,\xi^{\rho}\,\mathcal{H}_{\rho\mu\nu\,\textsf{m}}=&\,
 Z^{\Lambda, \sf m}\left(\xi^{\rho}\partial_{\rho}B_{\mu \nu\,\sf{m}}
  -2 \,\partial_{[\mu}\xi^\rho B_{\nu]\rho\,\textsf{m}}\right)\nonumber\\[.2ex]
  &\,
  +2\, Z^{\Lambda,\sf m} \mathcal{D}_{[\mu} ( \xi^\rho
  B_{\nu]\rho\, \sf m}) \nonumber\\
  &\,
  +2 \,g\, T_{(MN)}{}^\Lambda Z^{M,{\sf m}}
  \,\xi^\rho \,W_{\rho}{}^N B_{\mu\nu\,\sf m}\,.
\end{align}
The first two terms in the equations \eqref{eq:closure-tensor} denote
the expected general coordinate transformation, and the tensor gauge
transformations with parameters given in \eqref{eq:cov-transl-tensor}.
The third term in the first equations represents the appropriate gauge
transformation. The last terms in the two equations cancel directly,
so that the only terms in \eqref{eq:tensor-lin-comb-comm} that are
still unaccounted for, are given by
\begin{align}
  \label{eq:tensor-lin-comb-comm2}
  [\delta(\epsilon_1),\delta(\epsilon_2)] \big(Z^{\Lambda,\sf a}\,
  B_{\mu\nu\,\sf a}+ Z^{\Lambda,\sf
    m}\,B_{\mu \nu \,\sf m}\big) =&\,  -2\,T_{(MN)}{}^\Lambda
  W_{[\mu}{}^M\big(\xi^{\rho}\partial_{|\rho|}W_{\nu]}{}^N+
    \partial_{\nu]}\xi^\rho W_\rho{}^N\big)\nonumber\\
  &\,
   +2\,T_{(MN)}{}^\Lambda W_{[\mu}{}^M\xi^{\rho}
    (\mathcal{G}-\mathcal{H})_{\nu]\rho}{}^N +\cdots \,.
\end{align}
The first of these terms cancels against the general coordinate
transformation induced by the supersymmetry commutator on $W_\nu{}^N$
in \eqref{eq:susy-comm-tensor}, which we already referred to earlier,
and which is not required on the tensor fields in view of the fact
that the above equations \eqref{eq:closure-tensor} already account for
the general coordinate transformation. The second term
can be suppressed by virtue of the special invariance noted in
(\ref{eq:Delta-invariance}). To see this, we note that, up to the
first equation of motion \eqref{eq:eom-B}, we can write the induced
variation of $B_{\mu\nu{\sf a}}$ as,
\begin{align}
  \label{eq:2}
  Z^{\Lambda,{\sf a}}\, \delta B_{\mu\nu{\sf a}}\propto &\,
  T^{(\Lambda}{}_M{}^{\Sigma)} \,
  [4\, \xi^\rho W_{[\mu}{}^M  - \xi^\sigma W_\sigma{}^M
  \,\delta^\rho_{[\mu} ] (\mathcal{G} -\mathcal{H})_{\nu]\rho\Sigma}
  \nonumber\\
  &\,
  - T^{[\Lambda}{}_M{}^{\Sigma]} \,\xi^\sigma  W_\sigma{}^M \,
  (\mathcal{G} -\mathcal{H})_{\mu\nu\Sigma} \,.
\end{align}
This completes our discussion of the supersymmetry algebra.

Finally we summarize the modifications to the Lagrangian that are
required by the general gaugings. As usual these concern both masslike
terms for the fermions, which are proportional to the gauge coupling
$g$, and a scalar potential proportional to $g^2$. The masslike terms
independent of the gravitini follow directly from the rigid theory in
the presence of both electric and magnetic charges \cite{de
  Vroome:2007zd}. The terms that involve gravitini are generalizations
of the known results for the superconformal theory in the presence of
electric charges \cite{deWit:1980tn,deWit:1984px,deWit:1999fp}. The
result includes also a non-fermionic term which describes the coupling
of the auxiliary fields $Y_{ij}{}^\Lambda$ to the moments $\mu_{ijM}$,
\begin{align}
  \label{eq:Lag-order-g}
  e^{-1}\mathcal{L}_g =&\,-\ft12
  \mathrm{i} g \,\Omega_{MQ}
  T_{PN}{}^Q \, \varepsilon^{ij}\, \bar X^N \bar\Omega_i{}^M
  \big(\Omega_j{}^P  + \gamma^\mu \psi_{\mu j} X^P
  \big)+\text{h.c.}\nonumber\\
  &\,+ 2g\, k_{AM}
  \gamma^A_{i\bar\alpha} \varepsilon^{ij}\,{\bar\zeta}^{\bar\alpha}
  \big(\Omega_j{}^{M}+\gamma^\mu \psi_{\mu j}
    X^M\big)+\text{h.c.}\nonumber\\
  &\, +  g \, \mu^{ij}{}_M\,\bar{\psi}_{\mu i} \left(\gamma^\mu
  \Omega_j{}^M+\gamma^{\mu\nu} \psi_{\nu j} X^M \right)
   +\mbox{h.c.}\nonumber\\
   &\, +2g\left[ {\bar X}^M T_M{}^{\gamma}{}_{\alpha}\,\bar
     \Omega_{\beta\gamma}\,{\bar \zeta}^\alpha\zeta^\beta+
     X^M T_M{}^{\bar\gamma}{}_{\bar\alpha}\,
     \Omega_{\bar\beta\bar\gamma}\,{\bar\zeta}^{\bar\alpha}
     \zeta^{\bar\beta}\right]\nonumber\\
   &\,
   -\ft14 g\,\left[ F_{\Lambda\Sigma\Gamma}\,
     \mu^{ij\Lambda} \,\bar\Omega_i{}^\Sigma \Omega_j{}^\Gamma
     + \bar F_{\Lambda\Sigma\Gamma} \,
     \mu_{ij}{}^\Lambda\,\bar\Omega^{i\Sigma} \Omega^{j\Gamma}
   \right]
   \nonumber\\
  &\,+g\,Y^{ij\Lambda} \left[ \mu_{ij\Lambda}
     +\ft12 (F_{\Lambda\Sigma}+ \bar
     F_{\Lambda\Sigma})\,\mu_{ij}{}^\Sigma\right] \,.
\end{align}
Upon solving the auxiliary fields $Y_{ij}{}^I$ one obtains an
additional contribution to the scalar potential of order
$g^2$. Without this contribution the scalar potential reads,
\begin{align}
  \label{scalar-potential-wY}
  e^{-1}\mathcal{L}_{g^2} =&\,\mathrm{i}g^2\, \Omega_{MN} \, T_{PQ}{}^M X^P
  \bar X^Q  \;  T_{RS}{}^N \bar X^R X^S \nonumber\\
  &\,-2g^2k^A{}_M \,k^B{}_N  \,g_{AB}\,X^M{\bar
    X^N} - \ft12 g^2 \,N_{\Lambda\Sigma} \;\mu_{ij}{}^\Lambda \,
  \mu^{ij\Sigma}\,.
\end{align}
Upon eliminating the auxiliary fields, the last term in this
expression changes into
\begin{equation}
  \label{eq:Y-eliminate-g2}
  - \ft12 g^2 \,N_{\Lambda\Sigma} \;\mu_{ij}{}^\Lambda\,\mu^{ij\Sigma}
  \longrightarrow
  -2\,g^2 \left[\mu^{ij}{}_\Lambda +
    F_{\Lambda\Gamma}\,\mu^{ij\Gamma} \right] \,N^{\Lambda\Sigma}
    \left[\mu_{ij\Sigma} +\bar F_{\Sigma\Xi}\,\mu_{ij}{}^\Xi\right]
    \,.
\end{equation}

The above expressions are not of definite sign. From the Lagrangians
in section \ref{sec:lagrangian} one can deduce that
$\chi_\mathrm{vector}$, $\chi_\mathrm{hyper}$ and the metrics that
appear  in the kinetic terms of the physical scalar fields should be
negative. The latter metrics are proportional to two matrices,
$M_{\Lambda\Sigma}$ and $G_{AB}$, that should therefore be negative
definite. They are defined by
\begin{align}
  \label{eq:kinetic-matrices}
  M_{\Lambda\bar{\Sigma}}=&\, \chi_\mathrm{vector}^{-2}\left(N_{\Lambda
      \Sigma}N_{\Gamma\Xi}-N_{\Lambda \Gamma} N_{\Sigma \Xi}\right)
  \bar{X}^\Gamma X^\Xi
 \,,\nonumber\\
 G_{AB} = &\,\chi_\mathrm{hyper}^{-1}\left(g_{AB} -
   \chi_\mathrm{hyper}^{-1}(\ft12 \chi_A
   \chi_B+k_{Aij}k_B{}^{ij})\right)\,.
\end{align}
With these observations we can separate the terms in the potential in
positive and negative ones,
\begin{align}
  \label{scalar-potential}
  e^{-1}\mathcal{L}_{g^2} =&\,- g^2\,\chi_\mathrm{vector}\,
  M_{\bar\Lambda\Sigma} \, (T_{PQ}{}^\Lambda X^P \bar X^Q) \,
  (T_{RS}{}^\Sigma \bar X^R X^S)
  \nonumber\\
  &\,-4\,g^2\chi_\mathrm{vector}\, k^A{}_M \,k^B{}_N \,G_{AB}\,X^M{\bar
    X^N} \nonumber   \\
  &\, -2\, g^2\,\chi_\mathrm{vector}\, M_{\bar\Lambda\Sigma}\,
  N^{\Lambda\Gamma} \left[\mu^{ij}{}_\Gamma +
    F_{\Gamma\Omega}\,\mu{}^{ij\Omega}\right] \,N^{\Sigma\Xi}
  \left[\mu_{ij\Xi} +\bar
    F_{\Xi\Delta}\,\mu_{ij}{}^\Delta\right] \nonumber\\
  &\, -6\, g^2 \chi_\mathrm{vector}^{-1}\, X^M \bar X^N \, \mu_{ijM}\,
  \mu^{ij}{}_N \,,
\end{align}
where we used that $\chi_\mathrm{hyper} =2\,\chi_\mathrm{vector}$, as
is implied by the field equation associated with the field $D$. It
then follows that all contributions to $\mathcal{L}_{g^2}$ are
negative, with the exception of the last term which is positive. This
decomposition generalizes a similar decomposition known for purely
electric charges.

\section{Summary and some applications}
\label{sec:summary-applications}
\setcounter{equation}{0}
In this paper we presented Lagrangians and supersymmetry
transformations for general superconformal systems of vector
multiplets and hypermultiplets in the presence of both electric and
magnetic charges. The results were verified to all orders and are
consistent with results known in the literature based on both rigidly
supersymmetric theories and on superconformal systems without magnetic
charges. In the presence of magnetic charges the off-shell closure of
the superconformal algebra is only realized on the Weyl multiplet. The
results of this paper establish a general framework for studying gauge
interactions in matter-coupled $N=2$ supergravity.

In the remainder of this last section we discuss two specific
applications to demonstrate the consequences of this general
framework. The first one discusses full and partial supersymmetric
solutions in maximally symmetric space-times, and the second one deals
with full or partial supersymmetric solutions in $\mathrm{AdS}_2\times
S^2$ space-times.
\subsection{Maximally symmetric space-times and supersymmetry}
\label{sec:maxim-symm-space}
In this application we briefly consider the question of full or
partial supersymmetry in a maximally symmetric space-time. Hence one
evaluates the supersymmetry variations of the fermion fields in the
maximally symmetric background, where only $g_{\mu\nu}$,
$A_i{}^\alpha$, $X^\Lambda$ and $Y_{ij}{}^\Lambda$ can take non-zero
values, taking into account that the fermion fields transform under
both Q- and S-supersymmetry. In this particular background, it turns
out that the gravitino field strength, $R(Q)_{\mu\nu}{}^i$ (and the
related spinor $\chi^i$) is S-invariant. Since its Q-supersymmetry
variation is proportional to the field $D$, it immediately follows
that $D=0$, so that the special conformal gauge field takes the value (we
assume the gauge choice $b_\mu=0$, which leaves a residual invariance
under constant scale transformations),
\begin{equation}
  \label{eq:f-max-symm}
  f_\mu{}^a = \tfrac12 R(e,\omega)_\mu{}^a -\tfrac1{12}  e_\mu{}^a\,
  R(e,\omega) \,,
\end{equation}
where $R(e,\omega)_{\mu\nu}{}^{ab}$ denotes the space-time curvature.

In what follows it thus suffices to concentrate on the fermions
belonging to the vector multiplets and the hypermultiplets. We first
present their variations in the background, which follow directly from
\eqref{eq:conv-rules-hyp} and \eqref{eq:conv-rules},
\begin{align}
  \label{eq:max-symm-fermions}
  \delta\zeta^\alpha  =&\, 2gX^M
  \,T_{M}{}^\alpha{}_\beta A_i{}^\beta\,\varepsilon^{ij}\epsilon_j
  +A_i{}^\alpha \,\eta^i\,,  \nonumber\\
    \delta \Omega_i{}^M =&\,
  \hat{Z}_{ij}{}^M\epsilon^j
  -2g\,{T_{PN}}^M\bar{X}^{P}X^{N}\varepsilon_{ij}\epsilon^j
  +2\,\mathrm{i} g \Omega^{MN}\mu_{ijN}\epsilon^j
  +2X^M\eta_i\,.
\end{align}
Substituting the equations of motion for the auxiliary fields
$Y_{ij}{}^\Lambda$, the variation of the independent fermion fields
$\delta\Omega_i{}^\Lambda$ takes the following form,
\begin{equation}
  \label{eq:var-Omega-up}
    \delta\Omega_i{}^\Lambda =
    -2g\, T_{NP}{}^\Lambda \, \bar X^N X^P
    \,\varepsilon_{ij}\, \epsilon^j - 4\,gN^{\Lambda\Sigma}\big(\mu_{ij\Sigma}
    + \bar F_{\Sigma\Gamma} \,\mu_{ij}{}^\Gamma\big) \epsilon^j  +
    2\,X^\Lambda \eta_i \,,
\end{equation}

Following the strategy adopted by \cite{LopesCardoso:2000qm}, we
consider only combinations of fermion fields that are invariant under
S-supersymmetry. To construct S-invariant combinations of these
fermions, it is convenient to define the following two spinor fields,
\begin{align}
  \label{eq:Zeta-H-Omega-V}
  \zeta^\mathrm{H}_i  =&\,\chi_\mathrm{hyper}^{-1}  \bar\Omega_{\alpha\beta} A_i{}^\alpha\,
  \zeta^\beta  \nonumber\\
  \Omega^\mathrm{V}_i=&\, - \ft{1}{2}\mathrm{i} \chi_{\rm{vector}}^{-1}
  \Omega_{MN} \bar{X}^M \Omega_i{}^N = \tfrac12
  \chi_\mathrm{vector}^{-1} \,\bar X^\Lambda N_{\Lambda\Sigma}
  \Omega_i{}^\Sigma   \,,
\end{align}
which are both formally invariant under duality when
treating the embedding tensor as a spurion. Under supersymmetry these
two spinors transform equivalently in this background, provided we also
use the field equation of the field $D$, which yields $\chi_\mathrm{hyper}
=2\,\chi_\mathrm{vector}$. Indeed one easily derives,
\begin{equation}
  \label{eq:var-omega-h}
   \delta \Omega_i^{\mathrm{V}} =
  A_{ij}\,\epsilon^j+\eta_i = -\varepsilon_{ij}
  \,\delta\zeta^{\mathrm{H}\,j} \,,
\end{equation}
where the symmetric matrix $A_{ij}$ is given by,
\begin{equation}
  \label{eq:def-A}
  A_{ij} = -2\,g\, \chi_\mathrm{vector}^{-1} \,\bar{X}^M \mu_{ijM}\,.
\end{equation}
Here we made use of equations \eqref{eq:cov-der-invariance}.

To make contact with the terms appearing in the potential
\eqref{scalar-potential}, we consider the variations of three other
spinors, which are S-supersymmetry invariant and consistent with duality. As it
turns out, considering such variations gives important information
regarding the possible supersymmetric realizations, although it will
not yet fully determine whether the corresponding solutions will
actually be realized. The first two variations are,
\begin{align}
  \label{eq:4}
  g\big(\mu^{ij}{}_\Lambda + F_{\Lambda\Sigma}\, \mu^{ij\Sigma} \big)\,
  \delta[\Omega_j{}^\Lambda -2\, X^\Lambda \Omega_j^\mathrm{V}] =&\,
  -2\, g^2 \,\bar X^M X^N T_{MN}{}^P \mu^{ij}{}_P\,
  \varepsilon_{jk} \, \epsilon^k \nonumber\\
  &\,
  - 2\,g^2 (\mu^{kl}{}_\Lambda +F_{\Lambda\Sigma}\,\mu^{kl\Sigma})
  N^{\Lambda \Gamma}(\mu_{kl}{}_\Gamma
  +\bar{F}_{\Gamma\Xi}\,\mu_{kl}{}^\Xi)\epsilon^i  \nonumber\\
  &\,
  + \chi_\mathrm{vector}\, A^{ij} A_{jk} \epsilon^k \,,  \nonumber\\[.2ex]
  gN_{\Lambda\Sigma} \,T_{MN}{}^\Sigma X^M\bar X^N\,
  \delta[\Omega_i{}^\Lambda -2 \, X^\Lambda \Omega_i^\mathrm{V}] =&\,
   2\mathrm{i}\,g^2 \Omega_{MN}(T_{PQ}{}^M X^P \bar{X}^Q)\,
   (T_{RS}{}^N \bar{X}^R
  X^S)\, \varepsilon_{ij} \epsilon^j \nonumber\\
  &\,
   -4 \,g^2 \,X^M \bar{X}^N T_{MN}{}^P \mu_{ijP} \,\epsilon^j\,.
\end{align}
In deriving this result we made use of identities such as
\eqref{eq:derivative-invariance} and
\eqref{eq:cov-der-invariance}. Furthermore we used
$\Omega^{MN}\mu_{ijM} \,\mu_{kl N} = \mu_{ij\Lambda}
\,\mu_{kl}{}^\Lambda- \mu_{ij}{}^\Lambda \,\mu_{kl\Lambda} =0$, which
follows directly from \eqref{eq:locally-em-3}. The third spinor
variation is based on hypermultiplets,
\begin{align}
  \label{eq:spinor-comg-hyper}
  g\,  \bar X^M T_M{}^\alpha{}_\beta A_i{}^\beta \,
  \bar\Omega_{\alpha\gamma}
  \,\delta\big[\zeta^\gamma + \varepsilon^{jk} A_j{}^\gamma \,\zeta_k
  ^\mathrm{H} \big]
  =&\,
  -g^2 \bar X^MX^N \,k^A{}_M \,k^B{}_N \,g_{AB}\, \epsilon_i
  \nonumber\\
  &\, - 2\,g^2
  \bar X^M X^N\,T_{MN} {}^P\,\mu_{ijP} \,\varepsilon^{jk} \,\epsilon_k    \,
  \nonumber\\
  &\,
  +\chi_\mathrm{vector}\, A_{ij} A^{jk} \,\epsilon_k \,.
\end{align}
Here we made use of the identity,
\begin{align}
    \label{eq:T-A-squared}
    T_M{}^\alpha{}_\beta A_i{}^\beta \,\bar\Omega_{\alpha\gamma}\,
    T_N{}^\gamma{}_\delta A_j{}^\delta = \tfrac12 \varepsilon_{ij}\,
    k^A{}_M\,k_{AN} +T_{MN}{}^P \,\mu_{ij P}\,,
\end{align}
which follows from \eqref{eq:equivariance},
\eqref{eq:DA-eqs2}, \eqref{eq:repr-t-mu} and \eqref{eq:closure}.
Combining \eqref{eq:spinor-comg-hyper} with the two previous
identities gives,
\begin{equation}
  \label{eq:potential-AA}
   \big[ e^{-1}\mathcal{L}_{g^2} \,\delta^i{}_j +
   3\,\chi_\mathrm{vector} A^{ik}A_{kj} \big]\, \epsilon^j=0\,.
\end{equation}
This relation requires $e^{-1}\mathcal{L}_{g^2}$ to be non-negative,
confirming the known result that de Sitter space-times cannot be
supersymmetric.

According to \cite{LopesCardoso:2000qm} one must also consider the
symmetry variation of the supercovariant derivative of at least one of
these spinor fields. Let us, for instance, consider $D_\mu
\Omega^\mathrm{V}_i$, which transforms also under S-supersymmetry. The
following combination is then again S-invariant, and changes under
Q-symmetry according to,
\begin{equation}
  \label{eq:4s-var-D-zeta}
  \delta\big[D_\mu \Omega^\mathrm{V}_i - \tfrac12 A_{ij} \gamma_\mu
  \Omega^{\mathrm{V}\,j} \big] = f_\mu{}^a \gamma_a \epsilon_i -\tfrac12
  A_{ij} A^{jk}\,\gamma_\mu \epsilon_k \,.
\end{equation}
Therefore we must require that the supersymmetry parameters are
subject to the eigenvalue condition,
\begin{equation}
  \label{eq:eigenvalue}
 \big[ \delta^i{}_j\,\big(R(e,\omega)_\mu{}^a -\tfrac16 e_\mu{}^a
 R(e,\omega) \big)  - e_\mu{}^a\,A^{ik}A_{kj}\big]
 \epsilon^j=0\,.
\end{equation}
Combining this result with \eqref{eq:potential-AA} reproduces the
Einstein equation for the maximally symmetric space-time, irrespective
of whether supersymmetry is realized fully or partially. Observe that
full supersymmetry requires that $A^{ik}A_{kj}\propto \delta^i{}_j$.

The result \eqref{eq:potential-AA} can also be written as 
\begin{equation}
  \label{eq:5}
     \big[A^{ik}A_{kj} - \tfrac12 A^{kl}A_{kl} \,\delta^i{}_j\big]
     \epsilon^j =  - \frac{e^{-1}\mathcal{L}_{g^2}^-}
     {3\,\chi_\mathrm{vector}} \,\epsilon^i \,,
\end{equation}
where $\mathcal{L}_{g^2}^-$ pertains to the negative terms in
$\mathcal{L}_{g^2}$. For full supersymmetry we thus find that
$\mathcal{L}_{g^2}^-$ must vanish, while partial supersymmetry is
associated with the smallest eigenvalue of $A^{ik}A_{kj}$ and
$\mathcal{L}_{g^2}^-  \not=0$. We refrain from giving more explicit
details here, but we briefly consider the special case of Minkowski
space-time. For partial supersymmetry, the unbroken supersymmetry
parameter is subject to the condition $A_{ij}\epsilon^j =0$. In this
context one can consider the variation of yet another spinor, which is
invariant under S-supersymmetry, but no longer under duality,
\begin{align}
  \label{eq:nondual-spinor-var}
  X^\Lambda N_{\Lambda\Sigma}\, \delta[\Omega_i{}^\Sigma-
  2\,X^\Sigma \,\Omega_i^\mathrm{V}]  =&\, - 2\,g
  X^\Lambda N_{\Lambda\Sigma}\, \big[ T_{MN}{}^\Sigma \,\bar X^MX^N
  \,\varepsilon_{ij} -2 \mathrm{i} \mu_{ij}{}^\Sigma \big]
  \epsilon^j\nonumber\\
  &\,
  +2\, X^\Lambda N_{\Lambda\Sigma}\,\big[\bar X^\Sigma\,
  \varepsilon_{ik} \varepsilon_{jl}\,
  A^{kl} - X^\Sigma \, A_{ij}  \big]\epsilon^j\,.
\end{align}
In the absence of magnetic charges, the first term on the right-hand
side vanishes because $T_{MN}{}^\Sigma \bar X^M X^N$ can be replaced
by $T_{(MN)}{}^\Sigma\bar X^MX^N$ by virtue of the third equation of
\eqref{eq:cov-der-invariance}, which vanishes without magnetic
charges, and so does the moment map $\mu_{ij}{}^\Sigma$. Therefore
both $A_{ij}\epsilon^j$ and $A^{ij}\varepsilon_{jk} \epsilon^k$
vanish, which implies that $A_{ij}$ vanishes so that supersymmetry
must be fully realized. This is in accord with a known theorem
according to which $N=2$ supersymmetry can only be broken to $N=1$
supersymmetry in Minkowski space in the presence of magnetic charges
\cite{Cecotti:1984rk,Cecotti:1985sf,Ferrara:1995xi,Ferrara:1995gu,
  Fre:1996js,Louis:2009xd}. For the abelian gaugings the situation
simplifies, and one can show that Minkowski solutions with residual
$N=1$ supersymmetry are possible provided that,  
\begin{align}
  \label{eq:N1-mink}
  \bar X^M\, T_M{}^\alpha{}_\beta\, A_i{}^\beta \epsilon^i=&\, 0\,, \nonumber\\
  (\mu_{ij\Lambda} + \bar F_{\Lambda\Sigma}\,
    \mu_{ij}{}^\Sigma)\, \epsilon^j=&\, 0\,,
\end{align}
with the two terms of the abelian potential vanishing separately (this
follows from the first equation of \eqref{eq:4} and from
\eqref{eq:spinor-comg-hyper}),
\begin{align}
  \label{eq:pot-terms}
 \bar X^MX^N \,k^A{}_M \,k^B{}_N \,g_{AB}=&\,0\,, \nonumber \\
  (\mu^{kl}{}_\Lambda +F_{\Lambda\Sigma}\,\mu^{kl\Sigma})
  N^{\Lambda \Gamma}(\mu_{kl}{}_\Gamma
  +\bar{F}_{\Gamma\Xi}\,\mu_{kl}{}^\Xi)=&\,0\,. 
\end{align}
Without magnetic charges, one can easily verify that residual $N=1$
supersymmetric solutions are not possible. 

Apart from this latter result, the above analysis only indicates which
supersymmetric solutions can, in principle, exist. To confirm that
they are actually realized, one has to also examine the supersymmetry
variations of the remaining fermion fields. This can be done, but we
prefer not to demonstrate this here. Instead we will discuss this
explicitly in the application presented in the next subsection, which
is less straightforward, and where we will follow the same set-up as
in this subsection.

\subsection{Supersymmetry in $\mathrm{AdS}_2\times S^2$}
\label{sec:supersymm-s2xads2}
In this second application we consider an $\mathrm{AdS}_2\times S^2$
space-time background and analyze possible supersymmetric
solutions. Hence the space-time metric can be chosen equal to,
\begin{equation}
  \label{eq:ads2xs2}
    \mathrm{d}s^2 = g_{\mu\nu} \mathrm{d}x^\mu\mathrm{d}x^\nu =
  v_1\Big(-r^2\,\mathrm{d}t^2 + \frac{\mathrm{d} r^2}{r^2}\Big)
  + v_2 \Big(\mathrm{d} \theta^2 +\sin^2\theta
  \,\mathrm{d}\varphi^2\Big)\,,
\end{equation}
whose non-vanishing Riemann curvature components are equal to
\begin{equation}
  \label{eq:geo}
  R_{\underline{a} \underline{b}}{}^{\underline{c} \underline{d}}  =2\, v_1^{-1}
  \delta_{\underline{a}\underline{b}}{}^{\underline{c}\underline{d}}
  \,,\qquad
  R_{\hat a \hat b}{}^{\hat c \hat d} =-2 \,v_2^{-1} \delta_{\hat
    a\hat b}{}^{\hat c\hat d} \,,
\end{equation}
so that the four-dimensional Ricci scalar equals $R= 2
(v_1^{-1}-v_2^{-1})$. Observe that we used tangent-space indices
above, where $\underline{a}, \underline{b}, \ldots$ label the flat
$\mathrm{AdS}_2$ indices $(0,1)$ associated with $(t,r)$, and $\hat a,
\hat b, \ldots$ label the flat $S^2$ indices $(2,3)$ associated with
$(\theta,\varphi)$. Furthermore the non-vanishing components of the
auxiliary tensor field are parametrized by a complex scalar $w$,
\begin{equation}
  \label{eq:T-into-w}
  -T_{01}{}^{ij}\varepsilon_{ij}=-\mathrm{i}\,
  T_{23}{}^{ij}\varepsilon_{ij} = w\,.
\end{equation}
Using the previous results one finds the following expressions for the
bosonic part of the special conformal gauge field $f_a{}^b$,
\begin{align}
  \label{eq:K-gaugefield}
  f_{\underline{a}}{}^{\underline{b}}=&\, \big(\ft16 (2\,v_1^{-1}+
  v_2^{-1})  - \ft14
  D -\ft1{32} |w|^2\big) \delta_{\underline{a}}{}^{\underline{b}} +\ft12
  R(A)_{23}\, \varepsilon_{\underline{a}}{}^{\underline{b}}\,,
  \nonumber\\
  f_{\hat a}{}^{\hat b}=&\, \big(-\ft16 (v_1^{-1} +2\,v_2^{-1}) -
  \ft14D +\ft1{32} |w|^2\big )\delta_{\hat a}{}^{\hat b}+
  \ft12 R(A)_{01}\,\varepsilon_{\hat a}{}^{\hat b}\,,
\end{align}
where the two-dimensional Levi-Civita symbols are normalized by
$\varepsilon^{01}=\varepsilon^{23} =1$. The non-zero components of
the modified curvature $\mathcal{R}(M)_{ab}{}^{cd}$ are given by,
\begin{align}
  \label{eq:R(M)-values}
  \mathcal{R}(M)_{\underline{a} \underline{b}}{}^{\underline{c}
    \underline{d}} =&\,(D+\ft13 R)\, \delta_{\underline{a}\underline{b}}
  {}^{\underline{c}\underline{d}}  \,,\nonumber\\
  \mathcal{R}(M)_{\hat a \hat b}{}^{\hat c \hat d} =&\,(D+\ft13 R)\,
  \delta_{\hat a\hat  b}{}^{\hat c\hat d} \,,\nonumber\\
  \mathcal{R}(M)_{\underline{a} \hat b}{}^{\underline{c} \hat d}
  =&\,\ft12(D-\ft16 R)\, \delta_{\underline{a}}{}^{\underline{c}} \,
  \delta_{\hat b}{}^{\hat d} - \ft12
  R(A)_{23}\,\varepsilon_{\underline{a}}{}^{\underline{c}}\,
  \delta_{\hat b}{}^{\hat d}-\ft12
  R(A)_{01}\,\delta_{\underline{a}}{}^{\underline{c}}\, \varepsilon_{\hat
    b}{}^{\hat d}\, .
\end{align}
We refer to the appendices presented in \cite{deWit:2010za} for the
general definitions of these quantities, which appear in the
superconformal transformation rules of the Weyl multiplet fields and
are therefore needed below.

Motivated by the maximal symmetry of the two two-dimensional
subspaces, we expect the various fields to be invariant under the
same symmetry. Therefore we will assume that the scalars $X^M$ and
$A_i{}^\alpha$ are covariantly constant (for other fields the
covariant constancy will be discussed in due course). The
corresponding integrability condition then requires that the
$\mathrm{U}(1)$ and $\mathrm{SU}(2)$ R-symmetry curvatures are not
necessarily vanishing, and are related to the curvatures of the vector
multiplet gauge fields. This result is consistent with the field
equations for the R-symmetry gauge fields, $A_\mu$ and
$\mathcal{V}_\mu{}^i{}_j$, which lead to the expressions (we again
choose the gauge $b_\mu=0$),
\begin{align}
  \label{eq:R-symm-curv}
  R(A)_{\mu\nu}=&\, g\,\chi_\mathrm{vector}^{-1}
  \mathcal{H}_{\mu\nu}{}^M
    T_{MQ}{}^N \Omega_{PN} \bar{X}^Q X^P \,,  \nonumber \\
  R(\mathcal{V})_{\mu\nu}{}^i{}_j=&\, -4g  \chi_\mathrm{hyper}^{-1}
  \mathcal{H}_{\mu\nu}{}^M\mu^{ik}{}_M \,\varepsilon_{kj} \,.
\end{align}
Observe that the above equations only contribute for $\mu,\nu=t,r$, or
$\mu,\nu=\theta,\varphi$, in view of the space-time symmetry. We can
rewrite these equations in a different form, which is convenient later
on,
\begin{align}
  \label{eq:R-symm-curv-2}
  R(A)^-_{\mu\nu}=&\, g\,\chi_\mathrm{vector}^{-1}
  \hat{\mathcal{H}}_{\mu\nu}^-{}^\Lambda
  \big[T_{\Lambda Q}{}^N+F_{\Lambda\Sigma}\,
  T^\Sigma{}_{Q}{}^N\big]  \Omega_{PN} \bar{X}^Q X^P
  \,,  \nonumber \\[.4ex]
  R(\mathcal{V})^-_{\mu\nu}{}^i{}_j=&\, -4g \chi_\mathrm{hyper}^{-1}
  \hat{\mathcal{H}}^-_{\mu\nu}{}^\Lambda\big[
  \mu^{ik}{}_\Lambda+
  F_{\Lambda\Sigma}\,\mu^{ik\Sigma}\big]\,\varepsilon_{kj}
   + \tfrac14\,\varepsilon^{ik}A_{kj}\,
   T_{\mu\nu}{}^{mn}\varepsilon_{mn}    \,,
\end{align}
where we suppressed all the fermionic terms which vanish in the
background and made use of the field equations \eqref{eq:eom-B} of the
tensor fields $B_{\mu\nu\,{\sf a}}$ and $B_{\mu\nu\,{\sf m}}$, and of
\eqref{eq:cov-der-invariance}.

To study supersymmetry in this background, we present the
non-vanishing terms in the supersymmetry transformations of the
spinors $\Omega_i{}^\Lambda$ and $\zeta^\alpha$,
\begin{align}
  \label{eq:s2-ads2-symm-fermions}
    \delta\Omega_i{}^\Lambda =&\, \tfrac12
    \gamma^{\mu\nu}\hat{\mathcal{H}}^-_{\mu\nu}{}^\Lambda
      \,\varepsilon_{ij} \epsilon^j
    -2g\, T_{NP}{}^\Lambda \, \bar X^N X^P
    \,\varepsilon_{ij}\, \epsilon^j
    - 4\,gN^{\Lambda\Sigma}\big(\mu_{ij\Sigma}
    + \bar F_{\Sigma\Gamma} \,\mu_{ij}{}^\Gamma\big) \epsilon^j  +
    2\,X^\Lambda \eta_i \,,\nonumber\\[.2ex]
  \delta\zeta^\alpha  =&\,2gX^M
  \,T_{M}{}^\alpha{}_\beta A_i{}^\beta\,\varepsilon^{ij}\epsilon_j
  +A_i{}^\alpha \,\eta^i\,.
\end{align}
Note that $\delta\Omega_i{}^\Lambda$ has changed as compared to
\eqref{eq:var-Omega-up} by the presence of the field strength
\eqref{eq:hat-H} (suppressing the fermionic terms, so that
$\hat{\mathcal{H}}_{\mu\nu}^-{}^\Lambda =
\mathcal{H}_{\mu\nu}^-{}^\Lambda-\tfrac14 \bar X^\Lambda
T_{\mu\nu}{}^{ij}\varepsilon_{ij}$), while the expression for
$\delta\zeta^\alpha$ is identical to the one given in
\eqref{eq:max-symm-fermions}. Just as before, we make use of the two
spinors $\Omega_i^\mathrm{V}$ and $\zeta_i^\mathrm{H}$ defined in
\eqref{eq:Zeta-H-Omega-V}. The supersymmetry variation of these fields
in the given background are,
\begin{align}
  \label{eq:delta-omegaV-zetaH}
  \delta\Omega_i^\mathrm{V}=&\, \tfrac14 \chi_\mathrm{vector}^{-1}
  \,\bar X^\Lambda N_{\Lambda\Sigma} \hat{\mathcal{H}}_{\mu\nu}^-
  \gamma^{\mu\nu} \varepsilon_{ij} \epsilon^j +A_{ij} \epsilon^j
  +\eta_i \,, \nonumber\\
  \delta\zeta_i^\mathrm{H}=&\, \varepsilon_{ij} \big(A^{jk} \epsilon_k
  + \eta^j\big) \,,
\end{align}
where $A_{ij}$ was defined in \eqref{eq:def-A}. Supersymmetry
therefore implies that the terms proportional to $\gamma^{\mu\nu}$
must vanish. As it turns out, this condition is just the field
equation for $T_{ab}{}^{ij}$,
\begin{equation}
  \label{eq:T}
  \bar{X}^\Lambda N_{\Lambda\Sigma} \,\hat{\mathcal{H}}_{ab}^{-\Sigma}=0
  \,.
\end{equation}

Two additional fermionic variations are,
\begin{align}
  \label{eq:RQ-and-DOmega}
  \delta[ R(Q)_{ab}{}^i -\ft18 T_{cd}{}^{ij}\gamma^{cd}\gamma_{ab}
  \Omega_j^{\mathrm{V}}]=&\, R(\mathcal{V})_{ab}^-{}^i{}_j \epsilon^j
  -\ft12 \mathcal{R}(M)_{ab}{}^{cd} \gamma_{cd} \epsilon^i
  -\ft18 T_{cd}{}^{ij}\,\gamma^{cd}\gamma_{ab} A_{jk}\,\epsilon^k\,,
  \nonumber\\[.4ex]
  \delta\big[D_a \Omega^\mathrm{V}_i - \tfrac12 A_{ij} \gamma_a
  \Omega^{\mathrm{V}\,j} \big] =&\, f_a{}^b \gamma_b \epsilon_i
  +\tfrac14\mathrm{i}R(A)_{cd}^- \gamma^{cd}\gamma_a \epsilon_i
  -\ft18R(\mathcal{V})_{bci}^-{}^j \gamma^{bc}\gamma_a \epsilon_j
  \nonumber\\
  &\, +\tfrac1{16} A_{ij} T_{bc}{}^{jk}\gamma^{bc} \gamma_a\epsilon_k
  -\tfrac12 A_{ij}  A^{jk}\,\gamma_a\, \epsilon_k \,,
\end{align}
where we refer again to the appendices presented in
\cite{deWit:2010za} for more details. Observe that we have assumed,
motivated by the maximal symmetry of the two-dimensional subspaces,
that also $T_{ab}{}^{ij}$ and $A_{ij}$ are covariantly constant.

The consequences of \eqref{eq:RQ-and-DOmega} can be expressed as
follows,\footnote{ 
  There are also charge conjugated equations. For instance, the first
  equation reads,
  \begin{equation}
    \label{eq:conj}
    (D+\tfrac1{12} R)\epsilon_i + \big[R(\mathcal{V})_{23}^+{}_i{}^j
    +\mathrm{i} R(A)_{23}^+\,\delta_i{}^j \big]\gamma^{23}\,
    \epsilon_j  = 0\,.
  \end{equation}
  }  
\begin{align}
  \label{eq:consequences-RQ-DOmega}
  (D+\tfrac1{12} R)\epsilon^i + \big[R(\mathcal{V})_{23}^-{}^i{}_j
  -\mathrm{i} R(A)_{23}^-\,\delta^i{}_j \big]\gamma^{23}\,
    \epsilon^j  =&\,0\,,      \nonumber\\
  (D-\tfrac16 R)\epsilon^i - \big[2\mathrm{i}
  R(A)_{23}^- \, \delta^i{}_j
    +\tfrac12 \mathrm{i}w \,\varepsilon^{ik} A_{kj} \big]\gamma^{23}\,
    \epsilon^j  =&\,0\,, \nonumber\\
    \big[A^{ik} A_{kj}\epsilon^j +\tfrac14 \mathrm{i}w
    \,\varepsilon^{ik} A_{kj} \,\gamma^{23}\big]\,
    \epsilon^j  =&\,0\,, \nonumber\\
    (v_1^{-1}+v_2^{-1}-\ft18 |w|^2) \epsilon^i -
    \big[\ft12\mathrm{i}\bar{w}
    A^{ik}\varepsilon_{kj}  +
    2\,R(\mathcal{V})^+_{23}{}^i{}_j+2\mathrm{i}R(A)^+_{23}\,
    \delta^i{}_j\big]\gamma^{23} \epsilon^j
    =&\,  0\,.
\end{align}
Furthermore we note that the covariant constancy of $T_{ab}{}^{ij}$
and $A_{ij}$ implies the conditions,
\begin{align}
  \label{eq:cov-const-T-A}
   w\, R(A)_{\mu\nu} =0\,,\qquad R(\mathcal{V})_{\mu\nu}{}^k{}_{(i}
   \,A_{j)k} = - \mathrm{i}R(A)_{\mu\nu}\,A_{ij}  \,.
\end{align}
An important observation is that both
$\mathrm{i}R(\mathcal{V})_{\mu\nu}{}^i{}_j$ (for any $\mu,\nu$) and
$\varepsilon^{ik}A_{kj}$ are $2\times2$ matrices that take their value
in the Lie algebra of $\mathrm{SU}(2)$. However, while the matrices
$\mathrm{i}R(\mathcal{V})_{\mu\nu}{}^i{}_j$ are necessarily hermitian,
this is not the case with $\varepsilon^{ik}A_{kj}$, which is in general
complex-valued.

We now turn to possible supersymmetric solutions for this
background. We proceed in two steps. First we analyze the conditions
for supersymmetry, ignoring the transformations
\eqref{eq:delta-omegaV-zetaH}. This will reveal the possible existence
of three distinct classes of supersymmetric solutions, with four or
eight supersymmetries, depending on the values of
$R(\mathcal{V})_{\mu\nu}{}^i{}_j$ and $A_{ij}$. The corresponding
information is summarized in table~\ref{tab:susy-ads2-s2}. As a last
step we then analyze the transformations
\eqref{eq:delta-omegaV-zetaH}, which lead to additional
constraints. It then follows that one of the classes listed in
table~\ref{tab:susy-ads2-s2} is actually not realized. In what follows
we will decompose the equations \eqref{eq:consequences-RQ-DOmega} in
eigenstates of $\mathrm{i}\gamma^{23}$, denoted by $\epsilon^i_\pm=
\tfrac12(1\pm \mathrm{i}\gamma^{23})\epsilon^i$. Observe that these
spinors transform as a product representation of the $\mathrm{SU}(2)$
isometry group associated with $S^2$ and the $\mathrm{SU}(2)$
R-symmetry. This observation will be relevant shortly. Note also that
the spinors transform according to $\epsilon_\pm{}^i\to
\epsilon_{i\mp}$ under charge conjugation.

We start by noting that $w=0$ will only lead to a supersymmetric
solution provided $v_1^{-1}=0$. Discarding this singular solution, we thus
assume $R(A)_{\mu\nu}=0$. Then we consider two classes of solutions,
denoted by $A$ and $B$ in table~\ref{tab:susy-ads2-s2}, depending on
whether $D-\tfrac16R$ vanishes or not.

For $R(A)_{\mu\nu}=0$ and $D-\tfrac16R=0$, the equations
\eqref{eq:consequences-RQ-DOmega} imply,
\begin{align}
  \label{eq:D=R-6}
  w\,A_{ij}\,\epsilon_\pm^j=&\, 0\,, \nonumber\\
  \mathrm{i} R(\mathcal{V})_{23}^-{}^i{}_j \,\epsilon_\pm^j =&\,
  \pm \tfrac14 R \,\epsilon_\pm^i \,,\nonumber\\
  \big[\mathrm{i} R(\mathcal{V})_{23}^+{}^i{}_j -\tfrac14 \bar w
  A^{ik}\varepsilon_{kj} \big]\,\epsilon_\pm^j =&\, \mp
  \tfrac12(v_1^{-1} + v_2^{-2}-\tfrac18|w|^2)\,\epsilon_\pm^i  \,.
\end{align}
Let us now assume that $A_{ij}\not=0$. In that case
$\varepsilon^{ik}A_{kj}$ must have a single null vector in order that
a supersymmetric solution exists. On the other hand, it must commute
with the $\mathrm{SU}(2)$ curvatures, which in this case implies that
the $R(\mathcal{V})_{\mu\nu}{}^i{}_j$ must vanish. Supersymmetry then
requires that $v_1=v_2$ and
\begin{equation}
  \label{eq:7}
   w\,A_{ij}\,\epsilon_\pm^j=0\,, \qquad
   \bar w A^{ik}\varepsilon_{kj}\,\epsilon_\pm^j = \pm
   (4\,v_1^{-1} -\tfrac14|w|^2)\,\epsilon_\pm^i\,.
\end{equation}
These equations have no solution unless $A_{ij}=0$. When $A_{ij}=0$
and the $\mathrm{SU}(2)$ curvatures are non-vanishing, one can show
that \eqref{eq:D=R-6} implies,
\begin{equation}
  \label{eq:A-vanish}
  \mathrm{i} R(\mathcal{V})_{23}{}^i{}_j\,\epsilon_\pm^j = \pm
  \tfrac12 R\,\epsilon_\pm^i\,, \qquad  v^{-1}_1 = \tfrac1{16} \vert
  w\vert^2\,.
\end{equation}
This solution, denoted by $A_{[2]}$, has generically four
supersymmetries, two associated with two of the spinor parameters
$\epsilon_\pm^i$, and two related with the charge-conjugated spinors
$\epsilon_{i\mp}$. The two spinors of the $\epsilon_\pm^i$ must be
eigenspinors of both $\mathrm{i}\gamma^{23}$ and $\mathrm{i}
R(\mathcal{V})_{23}{}^i{}_j$ with related eigenvalues. Therefore the
supersymmetries of class $A_{[2]}$ (and also of class $B$, as we shall
see later) cannot transform consistently under the $\mathrm{SU}(2)$
isometry group. We will return to this aspect shortly.

In the special case where both $A_{ij}$ and the $\mathrm{SU}(2)$
curvatures vanish, we have $v_1^{-1}=v_2^{-1} = \tfrac1{16}
|w|^2$. Generically we then have eight supersymmetries. This class is
denoted by $A_{[1]}$. Here the supersymmetries act consistently under
the action of both $\mathrm{SU}(2)$ groups. This completes the
discussion of the type-$A$ solutions.

Subsequently we turn to the solutions of class $B$, where
$D-\tfrac16R\not=0$ and $R(A)_{\mu\nu}=0$. This class is denoted by
$B$. In that case the first two equations
\eqref{eq:consequences-RQ-DOmega} imply,
\begin{align}
  \label{eq:RV-A-epsilon}
  \mathrm{i}R(\mathcal{V})^-_{23}{}^i{}_j \epsilon_\pm^j =&\,
  \pm(D+\tfrac1{12} R)\,\epsilon_\pm^i \,,\nonumber\\
  \tfrac12 w\, \varepsilon^{ik}A_{kj} \,\epsilon_\pm^j =&\,
  \pm(D-\tfrac16 R)\,\epsilon_\pm^i \,.
\end{align}
With this result, the last two equations then yield the eigenvalue
equations,
\begin{align}
  \label{eq:RV-A-epsilon-cc}
  \mathrm{i}R(\mathcal{V})^+_{23}{}^i{}_j \epsilon_\pm^j =&\,
  \mp \tfrac12(v_1^{-1}+ v_2^{-1}-\tfrac14|w|^2)\,\epsilon_\pm^i
  \,,\nonumber\\
  \tfrac12 \bar w\, A^{ik}\varepsilon_{kj} \,\epsilon_\pm^j =&\,
  \pm \tfrac18 |w|^2 \,\epsilon_\pm^i \,.
\end{align}
Combining these equations leads to,
\begin{align}
  \label{eq:generic}
  \bar w\,A^{ij} =&\, -w\,\varepsilon^{ik}\,\varepsilon^{jl} \, A_{kl}
  \,, \nonumber\\
  R(\mathcal{V})_{23}^-{}^i{}_j = &\,
  R(\mathcal{V})_{23}^+{}^i{}_j =\tfrac12
  R(\mathcal{V})_{23}{}^i{}_j = -\frac{2\mathrm{i}}{v_2\,\bar w} \,
  \varepsilon^{ik} \,A_{kj}  \,,\nonumber\\
  \mathrm{i}R(\mathcal{V})_{23}{}^i{}_j \epsilon_\pm^j =&\,
  \mp  v_2^{-1}\,\epsilon_\pm^i    \,,\nonumber\\
  D=&\, -\tfrac16\big(v_1^{-1} +2v_2^{-1}\big)\,,
  \nonumber\\
  v_1^{-1} =&\, \tfrac14 |w|^2 \,.
\end{align}
Just as in class $A_{[2]}$, these solution have generically four
supersymmetries, which cannot transform consistently under the action
of the $\mathrm{SU}(2)$ isometry group. Furthermore, note that the
solutions become singular in the limit where
$\mathcal{V}_{\mu\nu}{}^i{}_j$ and $A_{ij}$ vanish, so that this class
is really distinct from the type-$A$ class.

\begin{table}[tb]
\begin{center}\begin{tabular}{|c||cccccc|}
 \hline class & $R(\mathcal{V})$ && $A_{ij}$ && $v_1,v_2$&  susy
 \\\hline\hline
  $A_{[1]}$ & $R(\mathcal{V})=0$ && $A_{ij}=0$&& $v_1^{-1}=v_2^{-1}
  =\tfrac1{16}|w|^2$   & ${\bf 4}+\bar {\bf 4}$
  \\[.1ex] \hline
  $A_{[2]}$ & $R(\mathcal{V})_{23}=\mathcal{O}(v_1^{-1}-v_2^{-1})$ &&
  $A_{ij}=0$ && $v_1^{-1}
  =\tfrac1{16}|w|^2\not =v_2^{-1}$ & ${\bf 2}+\bar {\bf 2}$  \\[.1ex]
  \hline
  $B$ &\multicolumn{3}{c}{
    $R(\mathcal{V})_{23}{}^i{}_j=-\tfrac{4\mathrm{i}}{v_2\,\bar w} \,
    \varepsilon^{ik}A_{kj}=\mathcal{O}(v_2^{-1}) $ }
  &&$v_1^{-1}=\tfrac14|w|^2$  &${\bf 2}+\bar {\bf 2}$  \\[.1ex] \hline
\end{tabular}\end{center}\caption{Three classes of supersymmetric
solutions. As shown in due course, only the classes $A_{[1]}$ and $B$
are actually realized. }
\label{tab:susy-ads2-s2}
\end{table}

In view of the fact that the supersymmetry spinors do not always seem
to transform consistently under the action of the $\mathrm{SU}(2)$
transformations associated with the $S^2$ isometries, let us now first
clarify this issue and turn to a discussion of the Killing spinor
equations (in gauge $b_\mu=0$) for each of the three classes. These
equations take the following form,
\begin{equation}
  \label{eq:var-gravitini}
  \delta\big(\psi_\mu{}^i + \gamma_\mu\,\Omega^{\mathrm{V}i}\big) =
  2 {\stackrel{\circ} {\nabla}}_\mu \epsilon^i +\mathrm{i} A_\mu
  \,\epsilon^i
  +\mathcal{V}_\mu{}^i{}_j  \,\epsilon^j - \varepsilon^{ik} \big[
  \tfrac14\mathrm{i}w\,\gamma^{23}\delta_k{}^j  + \varepsilon_{kl}
  A^{lj} \big]\gamma_\mu \, \epsilon_j\,.
\end{equation}
where ${\stackrel{\circ} {\nabla}}_\mu$ denotes the
$\mathrm{AdS}_2\times S^2$ covariant derivative. Obviously we may set
$A_\mu$ and $\mathcal{V}_{\underline a}=0$.

For class-$A$ solutions \eqref{eq:var-gravitini} leads to,
\begin{align}
  \label{eq:k-spinors}
  {\stackrel{\circ} {\nabla}}_{\underline a} \epsilon_\pm^i
  \mp\tfrac18 w\,\varepsilon^{ij} \gamma_{\underline
    a}\,\epsilon_{j\pm}=0 \,,\nonumber \\
  {\stackrel{\circ} {\nabla}}_{\hat a} \epsilon_\pm^i +\tfrac12
  \mathcal{V}_{\hat a}{}^i{}_j \,\epsilon_\pm^j
  \mp\tfrac18 w\,\varepsilon^{ij} \gamma_{\hat
    a}\,\epsilon_{j\mp}=0 \,,
\end{align}
where $v_1^{-1} = \tfrac1{16}|w|^{2}$. For the solution of class
$A_{[1]}$, we may take $\mathcal{V}_{\hat a}{}^i{}_j=0$, so that we
obtain the standard Killing spinor equations for $\mathrm{AdS}_2\times
S^2$. For the $A_{[2]}$ solutions, the Killing spinor equation on
$S^2$ is somewhat unusual, because of the presence of the R-symmetry
connection whose strength is not related to the size of the
$S^2$. Since we will show later that the type-$A_{[2]}$ solutions are
in fact not realized, we refrain from further discussion concerning
these solutions.

Hence we proceed to the class-$B$ solutions. In this case, the Killing
spinor equation \eqref{eq:var-gravitini} decomposes into,
\begin{align}
  \label{eq:k-spinors-other-class}
  {\stackrel{\circ} {\nabla}}_{\underline a} \epsilon_\pm^i
  \mp\tfrac14 w\,\varepsilon^{ij} \gamma_{\underline
    a}\,\epsilon_{j\pm}=&\,0 \,,\nonumber \\
  {\stackrel{\circ} {\nabla}}_{\hat a} \epsilon_\pm^i +\tfrac12
  \mathcal{V}_{\hat a}{}^i{}_j \,\epsilon_\pm^j
  =&\,0 \,.
\end{align}
Because $v_1^{-1} = \tfrac14|w|^{2}$, the first equation is the
standard $\mathrm{AdS}_2$ Killing spinor equation. However, the second
equation does not coincide with the standard Killing spinor equation
on $S^2$. We note that the strength of the R-symmetry connection is
proportional to $v_2^{-1}$, and is therefore also determined by the $S^2$ radius. To elucidate the situation, let us briefly
discuss the relevant equations for the unit sphere ($v_2=1$).

We use the standard coordinates $\theta$ and $\varphi$ on $S^2$, with
zweibeine $e^2 = \mathrm{d}\theta$ and $e^3=
\sin\theta\,\mathrm{d}\varphi$, and gamma matrices $\gamma_2$ and
$\gamma_3$ that satisfy the standard Clifford algebra relation with
positive signature. The spin connection field in our convention equals
$\omega=\omega^{23}=- \omega^{32} = \cos\theta
\,\mathrm{d}\varphi$. Consequently we have that ${\stackrel{\circ}
  {\nabla}}_\theta= \partial_\theta$ and ${\stackrel{\circ}
  {\nabla}}_\varphi= \big(\partial_\varphi -\tfrac12 \cos\theta
\,\gamma^{23}\big)$. Now we adopt an R-symmetry transformation to
bring $R(\mathcal{V})_{23}{}^i{}_j$ in diagonal form. In that case
we can assume $\mathcal{V}{}^i{}_j = - \mathrm{i} \lambda
\,(\sigma_3)^i{}_j \,\cos\theta\,\mathrm{d}\varphi$ with $\lambda$
some real constant and $\sigma_3$ the diagonal Pauli matrix. This
leads to the corresponding field strength $R(\mathcal{V})_{23}{}^i{}_j
= \mathrm{i} \lambda \,(\sigma_3)^i{}_j$. From the third equation of
\eqref{eq:generic} we conclude that $\vert \lambda\vert=1$ and by
an additional R-symmetry transformation we can ensure that
$\lambda=1$. In that case (remember that we put $v_2=1$) the
supersymmetries are parametrized by the parameters $\epsilon_+^1$ and
$\epsilon_-^2$. It is now straightforward to verify that these spinors
do not depend on the $S^2$ coordinates as a result of the second
equation \eqref{eq:k-spinors-other-class}.

Consequently the supersymmetries do not transform under the isometries
of $S^2$, which implies that they carry no spin! Along the same lines
one expects that also the fields in this background will change their
spin assignment. The reason that the spin assignments change in this
background, is that the spin rotations associated with the isometries
of $S^2$ become entangled with R-symmetry transformations, in a
similar way as in magnetic monopole solutions, where the rotational
symmetry becomes entangled with gauge transformations
\cite{Hasenfratz:1976gr}. In the superconformal context, where one has
R-symmetry connections (which in this solution live on $S^2$), the
geometric origin of the entanglement is clear. While such conditions
on the supersymmetry spinor have been obtained previously in the
literature for a variety of four- and five-dimensional supersymmetric
solutions (see,
e.g. \cite{Romans:1991nq,Caldarelli:1998hg,Gutowski:2004yv,Dall'Agata:2010gj,
  Hristov:2010ri}, this phenomenon seems not to have received special
attention.

Finally we must investigate the remaining variations based on
\eqref{eq:s2-ads2-symm-fermions}. Consider first the variation for the
fields $\Omega_i{}^\Lambda$, which we parametrize as
$\delta\Omega_i{}^\Lambda = A_{ij}{}^\Lambda \epsilon^j- 2X^\Lambda
\eta_i$, so that
\begin{align}
  \label{eq:defALambda}
  A_{ij}{}^\Lambda = 2\,
  \hat{\mathcal{H}}^-_{23}{}^\Lambda \varepsilon_{ij} \gamma^{23}
  -2g\, T_{NP}{}^\Lambda \, \bar X^N X^P \,\varepsilon_{ij}-
  4\,gN^{\Lambda\Sigma}\big(\mu_{ij\Sigma} + \bar F_{\Sigma\Gamma}
  \,\mu_{ij}{}^\Gamma\big) \,.
\end{align}
Then we consider the variation of two S-invariant combinations,
$\Omega_i{}^\Lambda -2 X^\Lambda \Omega_i^\mathrm{V}$, and $D_a
(\Omega^{i\,\Lambda} -2 \bar X^\Lambda \Omega^{i\mathrm{V}}) -
\tfrac12(A^{ij}{}^\Lambda -2\, \bar X^\Lambda A^{ij})
\gamma_a\Omega_j^\mathrm{V}$, whose vanishing under supersymmetry
imply the following identities,
\begin{align}
  \label{eq:Omega-DOmega}
  \big[A_{ij}{}^\Lambda -2\, X^\Lambda A_{ij}\big]\,\epsilon^j
  =&\,0\,,\nonumber \\
  \big(A^{ik}{}^\Lambda -2\,\bar X^\Lambda A^{ik}\big)\,\big( A_{kj}\,
    -\tfrac18 T_{bckj}\,\gamma^{bc}\big) \,\gamma_a\epsilon^j  =&\,0
    \,,
\end{align}
where we assumed that $\mathcal{D}_\mu A^\Lambda=0$ in line with our
earlier ans\"atze. Likewise we obtain two equations for the
hypermultiplets,
\begin{align}
  \label{eq:zeta-Azeta}
  \big[2\,g\bar X^M \bar T_M{}^{\bar\alpha}{}_{\bar\beta}
  \,A^{i\bar\beta}\,\varepsilon_{ij} -A^{i \bar\alpha}A_{ij}\big]
  \,\epsilon^j=&\,0\,, \nonumber\\
  \big( 2\,g X^M T_M{}^{\alpha}{}_{\beta}
  \,A_i{}^{\beta}\,\varepsilon^{ik} -A_i{}^{\alpha}A^{ik}\big)\, \big(A_{kj}
  -\tfrac18 T_{bckj} \,\gamma^{bc} \big) \,\gamma_a \epsilon^j
  =&\,0\,.
\end{align}
We note the presence of a universal factor on the right-hand side of
the equation in \eqref{eq:Omega-DOmega} and \eqref{eq:zeta-Azeta},
which is proportional to
\begin{equation}
  \label{eq:prefactor}
  A_{kj} -\tfrac18 T_{bckj} \,\gamma^{bc} =
  -\varepsilon_{kl}\big(\varepsilon^{lm} A_{mj}-\tfrac14\mathrm{i} \bar
  w \,\gamma^{23} \,\delta^l{}_j\big)   \,,
\end{equation}
which is the hermitian conjugate of the term that appears at the
right-hand side of \eqref{eq:var-gravitini}.

The equations \eqref{eq:Omega-DOmega} and \eqref{eq:zeta-Azeta} lead
to the following six conditions,
\begin{align}
  \label{eq:v-h-susy}
  \big[g\, T_{NP}{}^\Lambda \, \bar X^N X^P\,\delta^i{}_j -
  2\,gN^{\Lambda\Sigma} \,\varepsilon^{ik}\big(\mu_{kj\Sigma} + \bar
  F_{\Sigma\Gamma} \,\mu_{kj}{}^\Gamma\big) -
  X^\Lambda\,\varepsilon^{ik} A_{kj}
  \big]\,\epsilon_\pm^j =&\, \mp
  \mathrm{i}\hat{\mathcal{H}}^-_{23}{}^\Lambda
  \,\epsilon^i_\pm  \,, \nonumber\\[.4ex]
 \big[ g\, T_{NP}{}^\Lambda \, X^N \bar X^P \,\varepsilon^{ik}+
  2\,gN^{\Lambda\Sigma} \big(\mu^{ik}{}_\Sigma + F_{\Sigma\Gamma}
  \,\mu^{ik \Gamma}\big) + \bar X^\Lambda\,A^{ik}\big]\,A_{kj}
  \,\epsilon^j_\pm =&\, \tfrac14\mathrm{i} \bar
  w\,\hat{\mathcal{H}}^+_{23}{}^\Lambda \,\epsilon^i _\pm \,,
  \nonumber\\[.4ex]
   \bar w\big[g\,
   T_{NP}{}^\Lambda \, X^N \bar X^P \,\delta^i{}_j -
  2\,gN^{\Lambda\Sigma}\varepsilon^{ik}\big(\mu_{kj\Sigma} + F_{\Sigma\Gamma}
  \,\mu_{kj}{}^\Gamma\big) - \bar
  X^\Lambda\,A^{ik} \varepsilon_{kj} \big]
  \,\epsilon^j_\pm= &\,4\mathrm{i} \,
  \hat{\mathcal{H}}^+ _{23}{}^\Lambda \,\varepsilon^{ik} A_{kj}
   \,\epsilon^j_\pm   \,,  \nonumber\\[.4ex]
  \big[2\,g\bar X^M \bar T_M{}^{\bar\alpha}{}_{\bar\beta}
  \,A^{i\bar\beta}\,\varepsilon_{ij} -A^{i \bar\alpha}A_{ij}\big]
  \,\epsilon^j_\pm =&\,0\,, \nonumber\\[.4ex]
  \big[ 2\,g X^M T_M{}^{\alpha}{}_{\beta}
  \,A_i{}^{\beta}\,\varepsilon^{ik} -A_i{}^{\alpha} A^{ik}\big]\,
   A_{kj}\, \epsilon_\pm^j =&\,0\,,\nonumber\\[.4ex]
   \big[ 2\,g X^M T_M{}^{\alpha}{}_{\beta}
  \,A_i{}^{\beta}\,\varepsilon^{ik} -A_i{}^{\alpha} A^{ik}\big]   \,
  \varepsilon_{kj}\,\epsilon^j_\pm =&\, 0\,.
\end{align}

Let us now consider the various classes of solutions shown in
table~\ref{tab:susy-ads2-s2}. First of all the solutions of type $A$,
characterized by $A_{ij}=0$. From the second equation of
\eqref{eq:v-h-susy} it then follows that
$\hat{\mathcal{H}}_{\mu\nu}{}^\Lambda = 0$. Combining this result with
the equations \eqref{eq:R-symm-curv-2} shows that both $R(A)_{\mu\nu}$
and $R(\mathcal{V})_{\mu\nu}{}^i{}_j$ must vanish. This implies that
solution $A_{[2]}$ is not realized. Hence we are left with the fully
supersymmetric solution $A_{[1]}$. Therefore we proceed by determining the
additional restrictions for this solution.

The first, third, fourth and sixth equations of \eqref{eq:v-h-susy}
can be written as follows,
\begin{align}
  \label{eq:v-h-A}
  \mathrm{i}\varepsilon^{ik}\mu_{kj}{}^\Lambda\,\epsilon_\pm^j  =&\,-
  \tfrac12   T_{NP}{}^\Lambda (\bar
  X^NX^P-X^N\bar X^P) \,\epsilon_\pm^i\,, \nonumber\\[.4ex]
  \mathrm{i}N^{\Lambda\Sigma} \varepsilon^{ik}\big(
  2\,\mu_{kj\Sigma}+(F_{\Sigma\Gamma} + \bar F_{\Sigma\Gamma})
  \mu_{kj}{}^\Gamma\big)\,\epsilon_\pm^j =&\,\tfrac12 \mathrm{i}
  T_{NP}{}^\Lambda  (\bar X^NX^P+X^N\bar X^P)
  \,\epsilon_\pm^i\,,\nonumber\\[.4ex]
  \bar X^M \bar T_M{}^{\bar\alpha}{}_{\bar\beta}
  \,A^{i\bar\beta}\,\varepsilon_{ij}
  \,\epsilon^j_\pm =&\,0\,, \nonumber\\[.4ex]
  X^M T_M{}^{\alpha}{}_{\beta}
  \,A_i{}^{\beta}\,\epsilon^i_\pm =&\, 0\,.
\end{align}
Since a hermitian matrix must have real eigenvalues, it follows that
both sides of the first two equations should vanish. Also the factors
in the last two equations should vanish, so that
\begin{align}
  \label{eq:vector-eqs}
  \mu_{ij\Lambda} = \mu_{ij}{}^\Lambda =&\,0\,,\nonumber\\
  T_{NP}{}^\Lambda X^N\bar X^P =&\,0\,,\nonumber\\
  X^M T_M{}^{\alpha}{}_{\beta}
  \,A_i{}^{\beta}   =&\,0 =\bar X^M T_M{}^{\alpha}{}_{\beta}
  \,A_i{}^{\beta} \,.
\end{align}
Note that $\mathcal{L}_{g^2}$ is now vanishing. For electric charges
these solutions have already been identified in
\cite{Hristov:2009uj}. Without charges this is the well-known solution
that arises as a near-horizon geometry of BPS black holes. The fact
that the moment maps and certain combinations of Killing vectors are
vanishing does not warrant the conclusion that there is no
gauging. One can only conclude that the field equations require some
of these quantities to vanish for these solutions.

Now consider the type-$B$ solution where $A_{ij}$ is non-vanishing. In
that case the first three equations of \eqref{eq:v-h-susy} lead to two
independent equations,
\begin{align}
  \label{eq:v-h-susy-B}
  \big[g\, T_{NP}{}^\Lambda \, \bar X^N X^P\,\delta^i{}_j -
  2\,gN^{\Lambda\Sigma} \,\varepsilon^{ik}\big(\mu_{kj\Sigma} + \bar
  F_{\Sigma\Gamma} \,\mu_{kj}{}^\Gamma\big)
  \big]\,\epsilon_\pm^j =&\, \mp\big(
  \mathrm{i}\hat{\mathcal{H}}^-_{23}{}^\Lambda +
  \tfrac14 \bar wX^\Lambda\big)
  \,\epsilon^i_\pm  \,, \nonumber\\[.4ex]
   \big[g\,
   T_{NP}{}^\Lambda \, X^N \bar X^P \,\delta^i{}_j -
  2\,gN^{\Lambda\Sigma}\varepsilon^{ik}\big(\mu_{kj\Sigma} + F_{\Sigma\Gamma}
  \,\mu_{kj}{}^\Gamma\big) \big]
  \,\epsilon^j_\pm= &\,\mp\big( \mathrm{i} \,
  \hat{\mathcal{H}}^+_{23}{}^\Lambda - \tfrac14 w \bar
  X^\Lambda\big)  \,\epsilon^i_\pm   \,.
\end{align}
These equations can be analyzed in a similar way as the corresponding
equations in \eqref{eq:v-h-A}. The results are as follows,
\begin{align}
  \label{eq:vector-B}
  T_{NP}{}^\Lambda \bar X^N X^P =&\,0\,,\nonumber\\[.4ex]
  g \varepsilon^{ik} \mu_{kj}{}^\Lambda \, \epsilon^j_{\pm} = &\,\mp\tfrac12
  \big[\big(\hat{\mathcal{H}}^{-\Lambda}_{23}-\ft14 \mathrm{i}
      \bar{w} X^\Lambda\big) -
    \big(\hat{\mathcal{H}}^{+\Lambda}_{23}+\ft14 \mathrm{i} w
      \bar{X}^\Lambda\big)\big]\,\epsilon^i_\pm\,, \nonumber\\[.4ex]
      g \varepsilon^{ik} \mu_{kj\Lambda}\, \epsilon^j_{\pm} =&\, \pm
      \tfrac12 \big[F_{\Lambda \Sigma}
        \big(\hat{\mathcal{H}}^{-\Sigma}_{23}-\ft14 \mathrm{i}
          \bar{w} X^\Sigma\big) -\bar{F}_{\Lambda \Sigma}
        \big(\hat{\mathcal{H}}^{+\Sigma}_{23}+\ft14 \mathrm{i} w
          \bar{X}^\Sigma\big)\big]\,\epsilon^i_\pm\,.
\end{align}
From \eqref{eq:derivative-invariance}, it follows that the first
constraint of \eqref{eq:vector-B} can be generalized to $T_{MN}{}^P
\bar X^MX^N=0$. Using also the representation constraint
\eqref{eq:lin}, one reconfirms that $R(A)_{\mu\nu}$, as given in
\eqref{eq:R-symm-curv-2}, vanishes. The same argument applies to
solutions of type $A_{[1]}$.  Furthermore, as a check one may also
reconstruct the eigenvalue equation for $A_{ij}$ which shows once more
that \eqref{eq:T} must be valid.

One can use the same strategy and determine
$R(\mathcal{V})_{23}{}^i{}_j$ from \eqref{eq:R-symm-curv-2}, making use
of \eqref{eq:vector-B} with $T_{MN}{}^P \bar X^MX^N=0$. Evaluating
this curvature on the supersymmetry parameters, making use of the
eigenvalue condition for this curvature presented in
\eqref{eq:generic} as well as of \eqref{eq:T}, it follows that
\begin{equation}
  \label{eq:v2-eigenvalue}
  v_2^{-1}= -2 \chi_\mathrm{vector}^{-1} N_{\Lambda \Sigma}
  \hat{\mathcal{H}}^{-\Lambda}_{23}
  \hat{\mathcal{H}}^{+\Sigma}_{23}-\ft18 |w|^2\,.
\end{equation}
In the first expression on the right-hand side, one can verify,
replacing $N_{\Lambda\Sigma}$ by the negative definite metric
$M_{\Lambda \bar{\Sigma}}$ defined in \eqref{eq:kinetic-matrices} and
using \eqref{eq:T}, that this
expression must be positive, which yields an upper bound on $|w|^2$
for given field strengths $\hat{\mathcal{H}}_{23}{}^\Lambda$.

The last three equations of \eqref{eq:v-h-susy} lead to two equations,
\begin{align}
  \label{eq:hyper-B}
  X^M\big[ T_M{}^\alpha{}_\beta A_i{}^\beta + \chi_\mathrm{vector}^{-1}
  \varepsilon_{ij}\,\mu^{jk}{}_M \, A_k{}^\alpha\big]  =&\,0\,,
  \nonumber\\
  \bar X^M\big[ T_M{}^\alpha{}_\beta A_i{}^\beta + \chi_\mathrm{vector}^{-1}
  \varepsilon_{ij}\,\mu^{jk}{}_M \, A_k{}^\alpha\big]  =0
\end{align}
From these equations, one derives, upon using \eqref{eq:T-A-squared},
\begin{align}
  \label{eq:kk+AA-pot}
  g^2 \bar X^MX^N\, k^A{}_M \,k_{AN} = \tfrac1{16} \chi_\mathrm{vector}\,
  \vert w\vert^2 \,.
\end{align}
The scalar potential in the type-$B$ solutions thus takes the form,
\begin{align}
  \label{scalar-potential-B}
  e^{-1}\mathcal{L}_{g^2} =&\, -2\, g^2\,\chi_\mathrm{vector}\,
  M_{\bar\Lambda\Sigma}\, N^{\Lambda\Gamma} \left[\mu^{ij}{}_\Gamma +
    F_{\Gamma\Omega}\,\mu{}^{ij\Omega}\right] \,N^{\Sigma\Xi}
  \left[\mu_{ij\Xi} +\bar
    F_{\Xi\Delta}\,\mu_{ij}{}^\Delta\right] \nonumber\\
  &\, -\tfrac3{16} \chi_\mathrm{vector} \,\vert w\vert^2 \,,
\end{align}
where the first term is negative and the second one positive. We
refrain from giving further results.

For a single (compensating) hypermultiplet, which can only have
abelian gaugings, we expect that one of these type-$B$ solutions
describes the near-horizon geometry of the spherically symmetric
static black hole solution presented in
\cite{Dall'Agata:2010gj,Hristov:2010ri}. The result of this paper then
ensures that this black hole solution has supersymmetry enhancement at
the horizon.

\section*{Acknowledgement}
The authors thank Kiril Hristov, Stefanos Katmadas, Ivano Lodato and
Stefan Vandoren for valuable discussions, and Tetsuji Kimura for a
careful reading of the manuscript. The work of M.v.Z. is part
of the research programme of the `Stichting voor Fundamenteel
Onderzoek der Materie (FOM)', which is financially supported by the
`Nederlandse Organisatie voor Wetenschappelijk Onderzoek (NWO)'.  This
work is supported in part by the ERC Advanced Grant no. 246974, {\it
  ``Supersymmetry: a window to non-perturbative physics''}.
\\

\providecommand{\href}[2]{#2}


\begin{thebibliography}{10}
%
%
\bibitem{deWit:2001pz}
  B.~de~Wit, {\it Electric-magnetic duality in supergravity},
  {Nucl. Phys. Proc. Suppl.} {\bf 101} (2001) 154,
\href{http://xxx.lanl.gov/abs/hep-th/0103086}{{\tt
hep-th/0103086}}.
%
\bibitem{Andrianopoli:1996cm}
  L. Andrianopoli, M. Bertolini, A. Ceresole, R. D'Auria, S. Ferrara,
  P. Fr\'e and T. Magri, {\it N = 2 supergravity and N = 2 super
  Yang-Mills theory on general scalar  manifolds: Symplectic
  covariance, gaugings and the momentum map},
  J. Geom. Phys. {\bf 23} (1997) 111, {\tt hep-th/9605032}.
%
\bibitem{deWit:2005ub}
B.~de Wit, H.~Samtleben and M.~Trigiante, \emph{Magnetic charges
in local field theory}, JHEP {\bf 0509} 2005 016,
\texttt{hep-th/0507289}.
%
\bibitem{deWit:1980tn}
  B.~de Wit, J.~W.~van Holten and A.~Van Proeyen,
  {\it Structure of N=2 supergravity},
  Nucl.\ Phys.\ {\bf B184} (1981) 77
  [Erratum-ibid.\ {\bf B222} (1983) 516].
%
\bibitem{deWit:1984pk}
 B. de Wit and A. Van Proeyen, {\it Potentials and
    symmetries of general gauged N=2 supergravity-Yang-Mills
    theory}, Nucl. Phys. {\bf B245} (1984) 89.
%
\bibitem{deWit:1984px} B.~de~Wit, P.~G. Lauwers, and A.~Van~Proeyen,
  \emph{ Lagrangians of {N}=2 supergravity-matter systems}, { Nucl.
    Phys.} {\bf B255} (1985) 569.
%
\bibitem{de Vroome:2007zd}
  M.~de Vroome and B.~de Wit,
  {\it Lagrangians with electric and magnetic charges of N=2
  supersymmetric gauge theories},
  JHEP {\bf 0708} (2007) 064
  [arXiv:0707.2717 [hep-th]].
%
\bibitem{Louis:2002ny}
  J.~Louis and A.~Micu,
  {\it Type II theories compactified on Calabi-Yau threefolds in the
  presence  of  background fluxes},
  Nucl.\ Phys.\ {\bf B635} (2002) 395,
  {\tt hep-th/0202168}.
%
\bibitem{Dall'Agata:2003yr}
  G. Dall'Agata, R. D'Auria, L. Sommovigo and S. Vaula, {\it D = 4, N
  = 2 gauged supergravity in the presence of tensor multiplets},
  Nucl. Phys. {\bf B682} (2004) 243, {\tt hep-th/0312210}.
\bibitem{Sommovigo:2004vj} L.~Sommovigo and S.~Vaula, {\it {D} = 4,
    {N} = 2 supergravity with abelian electric and magnetic charge},
  { Phys. Lett.} {\bf B602} (2004) 130,
  \href{http://xxx.lanl.gov/abs/hep-th/0407205}{{\tt
      hep-th/0407205}}.
%
\bibitem{DAuria:2004yi} R.~D'Auria, L.~Sommovigo, and S.~Vaula, {\it
    ${N} = 2$ supergravity {L}agrangian coupled to tensor multiplets
    with electric and magnetic fluxes}, {JHEP} {\bf 11} (2004) 028,
  \href{http://xxx.lanl.gov/abs/hep-th/0409097}{{\tt
      hep-th/0409097}}.
%
\bibitem{Schon:2006kz}
J. Sch\"on and M. Weidner, {\it Gauged N=4 supergravities}, JHEP
{\bf 0605} (2006) 034, {{\tt hep-th/0602024}}.
%
\bibitem{de Wit:2007mt}
  B.~de Wit, H.~Samtleben and M.~Trigiante,
  {\it The maximal D=4 supergravities},
  JHEP {\bf 0706} (2007) 049
  [arXiv:0705.2101 [hep-th]].
%
\bibitem{Hartong:2009az}
  J.~Hartong, M.~Hubscher and T.~Ortin,
  {\it The supersymmetric tensor hierarchy of N=1,d=4 supergravity},
  JHEP {\bf 0906} (2009) 090
  [arXiv:0903.0509 [hep-th]].
%
\bibitem{Antoniadis:1996}
  I. Antoniadis, H. Partouche and T.R. Taylor,
  {\it Spontaneous breaking of N=2 global supersymmetry}, Phys.Lett. {\bf
  B372} (1996) 83, {{\tt hep-th/9512006}}.
%
%
\bibitem{Behrndt:2001mx}
     K. Behrndt, G.L. Cardoso and D. L\"ust,
     {\it Curved BPS domain wall solutions in four-dimensional N=2
     supergravity}, Nucl. Phys. {\bf B607} (2001) 391, {\tt
     hep-th/0102128}.
%
\bibitem{Grana:2005jc}
  M. Grana, {\it Flux compactifications in string theory: A
  comprehensive review}, Phys. Rept. {\bf 423} (2006) 91, {\tt
  hep-th/0509003}.
%
\bibitem{Louis:2009xd}
  J.~Louis, P.~Smyth and H.~Triendl,
  {\it Spontaneous N=2 to N=1 supersymmetry breaking in supergravity
    and type II string theory},
  JHEP {\bf 1002} (2010) 103
  [arXiv:0911.5077 [hep-th]].
%
\bibitem{Louis:2010ui}
  J.~Louis, P.~Smyth and H.~Triendl,
  {\it The N=1 Low-energy effective action of spontaneously broken N=2
  supergravities},
  JHEP {\bf 1010} (2010) 017
  [arXiv:1008.1214 [hep-th]].
%
\bibitem{Cecotti:1984rk}
  S.~Cecotti, L.~Girardello and M.~Porrati,
  {\it Two into one won't go},
  Phys.\ Lett.\  B {\bf 145} (1984) 61.
%
\bibitem{Cecotti:1985sf}
  S.~Cecotti, L.~Girardello and M.~Porrati,
  {\it An exceptional N=2 supergravity with flat potential and partial
    super-Higgs},
  Phys.\ Lett.\  B {\bf 168} (1986) 83.
%
\bibitem{Ferrara:1995xi} S.~Ferrara, L.~Girardello and M.~Porrati,
  {\it Spontaneous breaking of N=2 to N=1 in rigid and local
    supersymmetric theories}, Phys.\ Lett.\ B {\bf 376} (1996) 275
  [arXiv:hep-th/9512180].
%
\bibitem{Ferrara:1995gu}
  S.~Ferrara, L.~Girardello and M.~Porrati,
  {\it Minimal Higgs branch for the breaking of half of the
    supersymmetries in N=2 supergravity},
  Phys.\ Lett.\  B {\bf 366} (1996) 155
  [arXiv:hep-th/9510074].
%
\bibitem{Fre:1996js} P.~Fr\'e, L.~Girardello, I.~Pesando and
  M.~Trigiante, {\it Spontaneous $N=2 \to N=1$ local supersymmetry
    breaking with surviving compact gauge group},
  Nucl.\ Phys.\  B {\bf 493} (1997) 231
  [arXiv:hep-th/9607032].
%
\bibitem{Hristov:2009uj}
  K.~Hristov, H.~Looyestijn and S.~Vandoren,
  {\it Maximally supersymmetric solutions of D=4 N=2 gauged
    supergravity},
  JHEP {\bf 0911} (2009) 115
  [arXiv:0909.1743 [hep-th]].
%
\bibitem{Hasenfratz:1976gr} P.~Hasenfratz and G.~'t Hooft, {\it A
    Fermion-Boson puzzle in a gauge theory}, Phys.\ Rev.\ Lett.\ {\bf
    36 } (1976) 1119.
%
\bibitem{Dall'Agata:2010gj}
  G.~Dall'Agata and A.~Gnecchi,
  {\it Flow equations and attractors for black holes in N = 2 U(1)
    gauged supergravity},
  JHEP {\bf 1103 } (2011)  037.
  [arXiv:1012.3756 [hep-th]].
%
\bibitem{Hristov:2010ri}
  K.~Hristov and S.~Vandoren,
  {\it Static supersymmetric black holes in $AdS_4$ with spherical
    symmetry},
  JHEP {\bf 1104} (2011) 047
  [arXiv:1012.4314 [hep-th]].
%
\bibitem{deWit:2010za}
  B.~de Wit, S.~Katmadas, M.~van Zalk,
  {\it New supersymmetric higher-derivative couplings: Full N=2 superspace does not count!},
  JHEP {\bf 1101 } (2011)  007.
  [arXiv:1010.2150 [hep-th]].
%
\bibitem{de Wit:1983rz}
  B.~de Wit, P.~G.~Lauwers, R.~Philippe, S.~Q.~Su and A.~Van Proeyen,
  {\it Gauge and matter fields coupled to N=2 supergravity},
  Phys.\ Lett.\  B {\bf 134} (1984) 37.
%
\bibitem{Cecotti:1988qn} S.~Cecotti, S.~Ferrara and L.~Girardello,
  {\it Geometry of type II superstrings and the moduli of
    superconformal field theories}, Int.\ J.\ Mod.\ Phys.\ {\bf A4}
  (1989) 2475.
%
\bibitem{deWit:1999fp}
    B.~de Wit, B.~Kleijn and S.~Vandoren,
    {\it Superconformal hypermultiplets},
    Nucl.\ Phys.\ B {\bf 568} (2000) 475, {\tt hep-th/9909228}.
%
\bibitem{deWit:1998zg}
  B.~de Wit, B.~Kleijn and S.~Vandoren,
  {\it Rigid N=2 superconformal hypermultiplets},
  arXiv:hep-th/9808160.
%
\bibitem{Gibbons:1998xa}
  G.~W.~Gibbons and P.~Rychenkova,
  {\it Cones, triSasakian structures and superconformal invariance},
  Phys.\ Lett.\  B {\bf 443} (1998) 138
  [arXiv:hep-th/9809158].
%
\bibitem{deWit:2001bk}
  B.~de~Wit, M.~Ro\v{c}ek and S.~Vandoren, {\it Gauging
  isometries on hyperk\"ahler cones and quaternion-K\"ahler
  manifolds}, Phys. Lett. {\bf B511} (2001) 302, {{\tt
  hep-th/0104215}}.
%
\bibitem{Swann:1991} A. Swann, Math. Ann. {\bf 289} (1991) 421.
%
\bibitem{deWit:2006gn}
  B. de Wit and F. Saueressig, {\it Off-shell N=2 tensor
  supermultiplets}, JHEP {\bf 09} (2006) 062, {\tt hep-th/0606148}.
%
\bibitem{deWit:2003hr}
  B.~de Wit, H.~Samtleben and M.~Trigiante,
  {\it Gauging maximal supergravities},
  Fortsch.\ Phys.\  {\bf 52} (2004) 489
  [arXiv:hep-th/0311225].
%
\bibitem{deWit:2005hv} B.~de~Wit and H.~Samtleben, {\it Gauged maximal
    supergravities and hierarchies of nonabelian vector-tensor
    systems}, { Fortschr. Phys.} {\bf 53} (2005) 442,
  \href{http://xxx.lanl.gov/abs/hep-th/0501243}{{\tt hep-th/0501243}}.
%
\bibitem{deWit:2008ta}
  B.~de Wit, H.~Nicolai and H.~Samtleben,
  {\it Gauged Supergravities, Tensor Hierarchies, and M-Theory},
  JHEP {\bf 0802} (2008) 044
  [arXiv:0801.1294 [hep-th]].
%
\bibitem{D'Auria:1990fj}
  R.~D'Auria, S.~Ferrara and P.~Fr\'e, {\it Special and
  quaternionic isometries: General couplings in N=2 supergravity and
  the scalar potential}, Nucl. Phys. {\bf B359} (1991) 705.
%
\bibitem{LopesCardoso:2000qm}
  G.~Lopes Cardoso, B.~de Wit, J.~K\"appeli, T. Mohaupt,
  {\it Stationary BPS solutions in N=2 supergravity with $R^2$
    interactions},
  JHEP {\bf 0012} (2000) 019.
  [hep-th/0009234].
\bibitem{Romans:1991nq}
  L.J.~Romans,
  {\it Supersymmetric, cold and lukewarm black holes in cosmological
    Einstein-Maxwell theory},
  Nucl.\ Phys.\  {\bf B383 } (1992)  395-415.
  [hep-th/9203018].
%
\bibitem{Caldarelli:1998hg}
  M.M.~Caldarelli and D.~Klemm,
  {\it Supersymmetry of Anti-de Sitter black holes},
  Nucl.\ Phys.\  {\bf B545 } (1999)  434-460.
  [hep-th/9808097].
%
\bibitem{Gutowski:2004yv}
  J.B.~Gutowski and H.S.~Reall,
  {\it General supersymmetric AdS$_5$ black holes},
  JHEP {\bf 0404 } (2004)  048.
  [hep-th/0401129].
\end{thebibliography}
\end{document}